\documentclass[twocolumn,twocolappendix]{aastex63}

\include{aas_macros}

\received{June 22, 2021}
\revised{Sept 7, 2021}
\accepted{Sept 26, 2021}
\submitjournal{ApJ}

\shorttitle{Coorbiting M31 Satellites \& $\Lambda$CDM Simulations}
\shortauthors{Pawlowski \& Sohn}

\begin{document}

\title{On the Co-Orbitation of Satellite Galaxies Along the Great Plane of Andromeda:\\ NGC 147, NGC 185, and Expectations from Cosmological Simulations}

\correspondingauthor{Marcel S. Pawlowski}
\email{mpawlowski@aip.de}

\author[0000-0002-9197-9300]{Marcel S. Pawlowski}
\affiliation{Leibniz-Institut f\"ur Astrophysik Potsdam (AIP), An der Sternwarte 16, D-14482 Potsdam, Germany}

\author[0000-0001-8368-0221]{Sangmo Tony Sohn}
\affiliation{Space Telescope Science Institute, 3700 San Martin Drive, Baltimore, MD 21218, USA}

\begin{abstract}
Half of the satellite galaxies of Andromeda form a narrow plane termed the Great Plane of Andromeda (GPoA), and their line-of-sight velocities display correlation reminiscent of a rotating structure. Recently reported first proper motion measurements for the on-plane satellites NGC\,147 and NGC\,185 indicate that they indeed co-orbit along the GPoA. This provides a novel opportunity to compare the M31 satellite system to $\Lambda$CDM expectations. We perform the first detailed comparison of the orbital alignment of two satellite galaxies beyond the Milky Way with several hydrodynamical and dark-matter-only cosmological simulations (Illustris TNG-50, TNG-100, ELVIS, PhatELVIS), in the context of the Planes of Satellite Galaxies Problem. In line with previous works, we find that the spatial flattening and line-of-sight velocity correlation alone is already in substantial tension with $\Lambda$CDM, with none of the simulated analogs simultaneously reproducing both parameters. Almost none (3 to 4\%) of the simulated systems contain two satellites with orbital poles as well aligned with their satellite plane as indicated by the most-likely proper motions of NGC\,147 and NGC\,185. However, within current measurement uncertainties it is common ($\approx$70\%)  that the two \textit{best-aligned} satellites of simulated systems are consistent with the orbital alignment. Yet, the chance that \textit{any} two simulated on-plane satellites have as well aligned orbital poles as observed is low ($\approx$4\%). We conclude that confirmation of the tight orbital alignment for these two objects via improved measurements, or the discovery of similar alignments for additional GPoA members, hold the potential to further raise the tension with $\Lambda$CDM expectations.
\end{abstract}




\section{Introduction} \label{sec:intro}

The search and discovery of dwarf galaxies via modern observational surveys has resulted in an ever-increasing census of known satellites belonging to the the Milky Way \citep{2019ARA&A..57..375S, 2007ApJ...654..897B, 2015ApJ...807...50B, 2015ApJ...805..130K, 2015ApJ...813..109D, 2015ApJ...813...44L, 2018MNRAS.475.5085T},
 the neighboring Andromeda Galaxy M31 \citep{2009ApJ...705..758M, 2011ApJ...732...76R, 2013ApJ...772...15M,  2013ApJ...779L..10M,  2021arXiv210403859M}, 
 as well as around more distant host galaxies \citep{2014ApJ...787L..37M, 2014ApJ...795L..35C, 2015AstBu..70..379K, 2016A&A...588A..89J, 2017ApJ...847....4G, 2017A&A...602A.119M, 2017ApJ...848...19P, 2018ApJ...863..152S, 2018ApJ...867L..15T, 2018A&A...615A.105M, 2019A&A...629A..18M, 2020ApJ...891..144C,  2021MNRAS.500.3854D, 2021ApJ...907...85M}.

Considering the overall system of satellite galaxies around a host has revealed puzzling features in their distribution. Already \citet{1976MNRAS.174..695L} and \citet{1976RGOB..182..241K} have reported that the then-known satellites of the Milky Way are preferentially distributed along a flattened plane that is oriented perpendicular to the MW disk itself. This highly anisotropic distribution has since been further supported with the discovery of additional satellite galaxies in a similar distibution \citep{2005A&A...431..517K, 2009MNRAS.394.2223M, 2015MNRAS.453.1047P}. Survey footprint effects, while affecting the distribution of \textit{discovered} satellites, have been found to be insufficient to account for this pecularity \citep{2016MNRAS.456..448P}. 

\begin{figure}[ht!]
\plotone{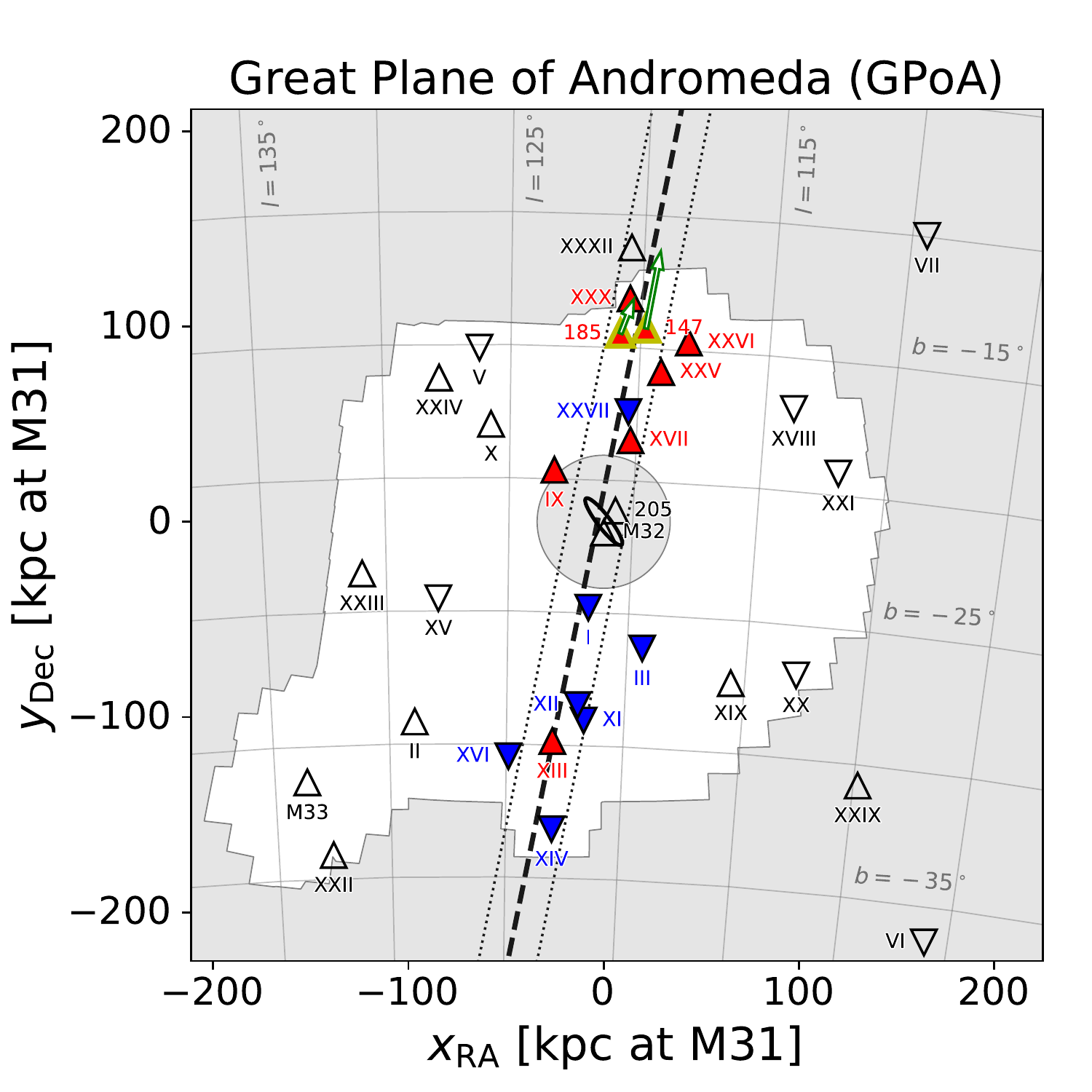}
\caption{Observed distribution of satellite galaxies around M31, with line-of-sight velocities relative to that of M31 indicated by upward (downward) facing triangles for satellites receding from (approach) us relative to M31's sytemic velocity. The on-plane satellites identified as GPoA members by \citet{2013Natur.493...62I} are plotted with filled and colored symbols (red is receding, blue approaching). The essentially edge-on GPoA is indicated with the dashed line, its width with dotted lines. The two GPoA members NGC\,147 and NGC\,185 for which proper motions have been measured in \citet{2020ApJ...901...43S} are highlighted in yellow, and their most-likely on-sky velocity direction relative to M31 is indicated with green arrows.
\label{fig:GPoA}}
\end{figure}

Similar structures of flattened satellite galaxy distributions were discovered around more distant host galaxies, most prominently for the Andromeda Galaxy M31 \citep{2013Natur.493...62I, 2013ApJ...766..120C} and Centaurus\,A \citep{2015ApJ...802L..25T,2018Sci...359..534M}. In both cases, the satellite galaxies display a high degree of correlation in their line-of-sight velocities along the edge-on orientation of their respective satellite plane: satellites on one side of the satellite plane preferentially recede, while those on the other side approach relative to the host, reminiscent of a rotating structure \citep{2013Natur.493...62I, 2018Sci...359..534M, 2021A&A...645L...5M}. Figure \ref{fig:GPoA} gives an overview of the observed Great Plane of Andromeda (GPoA).

Thus far, full three dimensional velocity information was only accessible for the Milky Way satellite galaxies, by combining spectroscopic measurements of their line-of-sight velocities with proper motion measurements, largely based on Hubble Space Telescope (HST, e.g. \citealt{2006AJ....131.1445P, 2013ApJ...764..161K, 2013ApJ...768..139S, 2015AJ....149...42P, 2017ApJ...849...93S}) data and -- in recent years -- data from the Gaia mission \citep{2018A&A...616A..12G, 2018ApJ...863...89S, 2018A&A...619A.103F, 2021arXiv210403974L}. This has revealed that among the 11 brightest, classical satellite galaxies of the Milky Way, 8 orbit along their spatial plane \citep{2008ApJ...680..287M,2013MNRAS.435.2116P,2019MNRAS.488.1166S,2020MNRAS.491.3042P}. Numerous fainter satellite galaxies have since also been found to co-orbit along the VPOS \citep{2018A&A...619A.103F, 2021arXiv210403974L}, some but not all of which can be associated as satellites to the Magellanic Clouds \citep{2018ApJ...867...19K, 2020ApJ...893..121P, 2020MNRAS.495.2554E}.

Recently, \citet{2020ApJ...901...43S} published the first proper motion measurements based on HST data for two M31 satellite galaxies that are members of the GPoA: NGC\,147 and NGC\,185. They report these two galaxies to have substantially different orbits, with NGC\,147 showing a recent, close pericenter passage with M31 which is consistent with the existence of tidal features around this object. Yet, both satellites' orbital planes were found to likely be well aligned with the spatial plane defined by the GPoA (see also Fig. \ref{fig:GPoA}), a striking similarity to the preferred orbital alignment of the classical Milky Way satellites with the VPOS.

The phase-space distribution of satellite galaxy systems can be considered a robust test of cosmological predictions. This is because the position and motion of a satellite galaxy is not strongly affected by baryonic effects internal to it. The overall spatial distribution and motion of the system of satellite galaxies surrounding a host is therefore robust against details in how baryonic physics are implemented and modelled in numerical simulations. 

Since satellite galaxy systems are well within the highly non-linear regime, most predictions relavant in this context are derived from cosmological simulations based on the $\Lambda$CDM paradigm \citep{2017ARA&A..55..343B}.
Comparing both the spatial flattening and the kinematic correlations of the observed planes of satellite galaxies with such expectations has revealed them to be exceedingly rare. Only of order 1 to 10 in 1000 simulated satellite systems are found to be both as flattened and as kinematically correlated as the respective observed systems \citep{2014ApJ...784L...6I, 2018Sci...359..534M, 2019MNRAS.488.1166S, 2020MNRAS.491.3042P, 2021MNRAS.504.1379S, 2021A&A...645L...5M}.
Those few sufficiently flattened planes that are identified in cosmological simulations are generally found to be transient alignments \citep{2016MNRAS.460.4348B,  2017MNRAS.466.3119A, 2019MNRAS.488.1166S, 2021MNRAS.504.1379S}. Their satellites tend to show large tangential velocities perpendicular to the plane, the planes are thus expected to be unstable and can not be considered to be rotationally supported. Even studies which specifically select for high line-of-sight velocity correlation among their simulated analogs of the apparently rotating GPoA typically find identified satellite structures to be transient.

\citet{2014MNRAS.438.2916B} have searched for satellite plane analogs in the Millennium II simulation. As a means to judge the stability of their identified satellite planes, they investigated the evolution of the plane height of the on-plane satellite sample before and after redshift $z=0$\ at which the planes were identified. The on-plane satellites were found to be in substantially wider distributions before and after $z=0$, which lead \citet{2014MNRAS.438.2916B} to conclude that the satellite galaxies can not align within a tight orbital plane, and that the planes they identified are predominantly statistical fluctuations of an approximately spherical satellite distribution.

Somewhat different conclusions were drawn by \citet{2015ApJ...800...34G}, who have focussed on an in-depth study of a single simulation within the CLUES project, containing two halos of approximate M31 mass, instead of a large sample of satellite systems. While not as significant as the observed GPoA, their identified flattened and kinematically coherent structures of satellite galaxies lead them to predict that such structures are dominated by a group of satellites with similar orbital directions. These constitute one half to two-thirds of their plane members, while the other one-third to half are aligned only by chance and will quickly fly out of the plane. We note that this scenario appears unlikely for NGC\,147 and NGC\,185, as \citet{2020ApJ...901...43S} demonstrate that they follow different orbits with different pericenters, which appears at odds with them both being a part of a common, coherent group of satellites.

\citet{2016MNRAS.460.4348B} follow an intermediate approach by studying 21 zoom simulations. Among their identified satellite planes, about one-third of the on-plane satellites display kinematics that make them chance alignments, and tracing the orbital evolution of the on-plane satellites reveals the planes to be transient features, in agreement with the findings by \citet{2015ApJ...800...34G} and \citet{2014MNRAS.438.2916B}, respectively. They conclude that some of the satellite galaxies of M31's GPoA should have high velocities out of the plane, and that line-of-sight velocities are insufficient to judge the kinematic coherence of a satellite plane.

While some of these studies have been critisized\footnote{Some sources of criticism were differences in how simulated satellites are selected compared to the observed sample, which can bias their spatial distribution, and identifying simulated analogs by requiring only an incomplete set of satellite plane properties to be matched \citep{2014MNRAS.442.2362P, 2014ApJ...784L...6I, 2019ApJ...875..105P}.}, they nevertheless provide a consistent prediction: even if flattened and possibly line-of-sight velocity correlated satellite planes resembling the GPoA are found in simulations, in general only some of the satellites are expected to orbit along the identified satellite plane, while a substantial fraction of satellites is expected to move out of the identified plane.

We here set out to further quantify this, by specifically comparing a range of satellite galaxy systems from several cosmological simulations to the information provided by the two first M31 satellites for which not only line-of-sight, but full 3D velocity information from proper motion measurements is available. We aim to quantify by how much cosmological expectations are in conflict with finding two highly correlated satellites whose orbits are well aligned with their satellite plane, and what is to be expected once proper motions for additional on-plane satellites become available. 
We will begin with quantifying the degree of alignment of the orbital poles of the two satellites with the spatially defined GPoA in Sect. \ref{sec:poles}. We will then present the analysed cosmological simulations and adopted methodology in Sect. \ref{sec:simulations}, and present our results in Sect. \ref{sec:simresults}. We conclude with a summary of our findings in Sect. \ref{sec:conclusion}.

\section{Proper Motions and Orbital Poles} 
\label{sec:poles}

For an accurate comparison of the GPoA with simulated satellite systems, we first evaluate the degree of orbital correlation for the two on-plane satellites for which proper motion measurements now exist. Following the standards set for the Milky Way satellite galaxy system, this analysis is based on the orientation of the orbital poles of satellite galaxies. In other words, we calculate the direction of their angular momentum vector around their host galaxy (in this case M31). This indicates the current orbital plane and direction of a satellite galaxy. It is, in contrast to orbit integrations, independent of assumptions such as on the host galaxy mass or shape of its gravitational potential. If an orbital pole is found to be close to the normal vector to the GPoA, then the satellite currently orbits along this plane defined by their spatial positions alone. Furthermore, if the orbital poles point into a common direction, the satellites co-orbit in the same orbital sense.

To ease comparison with the MW satellite plane and larger-scale structure of the Local Group \citep{2013MNRAS.435.1928P}, we work in a Galactic Cartesian coordinate system. The best-fit plane to the GPoA members has a normal vector pointing to $(l, b) = (205^\circ.8, 7^\circ.6)$\ \citep{2013MNRAS.435.1928P}. We adopt this as the orientation of the GPoA normal vector. A satellite with an orbital pole pointing in the same direction is orbiting exactly along this best-fit plane. Due to the finite height of the GPoA ($r_\perp = 12.6 \pm 0.6$\,kpc, \citealt{2013Natur.493...62I}), some scatter around this direction is expected. We adopt a distance of $779.0 \pm 18.5$\,kpc to M31 throughout this paper (see Table \ref{tab:observed}). 

\subsection{Measured proper motions and resulting alignment of orbital pole with GPoA normal}
\label{sec:observed}

\begin{figure*}
\gridline{\fig{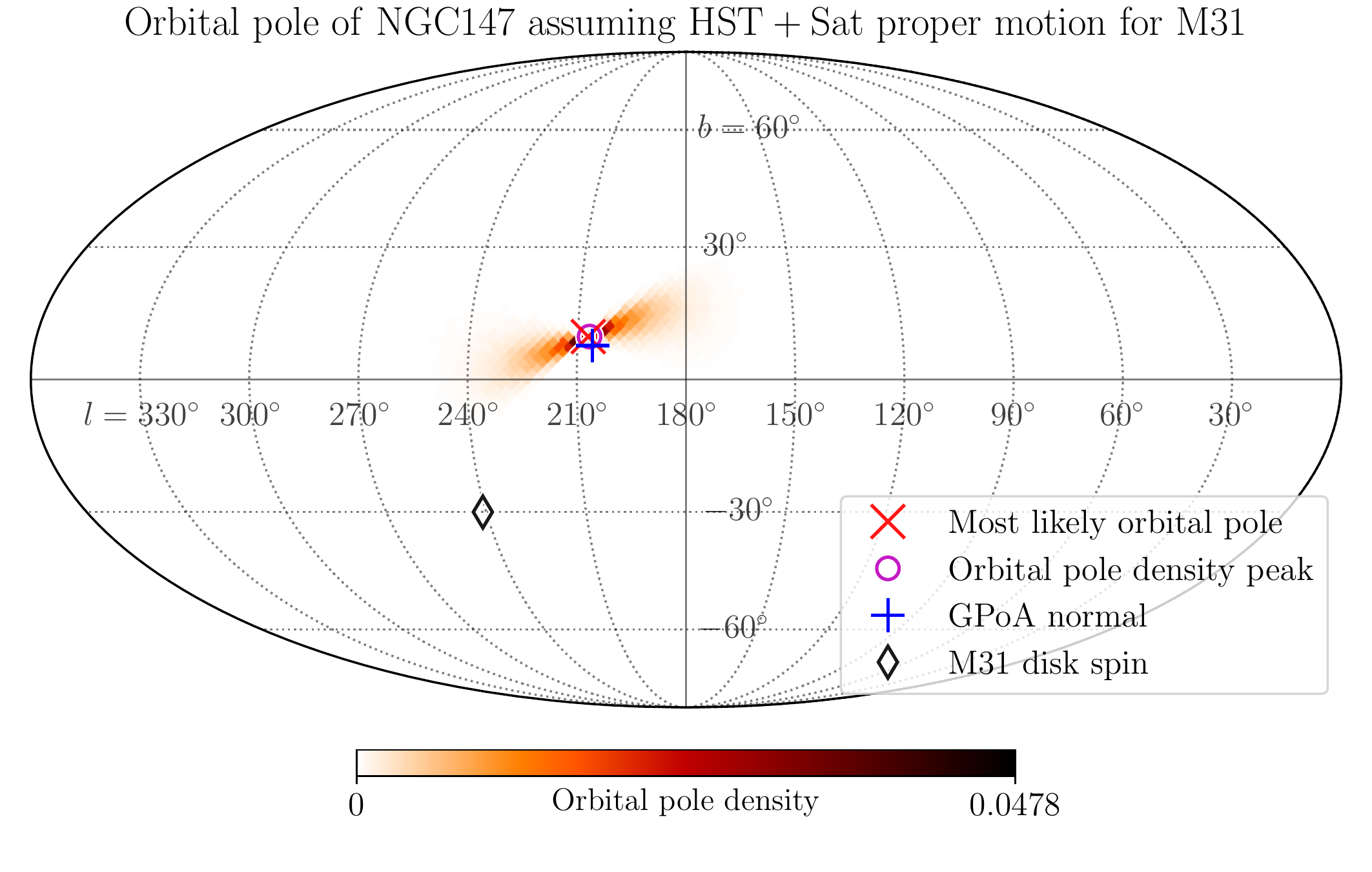}{0.5\textwidth}{(a)}
              \fig{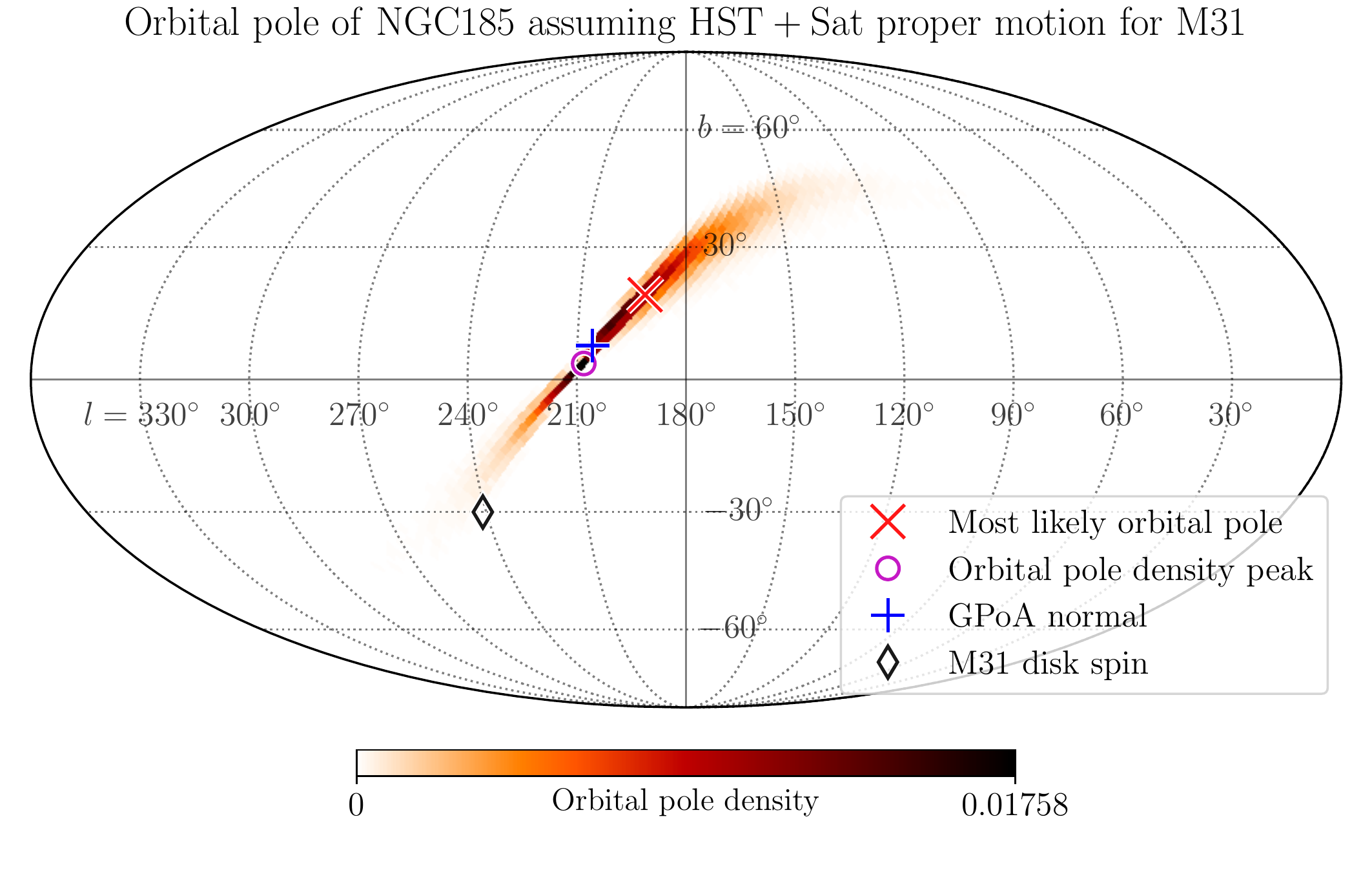}{0.5\textwidth}{(b)}
          }
\caption{Distribution of density of orbital poles in Galactic coordinates from 100,000 Monte-Carlo realizations drawing from the measurement uncertainties in the distances, line-of-sigh velocities, and proper motions of M31 and the two satellite galaxies NGC\,147 and NGC\,185. Also plotted are the orbital pole corresponding to the most-likely measured position and velocity of the satellites (red cross), the peak in the orbital pole density distribution (magenta circle), that normal vector to the GPoA which corresponds to the rotation direction inferred from the line-of-sight velocity coherence (blue plus sign), and the spin of the galactic disk of M31 itself (black diamond).
\label{fig:orbitalpoles}}
\end{figure*}

\begin{figure}[ht!]
\plotone{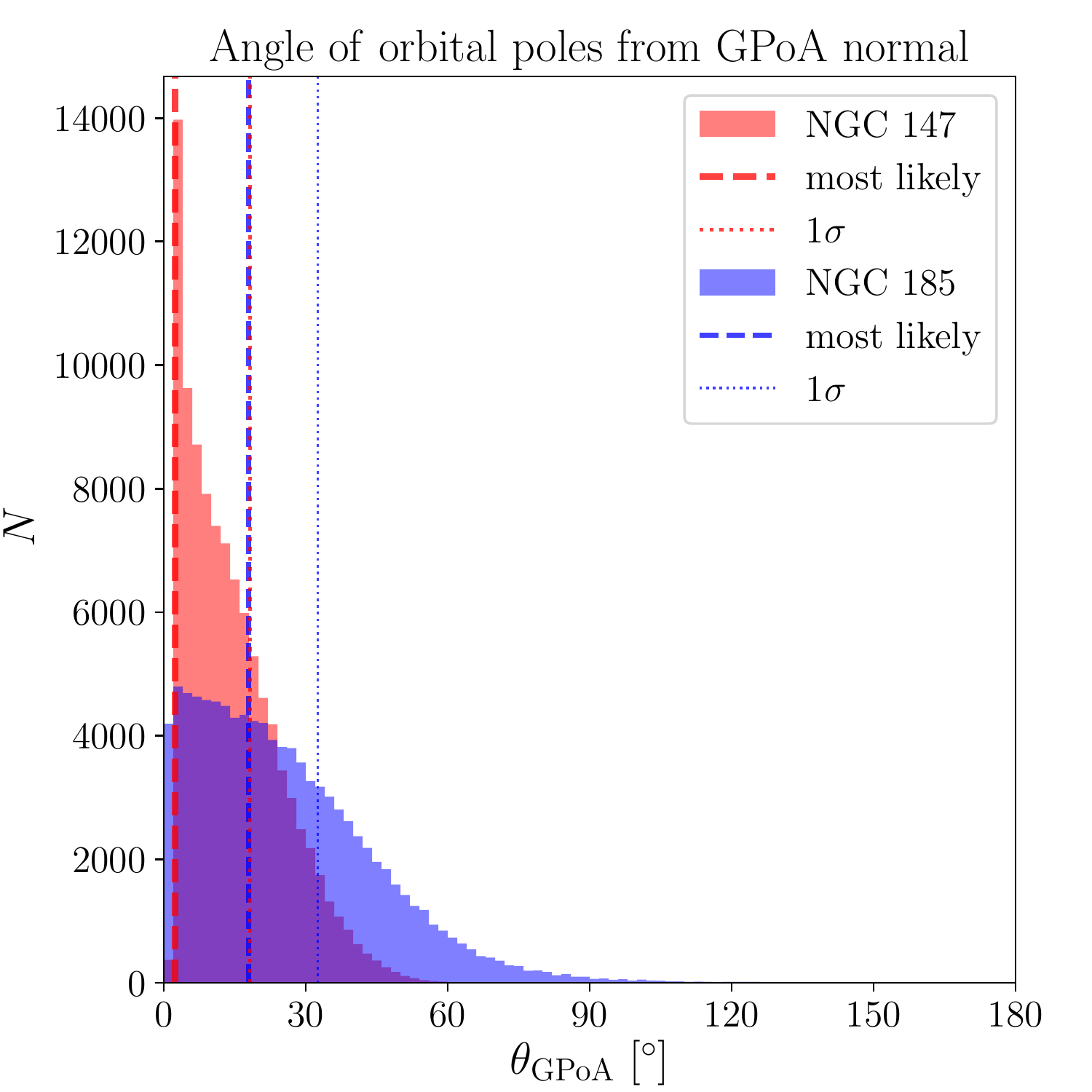}
\caption{Distribution of angle between the orbital pole of NGC\,147 (red) and NGC\,185 (blue) and the (co-orbiting) GPoA normal vector for 100,000 Monte-Carlo samplings from measurement uncertainties, assuming the HST+Sats. proper motion for M31. The dashed lines indicate the angle assuming the most-likely measured distance, line-of-sight velocity, and proper motion for the satellite and M31.
\label{fig:anglehist}}
\end{figure}

\begin{deluxetable*}{lccccccc}
\tablenum{1}
\tablecaption{Adoped observational measurements fot M31 and its satellite galaxies NGC147 and NGC185 \label{tab:observed}}
\tablewidth{0pt}
\tablehead{
\colhead{Object} & \colhead{$\alpha$} & \colhead{$\delta$} & \colhead{$d$} &
\colhead{$v_\mathrm{los}$} & \colhead{$\mu_\alpha \cos \delta$} & \colhead{$\mu_\delta$} & \colhead{References} \\
\colhead{} & \colhead{$(^\circ)$} & \colhead{$(^\circ)$} & \colhead{$(\mathrm{kpc})$} &
\colhead{$(\mathrm{km}\,\mathrm{s}^{-1})$} & \colhead{$(\mu\mathrm{as}\,\mathrm{yr}^{-1})$} & \colhead{$(\mu\mathrm{as}\,\mathrm{yr}^{-1})$}  & \colhead{}
}
\startdata
Satellites &  &  &  &  &  &  &  \\
~~NGC147 & 8.301 & 48.509 & $712.0 \pm 20.0$ & $-193.1 \pm 0.8$ & $23.2 \pm 14.3$ & $~37.8 \pm 14.6$ & (1), (2)  \\
~~NGC185 & 9.742 & 48.337 & $620.0 \pm 18.5$ & $-203.8 \pm 1.1$ & $24.2 \pm 14.1$ & $~~5.8 \pm 14.7$ & (1), (2)  \\
\hline
M31 & 10.685 & 41.269 & $779.0 \pm 18.5$ & $-300.1 \pm 3.9$ & & & (1) \\
~~HST+Sats. (fiducial) &  &  &  &  & $  34.3 \pm ~8.4$ & $-20.2 \pm ~7.8$ & (3) \\
~~HST+Gaia DR2 &  &  &  &  & $49.0 \pm 11.0$ & $-38.0 \pm 11.0$ & (4) \\
~~HST only &  &  &  &  & $45.0 \pm 13.0$ & $-32.0 \pm 12.0$ & (3) \\
~~Gaia eDR3 &  &  &  &  & $49.0 \pm 10.5$ & $-36.9 \pm ~8.0$ & (5) \\
~~HST+Gaia eDR3 &  &  &  &  & $47.4 \pm ~8.2$ & $-35.3 \pm ~6.7$ & weighted average \\
~~HST+Stats.+Gaia eDR3 &  &  &  &  & $40.1 \pm ~6.6$ & $-28.3 \pm ~5.6$ & weighted average \\
~~Sats. only &  &  &  &  & $~8.5 \pm 19.2$ & $~~5.2 \pm 16.4$ & (6) \\
~~Sats. only, no GPoA &  &  &  &  & $ 17.7 \pm 19.4$ & $-21.7 \pm 17.6$ & (6) \\
\enddata
\tablecomments{
References: (1): \citet{2012ApJ...758...11C}, (2): \citet{2020ApJ...901...43S}, (3): \citet{2012ApJ...753....8V},  (4): \citet{2019ApJ...872...24V},  (5): \citet{2020arXiv201209204S}, (6): \citet{2016MNRAS.456.4432S}. 
}
\end{deluxetable*}

\begin{deluxetable*}{lcccccccc}
\tablenum{2}
\tablecaption{Alignment of Orbital Poles with GPoA normal vector\label{tab:alignments}}
\tablewidth{0pt}
\tablehead{
\colhead{} & \multicolumn4c{NGC147} & \multicolumn4c{NGC185} \\
\colhead{Adopted M31 PM} & \colhead{$\theta_\mathrm{0}$} & \colhead{$\theta_\mathrm{peak}$} & \colhead{$\theta_\mathrm{median}$} & \colhead{$\theta_{1\sigma}$} & \colhead{$\theta_\mathrm{0}$} & \colhead{$\theta_\mathrm{peak}$} & \colhead{$\theta_\mathrm{median}$} & \colhead{$\theta_{1\sigma}$} \\
\colhead{} & \colhead{$(^\circ)$} & \colhead{$(^\circ)$} & \colhead{$(^\circ)$} & \colhead{$(^\circ)$} & \colhead{$(^\circ)$} & \colhead{$(^\circ)$} & \colhead{$(^\circ)$} & \colhead{$(^\circ)$}
}
\startdata
HST+Sat. (fiducial) & $2.4$ & $2.2$ & $12.5$ & $18.2$ & $17.9$ & $4.6$ & $22.5$ & $32.5$ \\
HST+Gaia DR2 & $9.1$ & $2.2$ & $12.9$ & $18.7$ & $23.2$ & $13.8$ & $23.6$ & $31.3$ \\
HST only & $7.3$ & $2.2$ & $13.7$ & $19.9$ & $23.0$ & $11.9$ & $24.3$ & $33.5$ \\
Gaia EDR3 only& $9.4$ & $2.2$ & $12.9$ & $18.7$ & $23.8$ & $15.6$ & $24.1$ & $31.5$ \\
HST+Gaia EDR3& $8.5$ & $2.2$ & $12.0$ & $17.4$ & $23.3$ & $17.4$ & $23.5$ & $30.6$ \\
HST+Sat+Gaia EDR3& $4.0$ & $2.2$ & $10.9$ & $15.8$ & $19.9$ & $11.9$ & $20.8$ & $28.7$ \\
Sats. only (Salomon et al.) & $32.9$ & $2.2$ & $36.6$ & $49.0$ & $79.5$ & $6.4$ & $83.3$ & $107.9$ \\
Sats. only, no GPoA (Salomon et al.) & $19.4$ & $2.2$ & $23.6$ & $33.4$ & $14.2$ & $6.4$ & $31.7$ & $47.0$ \\
\enddata
\tablecomments{
$\theta_\mathrm{0}$: angle for most-likely distances, line-of-sight velocities, and proper motions.
$\theta_\mathrm{peak}$: angle between GPoA normal and density peak in orbital pole distribution from Monte Carlo realizations.
$\theta_\mathrm{median}$: median of angle between GPoA plane normal and orbital poles from Monte Carlo realizations.
$\theta_{1\sigma}$: one-sigma limit of angle between GPoA plane normal and orbital poles from Monte Carlo realizations.
}
\end{deluxetable*}

\begin{deluxetable*}{lcccccccc}
\tablenum{3}
\tablecaption{Alignment of Orbital Poles for alternative distance measurements\label{tab:alignmentsGehaDist}}
\tablewidth{0pt}
\tablehead{
\colhead{} & \multicolumn4c{NGC147} & \multicolumn4c{NGC185} \\
\colhead{Adopted M31 PM} & \colhead{$\theta_\mathrm{0}$} & \colhead{$\theta_\mathrm{peak}$} & \colhead{$\theta_\mathrm{median}$} & \colhead{$\theta_{1\sigma}$} & \colhead{$\theta_\mathrm{0}$} & \colhead{$\theta_\mathrm{peak}$} & \colhead{$\theta_\mathrm{median}$} & \colhead{$\theta_{1\sigma}$} \\
\colhead{} & \colhead{$(^\circ)$} & \colhead{$(^\circ)$} & \colhead{$(^\circ)$} & \colhead{$(^\circ)$} & \colhead{$(^\circ)$} & \colhead{$(^\circ)$} & \colhead{$(^\circ)$} & \colhead{$(^\circ)$}
}
\startdata
HST+Sat. (fiducial) & $2.6$ & $2.2$ & $13.5$ & $19.7$ & $16.5$ & $4.6$ & $21.8$ & $31.5$ \\
HST+Gaia DR2 & $9.0$ & $2.2$ & $13.8$ & $20.3$ & $22.5$ & $14.1$ & $23.1$ & $30.9$ \\
HST only & $7.1$ & $2.2$ & $14.7$ & $21.4$ & $22.1$ & $4.6$ & $23.8$ & $32.8$ \\
Gaia EDR3 only & $9.3$ & $2.2$ & $13.9$ & $20.2$ & $23.2$ & $15.9$ & $23.7$ & $31.1$ \\
HST+Gaia EDR3 & $8.3$ & $2.2$ & $12.9$ & $18.8$ & $22.5$ & $15.9$ & $23.0$ & $30.1$ \\
HST+Sat+Gaia EDR3 & $3.6$ & $2.2$ & $11.8$ & $17.3$ & $18.9$ & $4.6$ & $20.3$ & $28.2$ \\
\enddata
\tablecomments{
Same as Table \ref{tab:alignments}, but with distances from \citet{2015ApJ...811..114G} instead of \citet{2012ApJ...758...11C}.
}
\end{deluxetable*}

All adopted parameters as well as their one-sigma uncertainties are compiled in Table \ref{tab:observed}. For consistency with the observed GPoA parameters as reported by \citet{2013Natur.493...62I}, we adopt the distances to M31, NGC147, and NGC185 from \citet{2012ApJ...758...11C}. Note that \citet{2020ApJ...901...43S} used slightly different distances from \citet{2015ApJ...811..114G}. However, these are consistent within their reported measurement uncertainties with those of \citet{2012ApJ...758...11C}, and we have checked that adopting the distances by \citet{2015ApJ...811..114G} does not significantly change our later findings (the alignment of the orbital pole with the GPoA normal in fact becomes consistently tighter by about 0.5 to $1^\circ$\ for NGC\,185 if the distance by \citet{2015ApJ...811..114G} is adopted).

A choice which affects the results more is that of the proper motion of M31 itself, as this sets the overall motion of the system. We here adopt the HST+Sats proper motion as used in \citet{2020ApJ...901...43S} as our fiducial value.
However, for completeness we have also repeated our analysis using only the HST-based proper motion \citep{2012ApJ...753....8V}, a combination of direct proper motion measurements of HST and Gaia DR2 (HST+GaiaDR2) from \citet{2019ApJ...872...24V}, the M31 proper motions measured from Gaia eDR3 by \citet{2020arXiv201209204S}, and error-weighted means of the Gaia EDR3 proper motions with the HST as well as with the HST+Sats proper motion. The latter is an error-weighted average of three different, independent measurement techniques, and thus can be argued to constitute the best current knowledge of the overall systemic proper motion of M31. Finally, we also used the proper motions purely inferred from the satellite galaxy line-of-sight velocities (Sats. only) from \citet{2016MNRAS.456.4432S}, as well as their analysis excluding the on-plane satellites. The respective proper motions are compiled in Table \ref{tab:observed}.

Uncertainties (in the distance, line-of-sight velocities, and proper motions of both the satellites and M31) are in the following accounted for with a Monte-Carlo approach, by drawing 100,000 random realizations from normal distributions around the most-likely observed values with widths given by the reported errors. 

We calculate the orbital poles as follows: for a given Monte-Carlo realization, we determine the three-dimensional Cartesian positions of M31, $\mathbf{r}_\mathrm{M31}$, and the satellite, $\mathbf{r}_\mathrm{sat}$, in Galactic coordinates. We also calculate their three-dimensional velocities in the same coordinate system, $\mathbf{v}_\mathrm{M31}$\ and $\mathbf{v}_\mathrm{sat}$, respectively. We subtract the M31 position and velocity from the respective position and velocity of the satellites, and then calculate the orbital pole as the normalized cross product between the resulting relative position and velocity: 
$$
\mathbf{L}_\mathrm{pole} = \frac{\left(\mathbf{r}_\mathrm{sat} - \mathbf{r}_\mathrm{M31} \right) \times \left(\mathbf{v}_\mathrm{sat} - \mathbf{v}_\mathrm{M31} \right)}{\left| \left(\mathbf{r}_\mathrm{sat} - \mathbf{r}_\mathrm{M31} \right) \times \left(\mathbf{v}_\mathrm{sat} - \mathbf{v}_\mathrm{M31} \right) \right|}.
$$

Figure \ref{fig:orbitalpoles} plots the resulting density of orbital pole direction in an all-sky plot (in Galactic coordinates), in comparison with the direction of the GPoA normal (blue plus sign), for the 100,000 Monte-Carlo drawings. In Galactic coordinates, the co-orbiting direction of the GPoA normal vector (as inferred from the coherence in line-of-sight velocities) points to $(l, b) = (205^\circ.8, 7^\circ.6)$\ \citep{2013MNRAS.435.1928P}. 

The orbital pole densities tend to peak very close to the GPoA normal vector for both satellites. This can be understood as follows: We see the GPoA close to edge-on. If the proper motion component perpendicular to the GPoA approaches zero, then the orbital pole points in the direction of the GPoA normal (modulo the slight offset between the position of the satellite and the best-fit GPoA plane). Any variation in the proper motion component along the GPoA, or in the distance of the satellite which is also oriented along the GPoA, from our Monte-Carlo approach then does not affect the direction of the orbital pole. As a consequence, many different realizations end up with almost the same orbital pole direction close to the GPoA normal, making a close alignment the most-likely one.

This can also be seen in Fig. \ref{fig:anglehist}, which plots the angle $\theta_\mathrm{GPoA}$\ between the GPoA normal vector and the orbital pole of the satellite galaxy. Both distributions peak at very small $\theta_\mathrm{GPoA}$. We determine the most-likely, the median and the one-sigma upper limit of the alignment angle between the orbital poles and the GPoA normal. Table \ref{tab:alignments} lists these angles as well as the one for the most-likely values of distance, line-of-sight velocities, and proper motions, while Table \ref{tab:alignmentsGehaDist} lists the alignments if we adopt the alternative distances from \citet{2015ApJ...811..114G}.

The direct measurements involving the HST and Gaia data for the M31 proper motion give similar results: both NGC\,147 and NGC\,185 are consistent with closely orbiting along the GPoA. The former tends to result in better aligned orbital poles if the most-likely proper motions are assumed. For all direct measurements, and also their respective combinations, NGC\,147 and NGC\,185 are well in agreement within the uncertainties to orbit as closely as possible along the GPoA. The smaller the uncertainties on the proper motion of M31 (e.g. from the uncertainty-weighted combinations of Gaia and HST), the smaller is the one-sigma upper limit of the alignement of the orbital poles of both NGC\,147 and NGC\,185. This again indicates how well the two satellites are apparently co-orbiting along the GPoA. The main limitation on knowing the exact degree of orbital alignment are the current proper motion uncertainties for both the satellites and M31 as their host. Future improved proper motion measurements can therefore be expected to result in an even tighter alignment.

In contrast to the direct proper motion measurements for M31, adopting the ``Sats. only'' values for M31's sytemic proper motion typically results in less clear orbital alignment of the two satellites with the GPoA. Especially for NGC\,185, the orbital pole direction is essentially unconstrained if adopting this systemic motion of M31, because within the uncertainties it could be on a radial orbit relative to M31. However, deriving the systemic proper motions of M31 from the line-of-sight velocities of its satellites requires the assumption that no systematic motion is present within the satellite population. This assumption is in conflict with the observed, strong line-of-sight velocity correlation for the GPoA members. Consequently, \citet{2016MNRAS.456.4432S} have repeated their analysis after excluding the GPoA members. The resulting proper motion of M31 (Sats. only, no GPoA) is in better agreement with the more direct measures, and the inferred orbital poles also align better with the GPoA normal. However, the estimate is highly uncertain due to the low number of considered satellite galaxies.

In the following, we adopt the alignment angles of the HST+Sat. proper motion of M31 for our comparison with cosmological simulations in Sect. \ref{sec:simulations}. One reason for this choice is to be consistent with \citet{2020ApJ...901...43S}. More importantly, this measurement results in the tightest alignment for NGC\,147 for the most-likely proper motion, but also in relatively large one-sigma limits for both satellites (especially compared to the combined measurements involving HST and Gaia EDR3 data for the proper motion of M31), thereby conveniently bracketing the range of alignment angles of the various tested M31 proper motions listed in Table \ref{tab:alignments}. We note that none of our results in Sect. \ref{sec:simulations} depends strongly on these choices.

\subsection{Predicting proper motions by assuming orbital pole alignment}
\label{sec:predicted}


\begin{figure*}
\gridline{\fig{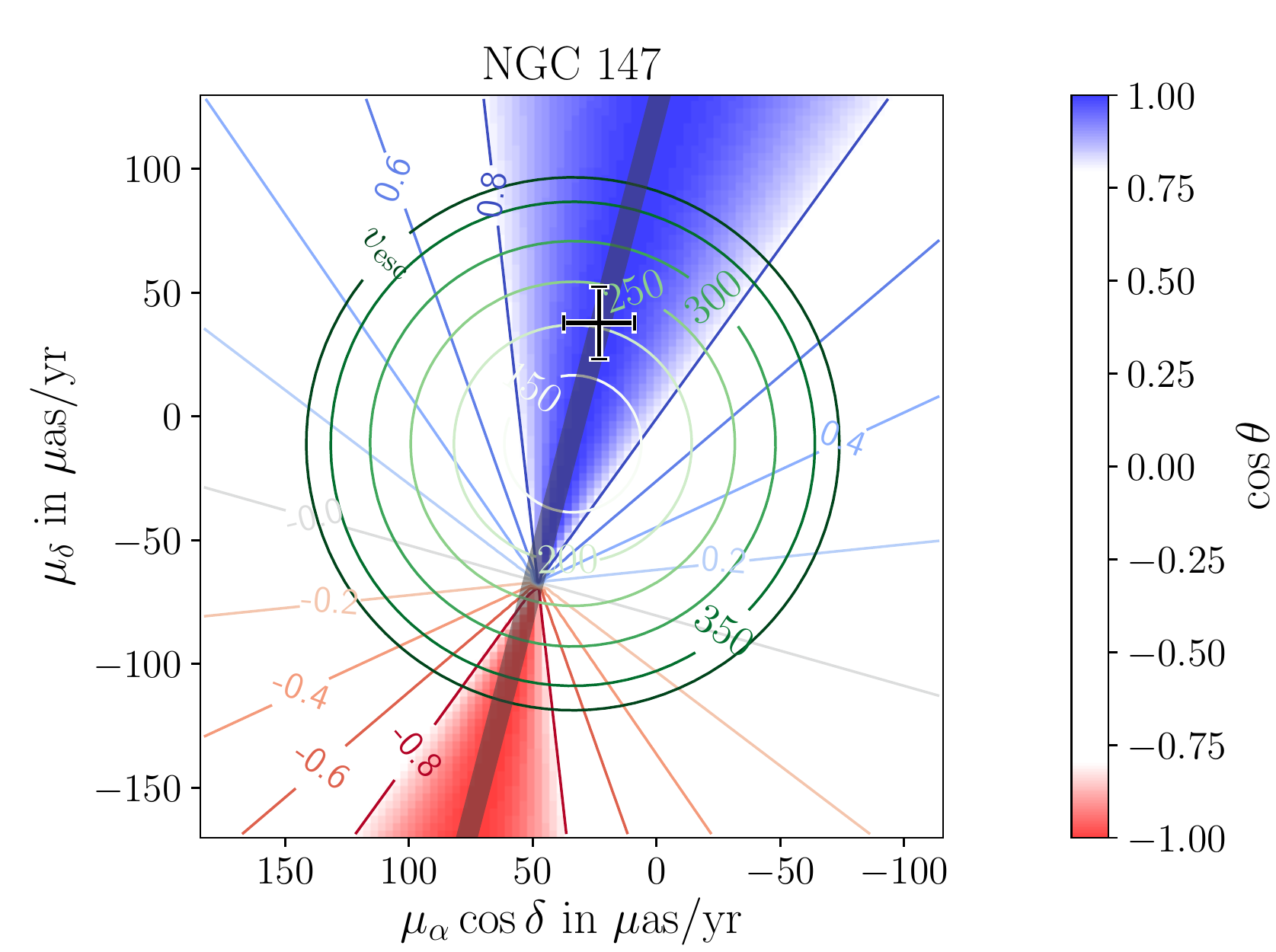}{0.5\textwidth}{(a)}
              \fig{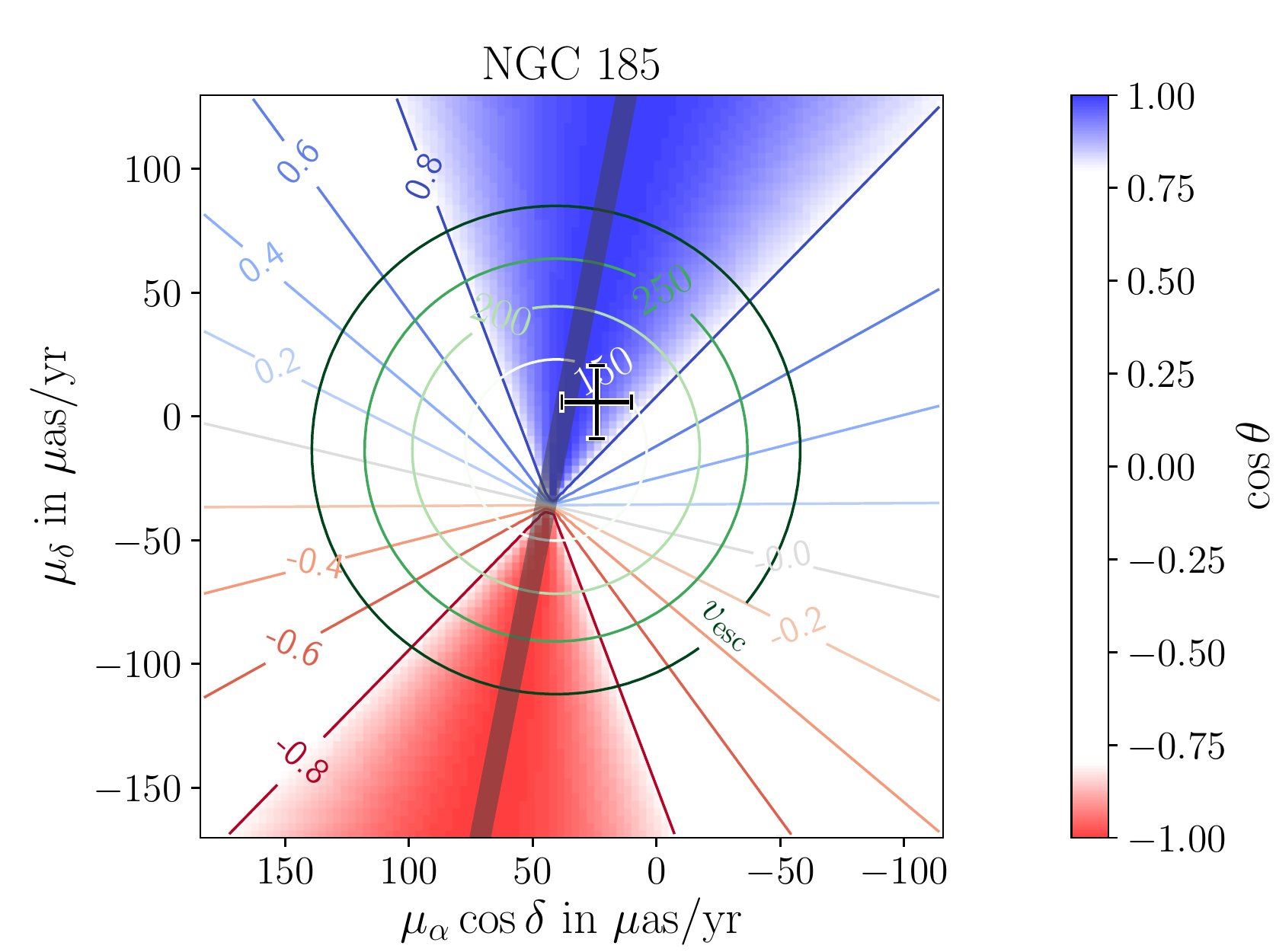}{0.5\textwidth}{(b)}
          }
\caption{Predicted proper motions for NGC\,147 (left panel a) and NGC\,185 (right panel b) if they are orbiting along the GPoA. These predictions assume the HST+Sats. proper motion for M31. Radial contours indicate the cosine of the angle $\theta$\ between the resulting orbital pole with the GPoA normal vector, where co-orbiting regions are shaded in blue and counter-orbiting regions in red. Here, co-orbiting is defined as following the rotational direction inferred from the line-of-sight velocity trend observed for 13 out of the 15 original GPoA satellites in \citet{2013Natur.493...62I}, as determined in \citet{2013MNRAS.435.1928P}.
The green circles indicate absolute velocities in the M31 rest frame. They are centered on those proper motions which in the M31-frame correspond to a motion along our line-of-sight. This is offset from the convergence point of the radial contours indicating the orbital alignment with the GPoA, because of the substantial measured line-of-sight velocities of the satellite galaxies relative to M31, which ensures a minimum absolute velocity. The outermost, dark-green circle indicates the escape velocity for a point mass of mass $m = 2 \times 10^{12}\,M_\sun$. This constitutes an estimate for the maximum predicted absolute velocity under the requirement that the satellites are gravitationally bound to M31.
 The measured proper motions for NGC\,147 and NGC\,185 from \citet{2020ApJ...901...43S} are indicated with black error bars. They are well within the predicted, co-orbiting regions. The satellites are thus both consistent with co-orbiting along the GPoA.
\label{fig:PMprediction}}
\end{figure*}

\begin{figure}[ht!]
\vspace{7mm}
\plotone{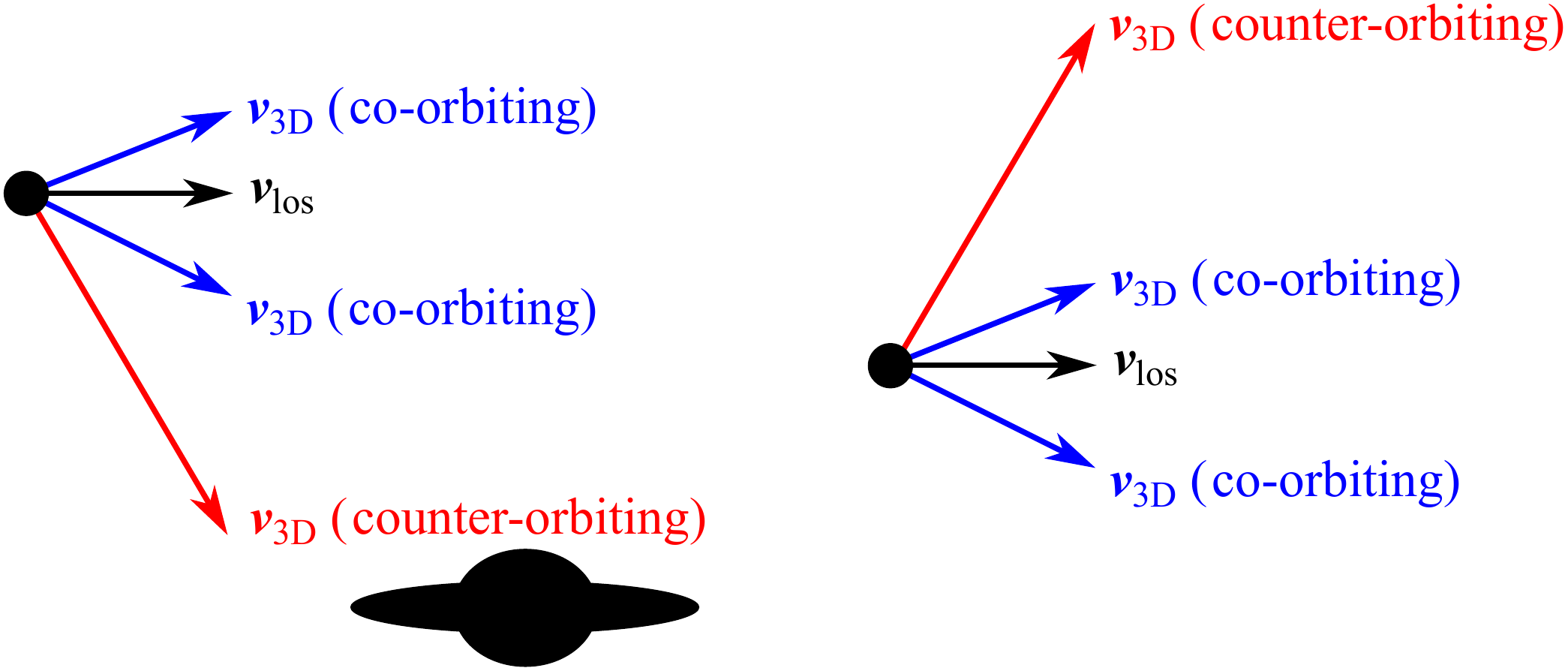}
\caption{
Sketch of two satellite galaxies (black dots) around their host (black shape), assuming a view onto their orbital plane. The line-of-sight components of the satellites (black arrows) are known and indicate them to co-orbit in the same direction (clockwise). However, this depends on their measured proper motions (assuming no motion out of the plane of the figure): while co-orbiting directions (blue arrows) are more probable as a larger number of possible proper motions would result in this orbital sense, counter-orbiting motions (red, counter-clockwise) are also possible. Though the latter could in some cases also imply unrealistically large velocities exceeding the escape speed from the host. This illustrates that apparently co-orbiting satellites, as judged from their line-of-sight velocities, could in fact be found to counter-orbit once full 3D velocities are measured.
\label{fig:sketch}}
\end{figure}

\citet{2013MNRAS.435.2116P} have predicted the proper motions of the then-known satellite galaxies of the Milky Way, based on the assumption that they are orbiting along the MW satellite galaxy plane. This purely geometric prediction was since confirmed for a number of MW satellite galaxies \citep{2018A&A...619A.103F, 2021arXiv210403974L}. Here, we adopt this method to predict the proper motions of those M31 satellite galaxies that are possibly associated with the GPoA. In this section we focus on NGC147 and NGC185, while the corresponding predictions for other satellites are compiled in appendix \ref{sect:appendix}, where Table \ref{tab:prediction} compiles, for all on-plane satellites, the best-possible alignment of the angle between a satellite galaxy orbital pole and the GPoA normal, given their positions.

We start by adopting the positions, most-likely distances and line-of-sight velocities for the satellite and M31 from Table \ref{tab:observed}. We also adopt the most-likely HST+Sat proper motion for M31. We then scan a grid of resonable proper motions in $\mu_\delta$\ and $\mu_\alpha \cos{\delta}$\ for the satellite galaxies, calculate the resulting 3D velocity relative to M31, absolute velocity $v_\mathrm{abs}$\ relative to M31, orbital pole around M31, and angle $\theta$\ between the orbital pole and our adopted GPoA normal vector. Assuming that the satellite galaxy co-orbits along the GPoA, the predicted proper motion is then those combinations of $\mu_\delta$\ and $\mu_\alpha \cos{\delta}$\ which result in small $\theta$ (we adopt $\left|cos \theta \right| > 0.8$). This method thus only relies on geometric considerations. In contrast to the proper motion predictions presented by \citet{2019MNRAS.488.3231H}, our approach does not rely on orbit integrations, and thus is independent of assumptions on the total mass and potential of M31. However, to ensure that the dwarf galaxies are indeed bound satellites of M31, we also require that their absolute velocities are below the escape speed for a point mass of $10^{12}\,M_\sun$ ($v_\mathrm{abs} < v_\mathrm{esc}$). Our predictions agree well with those by \citet{ 2019MNRAS.488.3231H}, despite the different techniques.

Figure \ref{fig:PMprediction} illustrates the predicted proper motions. The plots show all possible combination of proper motion components $\mu_\delta$\ and $\mu_\alpha \cos{\delta}$. The green circles indicate the absolute velocity of the satellites, with the largest one indicating the escape speed from a $10^{12}\,M_\sun$\ central mass at the satellite's distance from M31, an estimate for the maximum expected proper motions. The circles are centered on the proper motion of M31: the minimum absolute space motion for the satellite is acheived if its velocity difference with M31 is minimized, which happens if it follows the systemic proper motion of M31 (given that their line-of-sight velocity difference is fixed observationally for the prediction).

The radial contours in the plots indicate the cosine of the angle $\theta$\ between the orbital pole (for this proper motion combination) and the GPoA normal vector. The sign here is defined by the orbital sense of the GPoA as derived from the dominant line-of-sight velocity trend.
 If $\cos(\theta) > 0.8$, we consider the satellite to co-orbit along the GPoA (blue shaded regions, the orbital poles align to better than $\theta = 37^\circ$), while the red shaded region ($\cos(\theta) < -0.8$) indicates that a satellite would counter-orbit but also move along the GPoA plane. The blue shaded co-orbiting regions are slightly larger than the red counter-orbiting regions for NGC147 and NGC185, because their known line-of-sight velocities already indicate them to apparently co-orbit.

The black error bars in Fig. \ref{fig:PMprediction} are the observed proper motions of NGC147 and NGC185 from \citet{2020ApJ...901...43S}. They agree very well with the predicted range of proper motions for co-orbiting along the GPoA. They remain consistent with the predicted range for other direct M31 proper motion measurements, though the agreement becomes weaker, and essentially unconstrained for NGC\,185, when adopting the satellite-galaxy based proper motion estimates from \citet{2016MNRAS.456.4432S}.

It is also apparent from Fig. \ref{fig:PMprediction} that, even though the line-of-sight velocities of the two satellites indicate that they seem to co-orbit in the GPoA, in principle the proper motion measurements could also have resulted in counter-orbiting of the satellites. This is especially the case for NGC\,185, which is further in front of M31 than NGC\,147 such that a southern-pointing proper motion could more easily change its relative orbital sense compared to a case where it's velocity is oriented exclusively along the line-of-sight. This would have been the case if the $\mu_\delta$\ component  of the PM would have been more negative (red region in Fig. \ref{fig:PMprediction}). 
The underlying reason is that only the line-of-sight velocity was known, which, due to the edge-on orientation of the GPoA, is only one component within the putative orbital plane of the on-plane satellites. The proper motion can thus be split into one component along and one perpendicular to the GPoA\footnote{Since the GPoA is in an approximate north-south orientation in both Galactic and Heliocentric coordinates, the $\mu_\delta$\ component is largely along the plane, while the $\mu_\alpha$\ component is perpendicular to the plane.}. If the perpendicular component is close to zero the satellite moves along the GPoA. However, only for a close-to circular orbit does the line-of-sight velocity clearly indicate its orbital direction (see sketch in Fig. \ref{fig:sketch}).
Finding that both satellites indeed follow the orbital sense as hinted at by the line-of-sight velocities was thus not guaranteed. Consequently, this result alone already confirms an expectation based on interpreting the GPoA as a co-orbiting plane.

Note that the preceding discussion also imples that Andromeda\,XIII and XXVII, the two satellites of the spatial GPoA sample of \citet{2013Natur.493...62I} which do not follow the common line-of-sight velocity trend, might, in principle, be co-orbiting depending on what their proper motions are. See Table \ref{tab:prediction} and Figures \ref{fig:PMprediction_others} and \ref{fig:PMprediction_AndXXVII} in appendix \ref{sect:appendix} for a prediction of the required proper motions measurements. However, depending on the satellite's distance from M31 and its known line-of-sight velocity component, the required proper motion might imply that the full 3D velocity relative to M31 exceeds the escape speed. 
Since the dwarf galaxy would in this case not be bound to M31, we reject such predictions because we consider it unlikely that an unbound object would partake in the GPoA's coherent motion. This is the case for Andromeda\,XIII, which has a line-of-sight velocity counter to the overall velocity trend, and no possible proper motion which could result in a co-orbiting 3D velocity while simulataneously being bound to M31. For Andromeda\,XXVII, the feasibility of this scenario depends on the adopted distance of the dwarf to M31.

\section{Comparison to Simulations} 
\label{sec:simulations}

\subsection{The Simulations} 
\label{subsec:simintro}

The current orbits of both NGC\,147 and NGC\,185 have been found to align well with the GPoA. In contrast, satellite planes identified in cosmological simulations generally display a tendency to quickly disperse, because their satellites are predominantly moving out of the identified planes \citep{2015ApJ...800...34G, 2016MNRAS.460.4348B, 2021MNRAS.504.1379S, 2019MNRAS.488.1166S}.

We base our comparison on the publicly available hydrodynamic Illustris TNG50 \citep{2019ComAC...6....2N, 2019MNRAS.490.3234N, 2019MNRAS.490.3196P} and Illustris TNG100 simulations \citep{2019ComAC...6....2N,  2018MNRAS.475..648P,  2018MNRAS.475..676S}. These simulations are a compromise between sufficiently large box size ($L = 51.7$\ and $ 110.7$\,Mpc side length, respectively) to provide a large number of host galaxy systems and thus allow a statistical comparison, and resolution (dark matter particle mass of $m_\mathrm{dm} = 4.5 \times 10^5\,M_\sun$\ and $7.5 \times 10^6\,M_\sun$, respectively).

To ensure that resolution effects do not dominate our results, we have also investigated the dark-matter-only ELVIS \citep{2014MNRAS.438.2578G} and PhatELVIS \citep{2019MNRAS.487.4409K} zoom simulations ($m_\mathrm{dm} = 1.9 \times 10^5\,M_\sun$ and  $m_\mathrm{dm} = 3.0 \times 10^4\,M_\sun$, respectively). We here ignore the 12 PhatELVIS runs incorporating an analytical MW disk to ensure comparability with the ELVIS simulations, such that we consider a total of 60 host galaxies of MW and M31 mass.

\subsection{Selection of Satellite Systems} \label{subsec:simselection}

We first identify potential host galaxies in TNG50 (TNG100) by selecting all dark matter halos with virial mass in the range $m_{200} = 0.8\ \mathrm{to}\ 3.0 \times 10^{12}\,M_\sun$. There are 186 (1741) candidate hosts in this mass range. We further require these to be somewhat isolated by excluding any halo which has another halo of with $m_{200} > 0.5 \times 10^{12}$\ within $700\,\mathrm{kpc}$. We also exclude hosts with a subhalo that has a stellar mass exceeding 20 per cent of that of the host, to exclude ongoing major mergers, and finally we also exclude hosts too close to the simulation box boundary. This leaves us with 146 (1483) hosts.

We then extract all galaxies from the respective sub-halo catalog which lie within 350\,kpc of the host, and mock-observe them from a random direction and from a distance of 779\,kpc (our adopted distance to M31). To be directly comparable to the analysis by \citet{2013Natur.493...62I}, we identify all subhalos within the PAndAS footprint around the host, but outside of the inner $2^\circ.5$\ of the host. We rank them by their stellar mass, and then by their dark matter mass if they do not have a stellar component, and select the top 27 of these as the satellites of this particular system. Note that due to the limited resolution, we have to include dark subhalos in this analysis. However, since baryonic effects internal to the dwarf galaxies are not expected to affect their overall position and movement, this will not substantially affect our results. Nevertheless, it will be interesting to repeat this analysis for higher-resolution large-scale simulations in the future.

If, for a given random view direction, fewer than 27 satellites are identified in the PAndAS footprint, another random view is chosen until a sufficient number of satellites are found, or 1,000 random view directions have been tried. In the latter case, the host is discarded. In TNG100, this happens for 780 hosts, such that in the end we have 703 mock-observed satellite galaxy systems consisting of 27 satellite galaxies each. The superior resolution of TNG50 ensures that we find analogs in all 146 systems, with two thirds not even having to rely on dark subhalos.

To determine how much of an effect anisotropies and correlations within the $\Lambda$CDM simulation have versus what is introduced merely by the restrictions of mock-observing and being confined to the specific PAndAS footprint, we also generate a randomized sample from the selected satellite system. For each mock system, we randomize the satellite positions by drawing a new angular position around the host while keeping their radial distance constant. Similarly, each satellite is assigned a new velocity directions by drawing from an isotropic distribution. If the new projected position of a satellite falls outside of the PAndAS footprint or within $2^\circ.5$ of the host, we repeat the randomization until the new position is within the selection volume.

We have additionally  tested an alternative randomization procedure on the TNG50 simulation, in which the full satellite distribution around a simulated host is randomized first, and only then the sample selection based on the PAndAS footprint is applied to identify the set of 27 mock-observed satellites. These will then generally not all be identical to the satellites in the original mock-observed simulation sample, and thus the random sample's radial distribution will in differ from the mock-observed one, too. However, we find that this alternative randomization procedure provides very comparable results to the previously described randomization method. For optimal comparablity, we therefore opt to focus on the first approach, as it ensures an identical radial satellite distributions (which could otherwise biase the resulting plane parameters, see \citealt{2019ApJ...875..105P}).

For the ELVIS and PhatELVIS simulations, all 60 hosts are used and the satellites are selected following the same prescriptions.

\subsection{Analysis} 
\label{subsec:simanalysis}

We first have to identify the best satellite planes for each system with an algorithm that closely resembles that applied to the observed M31 satellites. This implies a two-step process, which first identifies the most narrow plane of 15 out of 27 satellite galaxies and then determines their velocity coherence\footnote{We here opt to select these numbers in order to be directly comparable to the observatonal study by \citet{2013Natur.493...62I} and the satellite plane parameters measured by them. This is because the number of satellites in a system to select plane members from, and the number of satellites required to be in a plane, biases the expected plane properties such as $r_\perp$\ \citep{2014MNRAS.442.2362P, 2017AN....338..854P, 2019ApJ...875..105P}. }.

To find the most narrow plane for each mock satellite system, we generate 10,000 trial plane normal vectors evenly distributed on the sphere. For each of these, we measure the offset of each of the 27 satellites from the plane defined by the corresponding normal vector direction and running through the center of the host. We rank the satellites by this offset, and measure the root-mean-square of the 15 smallest offsets. This is the rms height of the corresponding trial plane orientation. The trial plane with the smallest root-mean-square height is identified as the most-narrow satellite plane. We then use the common tensor-of-intertia method \citep{2015ApJ...815...19P} to fit a plane to the 15 closest satellites of this most-narrow trial plane, and measure the plane's rms height as this satellite system's satellite plane height $r_\perp$, their minor-to-major axis ratio $c/a$, and the angle of best-fit plane relative to the line-of-sight of the mock-observation $\phi_\mathrm{los}$.
For reference, for the observed M31 satellite galaxy system the corresponding root-mean-square height is $r_\perp(\mathrm{M31}) = 12.6 \pm 0.6 \,\mathrm{kpc}$\ \citep{2013Natur.493...62I}.

Once the most-narrow spatial plane is identified, we use it to split the volume around the host into two hemispheres. To determine the line-of-sight velocity coherence of the 15 on-plane satellites, we then only consider the mock-observed line-of-sight component $v_\mathrm{los}$\ of the satellite galaxy velocities relative to their host. Specifically, we calculate the cross product of the satellite position and line-of-sight velocity vectors relative to the host. We then count how many of the resulting angular momentum vectors point into each of the two hemispheres around the host. The number of velocity-correlated satellites $N_\mathrm{corr}$\ is then the larger of the two. For the observed M31 satellite system, this number is $N_\mathrm{corr}(\mathrm{M31}) = 13$\ out of 15 on-plane satellites \citep{2013Natur.493...62I}. 

Finally, using the full three-dimensional satellite galaxy velocities relative to their host, we calculate their full orbital poles and calculate for each on-plane satellite the angle $\theta_\mathrm{L}$\ between their orbital pole and the plane normal vector. Since the normal vector can have one of two opposite directions, we chose the one indicative of co-rotation as deduced from the line-of-sight velocity correlation. Thus, if all 15 satellites would co-orbit in the same sense along a common plane, they would have  $\theta_\mathrm{L} = 0^\circ$, whereas any perfectly counter-orbiting satellite would have  $\theta_\mathrm{L} = 180^\circ$.

\section{Results} 
\label{sec:simresults}

\begin{figure*}
\gridline{\fig{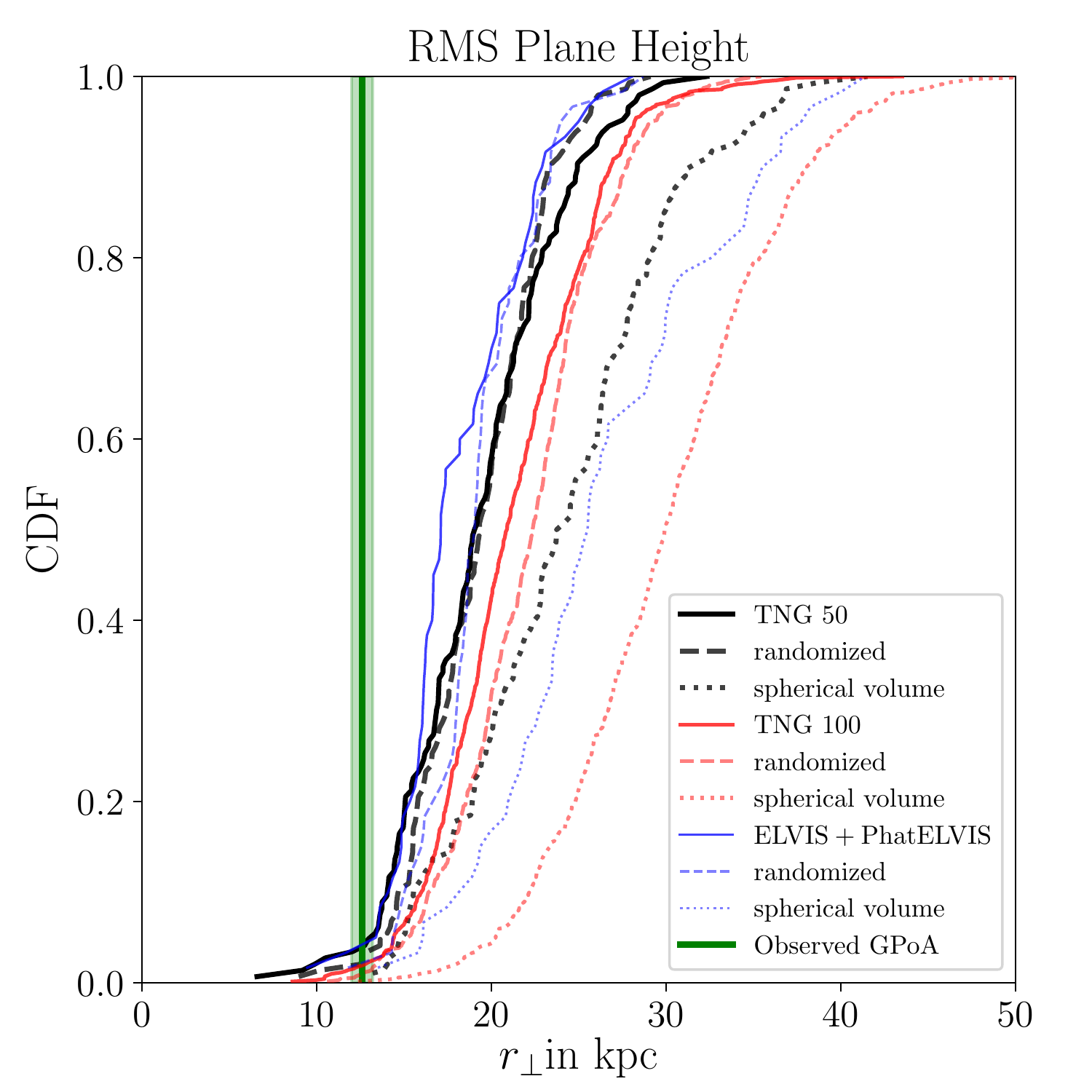}{0.33\textwidth}{(a)}
          \fig{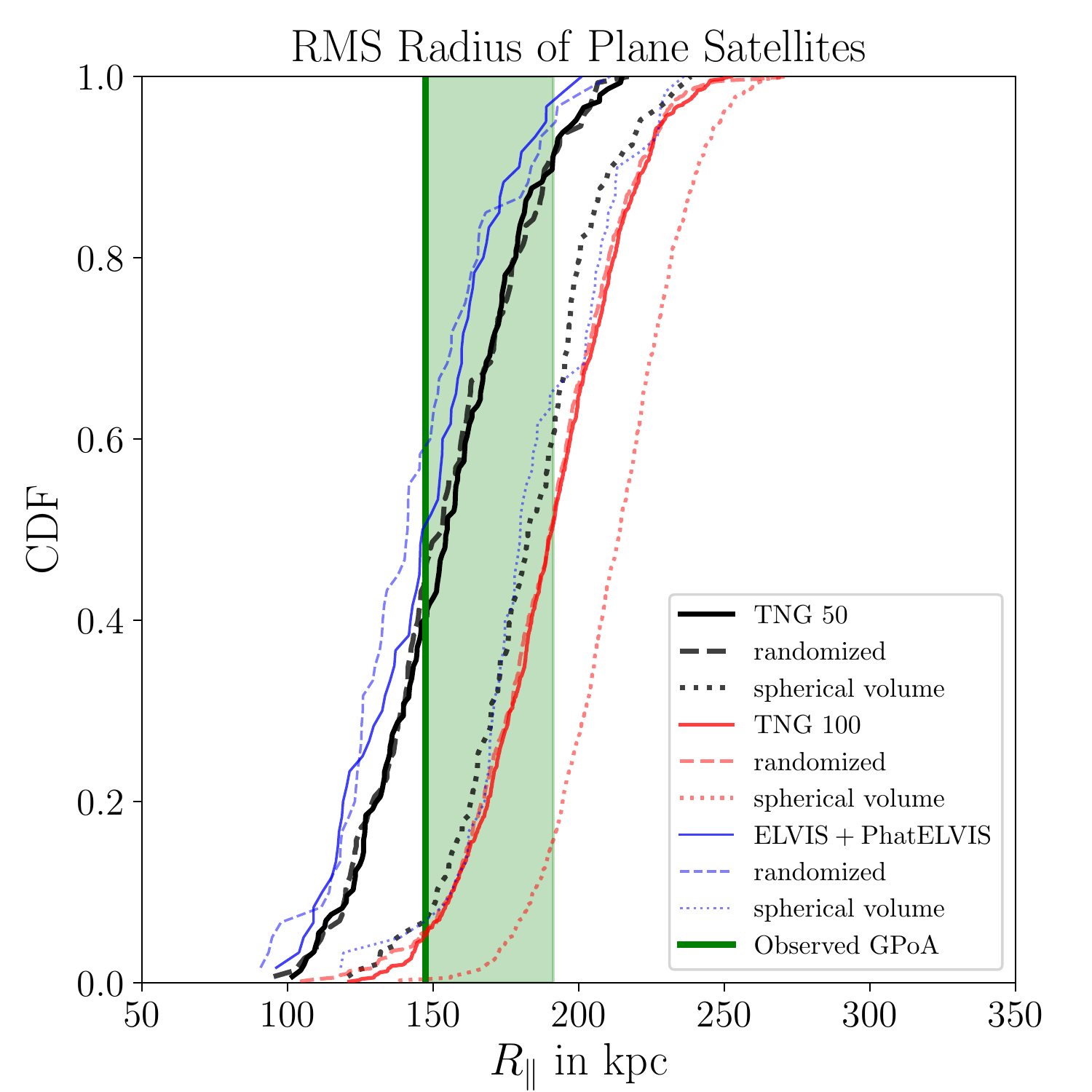}{0.33\textwidth}{(b)}
          \fig{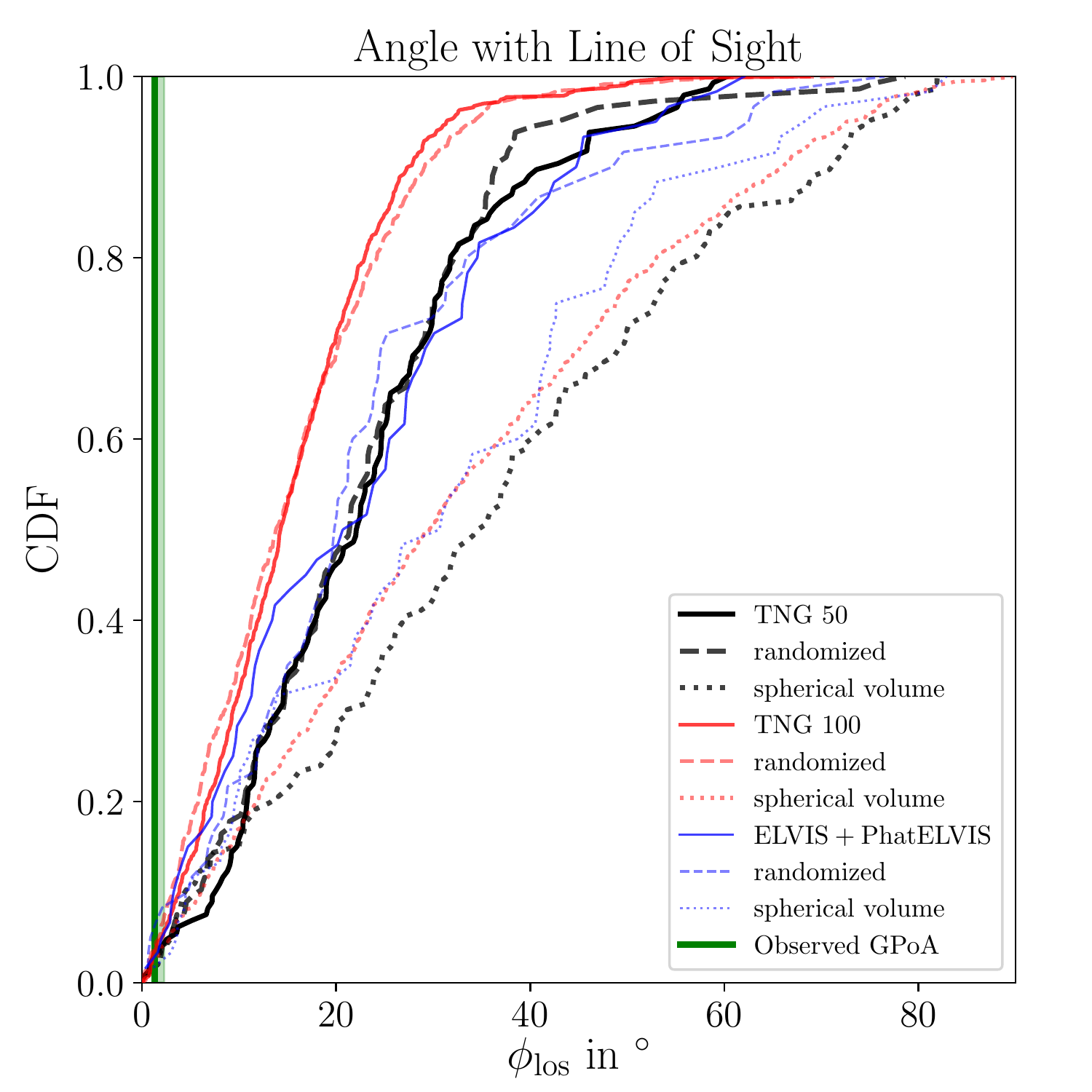}{0.33\textwidth}{(c)}
          }
\caption{Cumulative distributions of the properties of the most-narrow planes of 15 out of 27 satellites in the Illustris TNG 50, TNG 100, and ELVIS+PhatELVIS simulations. The solid and dashed lines are for satellite systems selected from within  within a mock PAndAS survey footprint, the dotted lines are for randomized satellite distributions within a spherical volume of 350\,kpc radius. The green band indicates the corresponding values for the observed satellite galaxy system within the observational uncertainties.
\label{fig:cumulativeplaneproperties}}
\end{figure*}

We now investigate the frequencies of simultaneously obtaining satellite plane parameters as extreme as those of the observed GPoA. We begin with the spatial distribution and line-of-sight velocities only, and then focus on the novel information provided by the close alignment of the orbital poles of NGC147 and NGC185 in the observed M31 system.

The cumulative distributions of plane heights $r_\perp$, root-mean-square radius of the on-plane satellites $r_\parallel$, and the angle of best-fit plane relative to the line-of-sight $\phi_\mathrm{los}$\ are plotted in Fig. \ref{fig:cumulativeplaneproperties}. Shown are the curves for the simulated satellite systems (solid lines) and for the corresponding randomized realizations (dashed lines), for the Illustris TNG50 (black), TNG100 (red), and the ELVIS+PhatELVIS (blue) simulations.

\subsection{Properties of the identified GPoA analogs}

As expected, the plane heights $r_\perp$\ of the simulated satellite systems are typically considerably wider than the observed value of $r_\perp = 12.6 \pm 0.6\ \mathrm{kpc}$\ \citep{2013Natur.493...62I}. Only 7 of the 146 TNG50 satellite systems, 17 of the 703 TNG100 satellite systems, and 2 of 60 for ELVIS+PhatELVIS satellite systems have a more narrow satellite plane than 13.2\,kpc. As has been demonstrated in many previous studies, the randomized satellite distributions, for which any intrinsic spatial correlations are removed, result in somewhat wider $r_\perp$, though the typical difference is only 2\,kpc at most. Selecting from a much wider spherical volume (as opposed to one restricted by the PAndAS footprint), here with a radius of 350\,kpc (dotted lines in Fig. \ref{fig:cumulativeplaneproperties}), results in even more radially extended satellite systems and correspondingly wider identified satellite planes. This illustrates the importance of considering the observational selection volume when comparing with simulations.

While $r_\perp$\ measures the height of a satellite plane, $r_\parallel$\ measures the radial extent of the on-plane satellite distribution as the root-mean-square of the satellite distances from the host, projected into the best-fit plane. Since the observed GPoA is seen edge-on, its radial extent is more affected by uncertainties in the measured distances to satellite galaxies than its height, and consequently $r_\parallel$\ is not very well constrained. This is in large part caused by the large distance uncertainty of Andromeda XXVII (see discussion in \citealt{2014MNRAS.442.2362P}).  Using the most-likely distance from the homogeneous set of distance measurements by \citet{2012ApJ...758...11C} we obtain $r_\parallel = 191\,\mathrm{kpc}$. If we instead adopt the lower limit of its distance from \citet{ 2011ApJ...732...76R},  $r_\parallel$\ becomes 147\,kpc. This range is indicated by the green area in the middle panel of Fig. \ref{fig:cumulativeplaneproperties}. 
From Fig. \ref{fig:cumulativeplaneproperties}, it is evident that the TNG100 satellite systems tend to have a larger radial extent, while the TNG50 and ELVIS simulations are in better mutual agreement. The TNG50 systems are on average slightly more radially extended, as expected due to the additional tidal disruption of subhalos in the inner halo region due to the additional presence of the central galaxy's potential, which is absent in the dark-matter-only ELVIS simulations \citep{2017MNRAS.471.1709G, 2019MNRAS.487.4409K}. However, all three sets of simulated systems have $r_\parallel$ that are in general agreement with the measured range of radial extent of the GPoA. This indicates that it is indeed the flattening of the observed satellite system, rather than its radial extent, which drives the planes of satellites galaxies problem. Despite the differences in the three sets of simulations, it will become clear in Sect. \ref{subsec:anglecomparison} that this does not affect the orbital alignment of the satellites with their satellite planes, for which all three sets give essentially identical results.

From our position, we see the GPoA edge-on, the satellite plane is aligned to within one degree with the Milky Way \citep{2013Natur.493...62I}. Such a near-perfect edge-on orientation is rare for the simulated systems, with only about 5 per cent of the mock-observed satellite systems being similarly close to an edge-on orientation. 
The observed, extreme orientation thus remains unlikely despite the survey-volume motivated spatial constraints that result in a preference towards more edge-on orientations. Because of this, we consider it improbable that the extreme observed alignment is merely driven by an observational effect. Instead, the extreme alignment can be interpreted as indicating that the M31 satellite plane is a real feature, and not a mere observational artifact.

While very extreme edge-on orientations are rare in the simulated systems, a majority of satellite planes identified in the simulation are nevertheless seen approximately edge-on. A total of 107 out of 146 (TNG 50), 657 out of 703 (TNG 100) and 42 out of 60 (ELVIS+PhatELVIS) simulated systems have $\phi_\mathrm{los} \leq 30^\circ$\ 
and almost all have $\phi_\mathrm{los} < 60^\circ$. 
In Appendix \ref{sect:appendix2}, we demonstrate that one can meaningfully infer the rotational signal of a satellite plane from the line-of-sight velocities of its satellites even at rather large inclinations (up to at least $\phi_\mathrm{los} > 60^\circ$) if a dynamically stable, co-orbiting satellite plane is assumed\footnote{If the satellites are not in a stable, rotating plane, then the line-of-sight velocities are not expected to show a meaningful correlation anyway.}. Consequently, the line-of-sight velocity components of the mock-observed simulated satellite systems can be meaningfully used to infer potential co-rotation signatures of the satellite planes.

The differences in $\phi_\mathrm{los}$\ between the TNG100 in contrast to the more similar typical orientation in the TNG50 and ELVIS+PhatELVIS simulations can be understood from their different radial distribution.
The TNG100 systems are more radially extended, in part due to the simulation's limited resolution and in part due to the inclusion of baryonic effects, predominantly the additional potential of the central host galaxy which will enhance tidal disruption, causing subhalos close to the host to be disrupted more easily. As a result, the satellites identified in the $z=0$\ simulation snapshot tend to be at larger distance from their host than for the higher-resolution simulations. A more extended satellite system tends to result in wider satellite planes \citep{2019ApJ...875..105P}. In addition, the shape of a more radially extended distribution is more defined by the mock survey footprint volume, which is essentually a cone elongated along the line-of-sight. Consequently, the spatial distribution of the identified satellite systems are also more defined by this volume, and identified satellite planes thus tend to be more closely aligned with selection volume, resulting in smaller $\phi_\mathrm{los}$. 

This is confirmed by comparing with randomized satellite systems selected from a spherical volume (dotted lines in Fig. \ref{fig:cumulativeplaneproperties}). These distributions do not have any preferred alignment, their cumulative distribution of $\phi_\mathrm{los}$\ thus follows a $\sin(\phi_\mathrm{los})$-behaviour as expected for random alignment angles, with only 50 per cent of all systems having $\phi_\mathrm{los} \leq 30^\circ$.

\begin{figure*}
\gridline{\fig{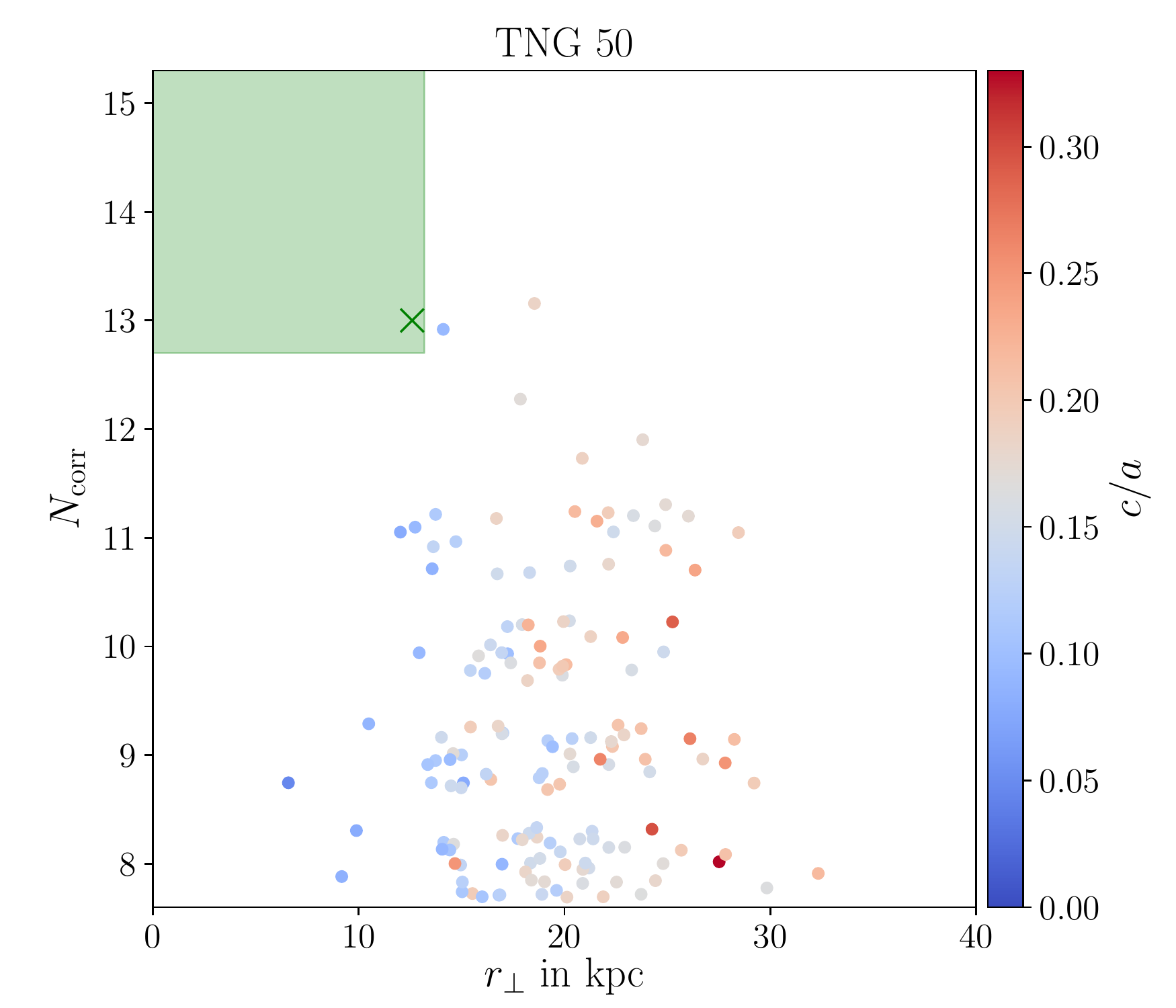}{0.33\textwidth}{(a)}
          \fig{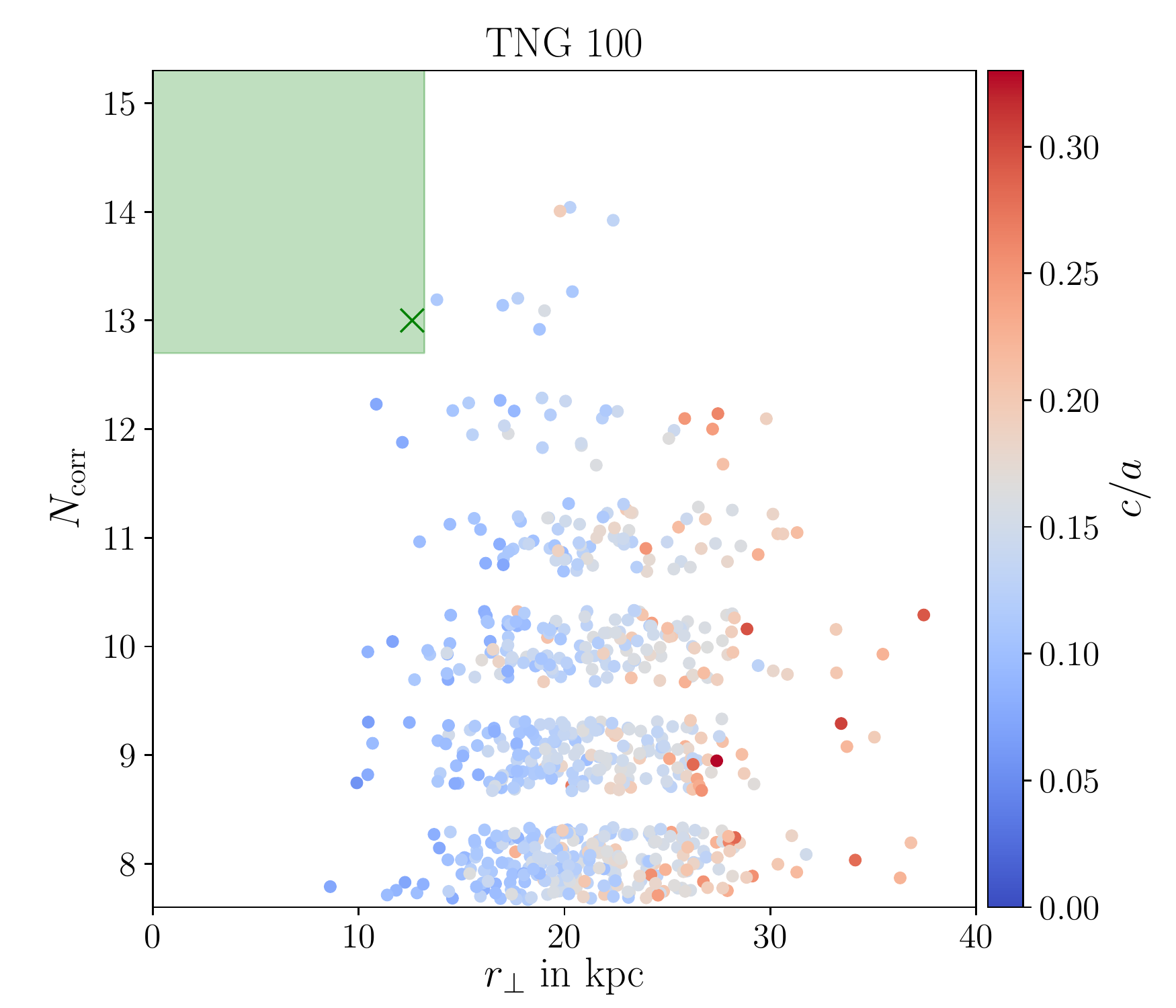}{0.33\textwidth}{(b)}
          \fig{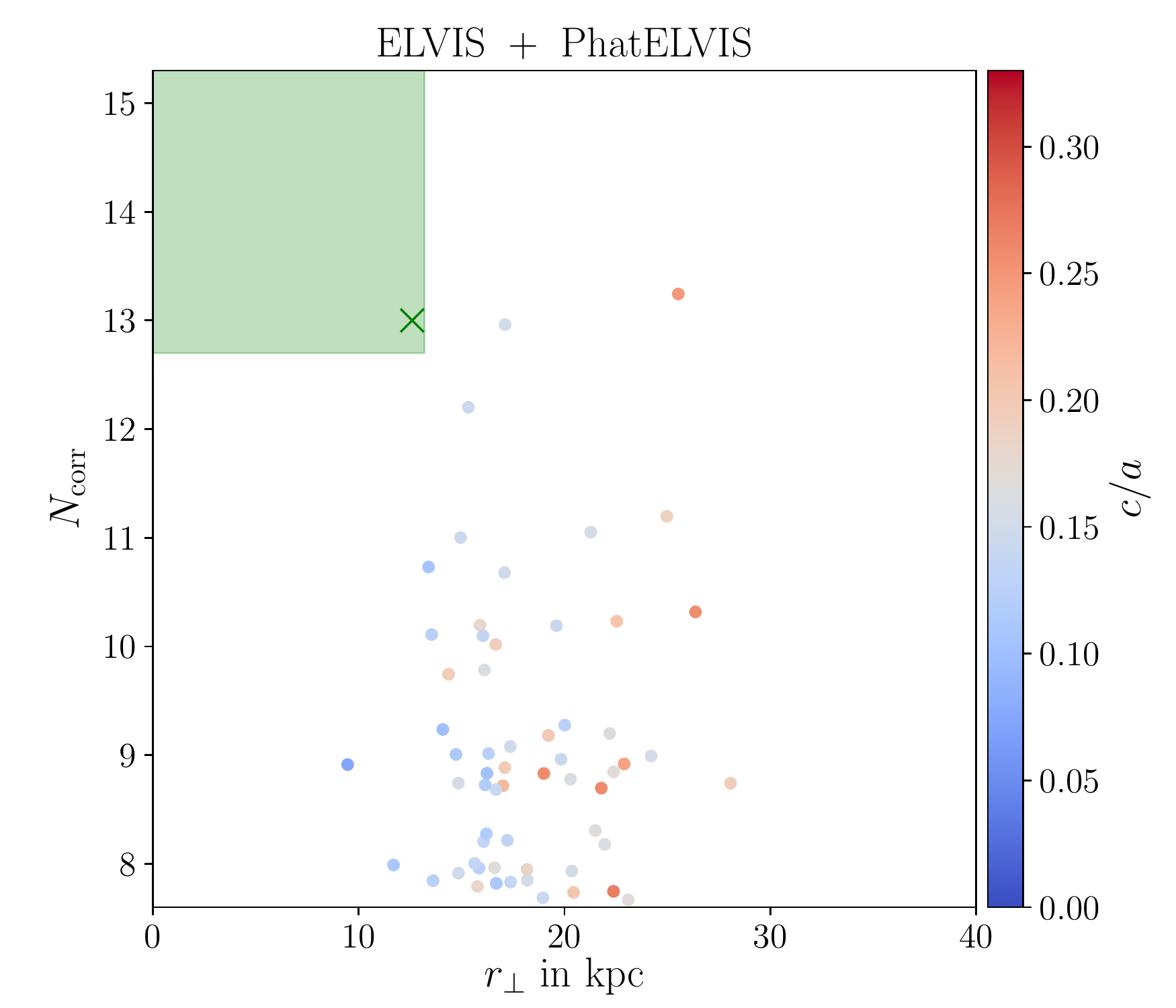}{0.33\textwidth}{(c)}
          }
\gridline{\fig{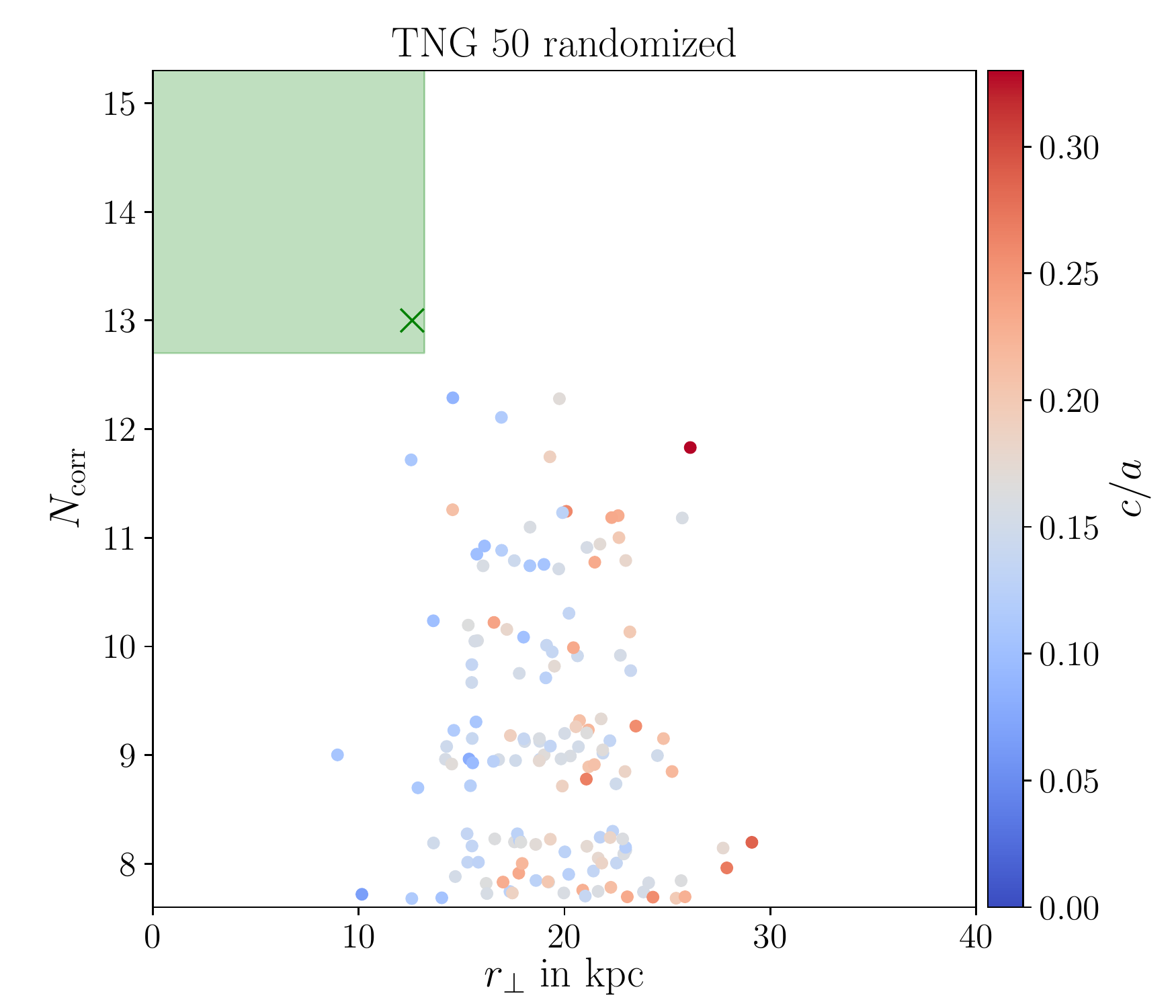}{0.33\textwidth}{(d)}
          \fig{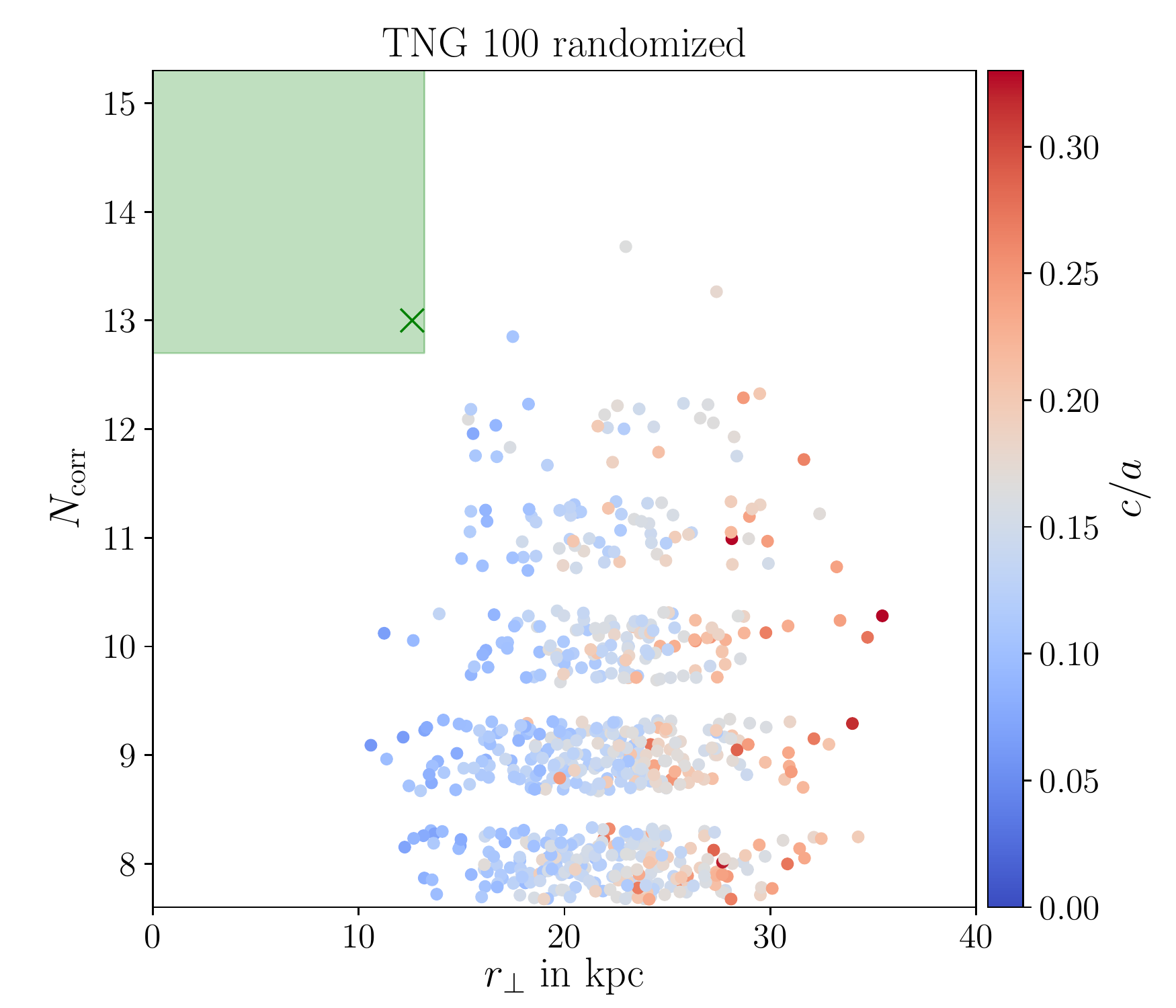}{0.33\textwidth}{(e)}
          \fig{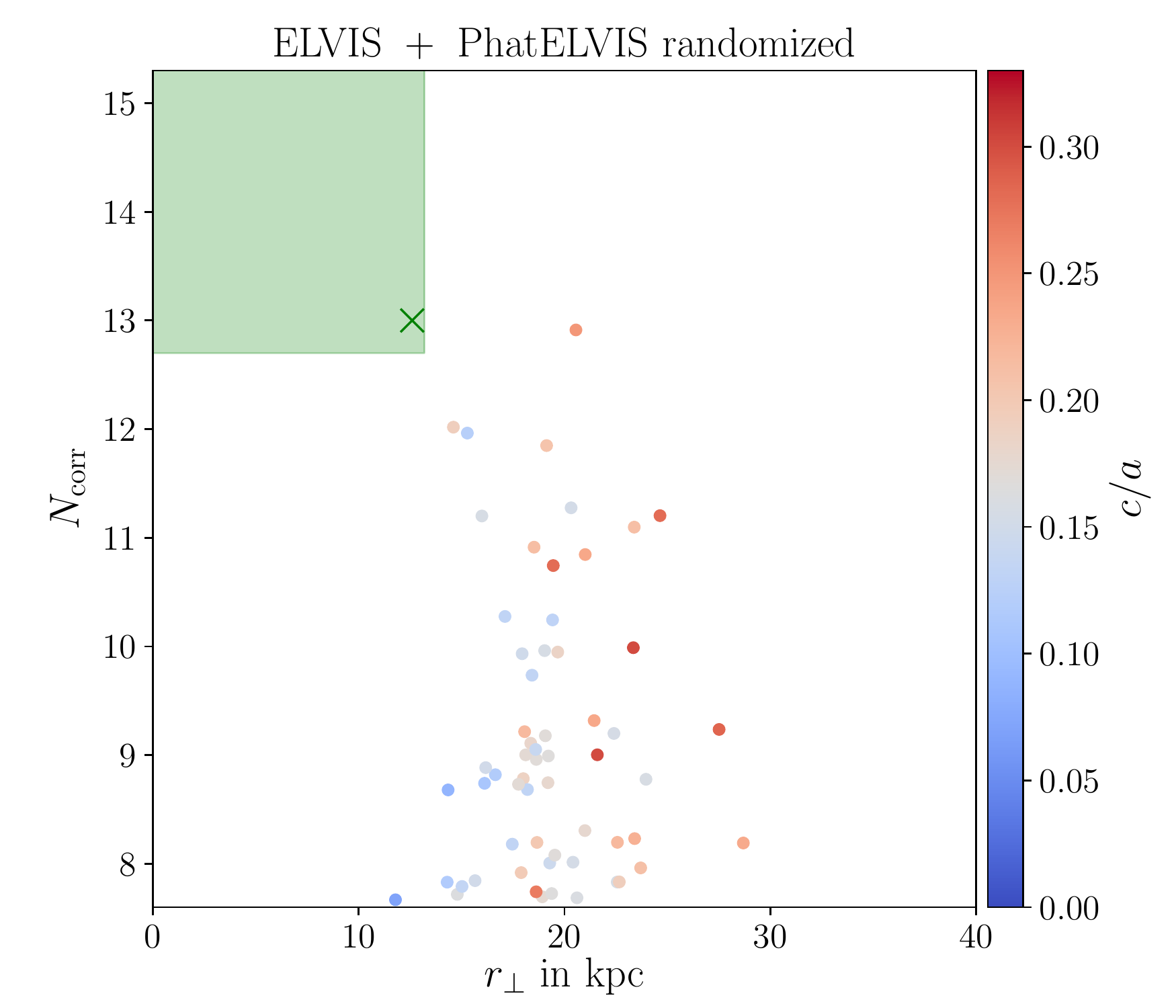}{0.33\textwidth}{(f)}
          }
\caption{Flattening of the satellite planes as measured by the root-mean-square height from the best-fit plane $r_\perp$\ is plotted against the number $N_\mathrm{corr}$\ of satellites with correlated line-of-sight velocities. For better visibility despite the discrete nature of $N_\mathrm{corr}$, we add a random scatter in the range of $\pm 0.3$\ from the measured value. The colors of the points indicate the minor-to-major axis ratio $c/a$ of the corresponding satellite plane, which correlates well with $r_\perp$. Each point corresponds to one simulated satellite system out of the 146 TNG 50 (panel a), 703 TNG 100 (panel b) or 60 ELVIS+PhatELVIS (panel c) hosts, or their randomized analogs (panels d to f). Satellites are in all cases confined to the mock PAndAS survey volume. The region with values as or more extreme than the observed GPoA is indicated by the green shaded area, the most-likely observed values for the GPoA are indicated by the green cross. None of the simulated systems is simultaneously as flattened and has as many correlated line-of-sight velocities as the observed GPoA ($r_\perp = 12.6 \pm 0.6\,\mathrm{kpc}$, $N_\mathrm{corr}=13$).
\label{fig:planeproperties}}
\end{figure*}

Figure \ref{fig:planeproperties} plots the plane heights and degree of kinematic correlation of the simulated satellite galaxy systems, as well as their randomized realizations. It is clear that the simulated satellite systems preferentially populate regions well removed from the observed GPoA in this parameter space -- satellite planes as extreme as the observed GPoA are rare. While 7 of the 146 (17 of the 703) Illustris TNG50 (TNG100) satellite systems have $r_\perp \leq 13.2$, only nine have $N_\mathrm{corr} \geq 13$. None fulfill both criteria simultaneously, but one each of the TNG50 and TNG100 halos comes reasonably close to the observed plane parameters: Halo 99 with $r_\perp = 14.1\,\mathrm{kpc}$\ and $N_\mathrm{corr}=13$\ in TNG50, and Halo 863 with $r_\perp = 13.8\,\mathrm{kpc}$\ and $N_\mathrm{corr}=13$\ in TNG100. For the 60 ELVIS+PhatELVIS satellite systems, two systems each have have $r_\perp \leq 13.2$\ or $N_\mathrm{corr} \geq 13$, but none meet both criteria simultaneously. These results are in agreement with similar previous studies such as \citet{2014ApJ...784L...6I} and \citet{2014MNRAS.442.2362P} who found that GPoA-like systems are rare in simulated satellite systems extracted from the Millennium-II simulation. The randomized systems tend to show even less extreme properties, with less systems of large $N_\mathrm{corr}$\ as well as typically larger $r_\perp$, though overall the randomized systems cover a similar region in this parameter space as the simulated systems.

Alternatively, the minor-to-major axis ratio $c/a$\ can be used as a measure of relative plane flattening (as opposed to the absolute plane flattening measured with $r_\perp$), with essentially identical results. This is unsurprising, because as indicated by the color-coding in Fig. \ref{fig:planeproperties}, $c/a$\ correlates well with the root-mean-square plane height $r_\perp$.

\subsection{Chance to find any two satellites as closely aligned}
\label{subsec:anglecomparison}

\begin{figure*}
\gridline{\fig{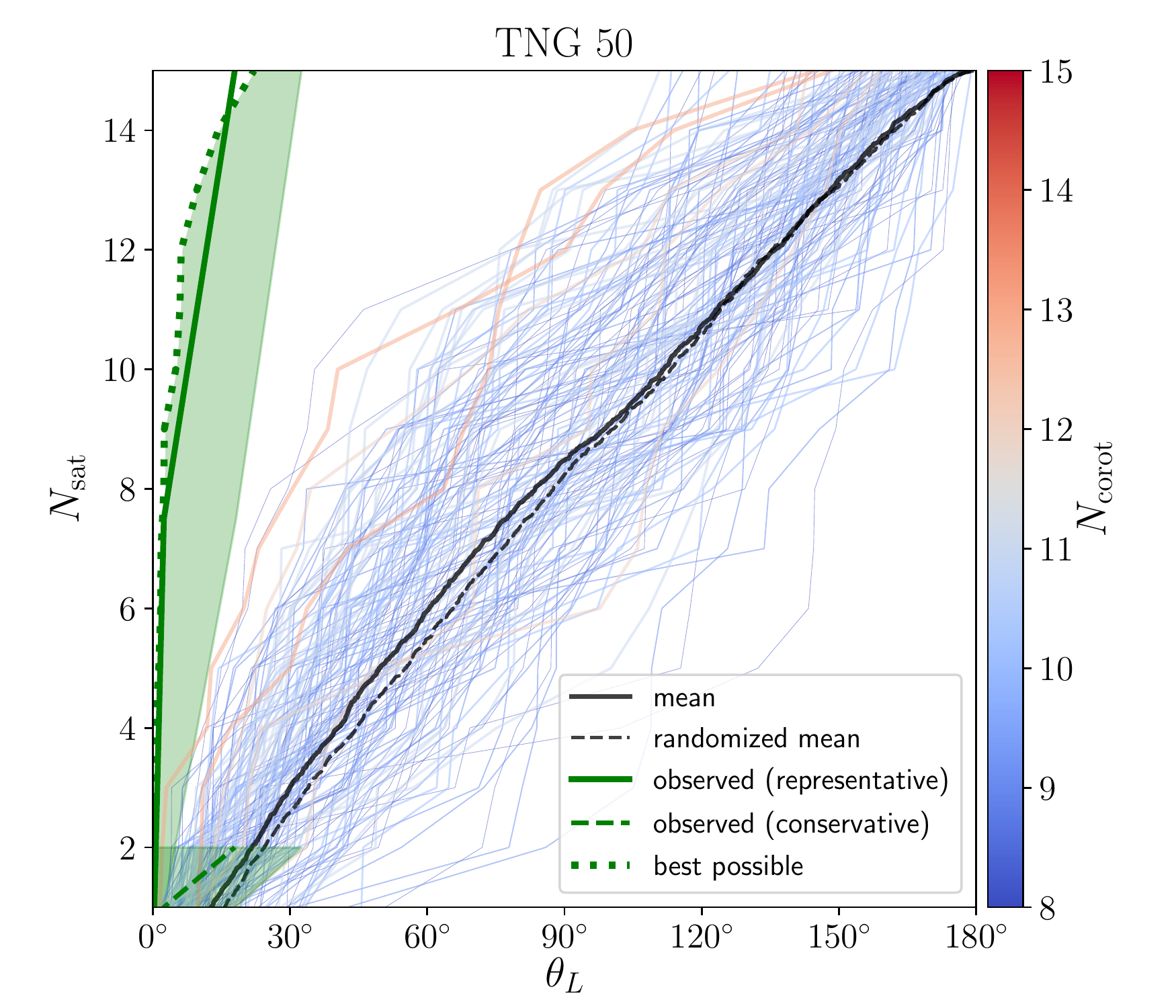}{0.33\textwidth}{(a)}
          \fig{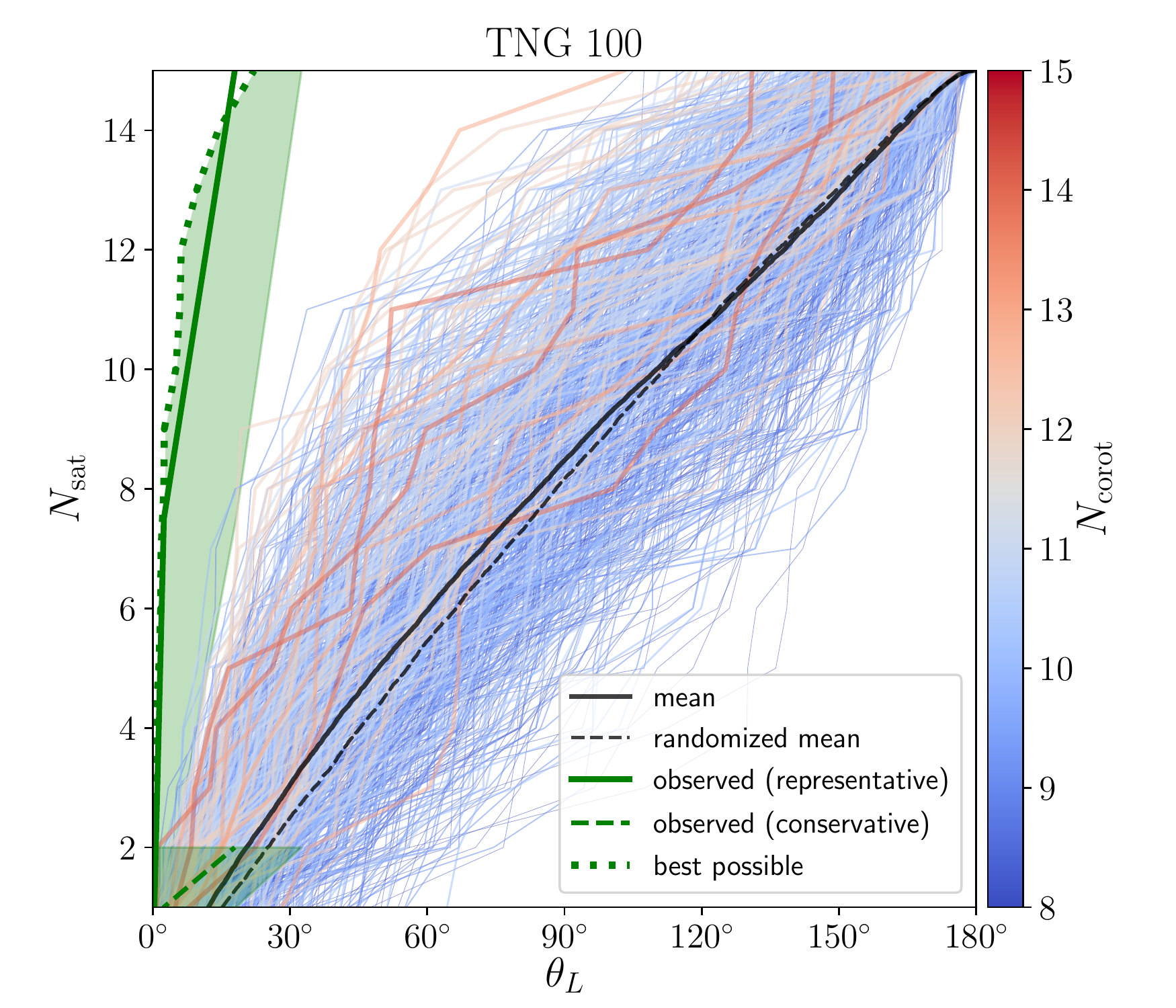}{0.33\textwidth}{(b)}
          \fig{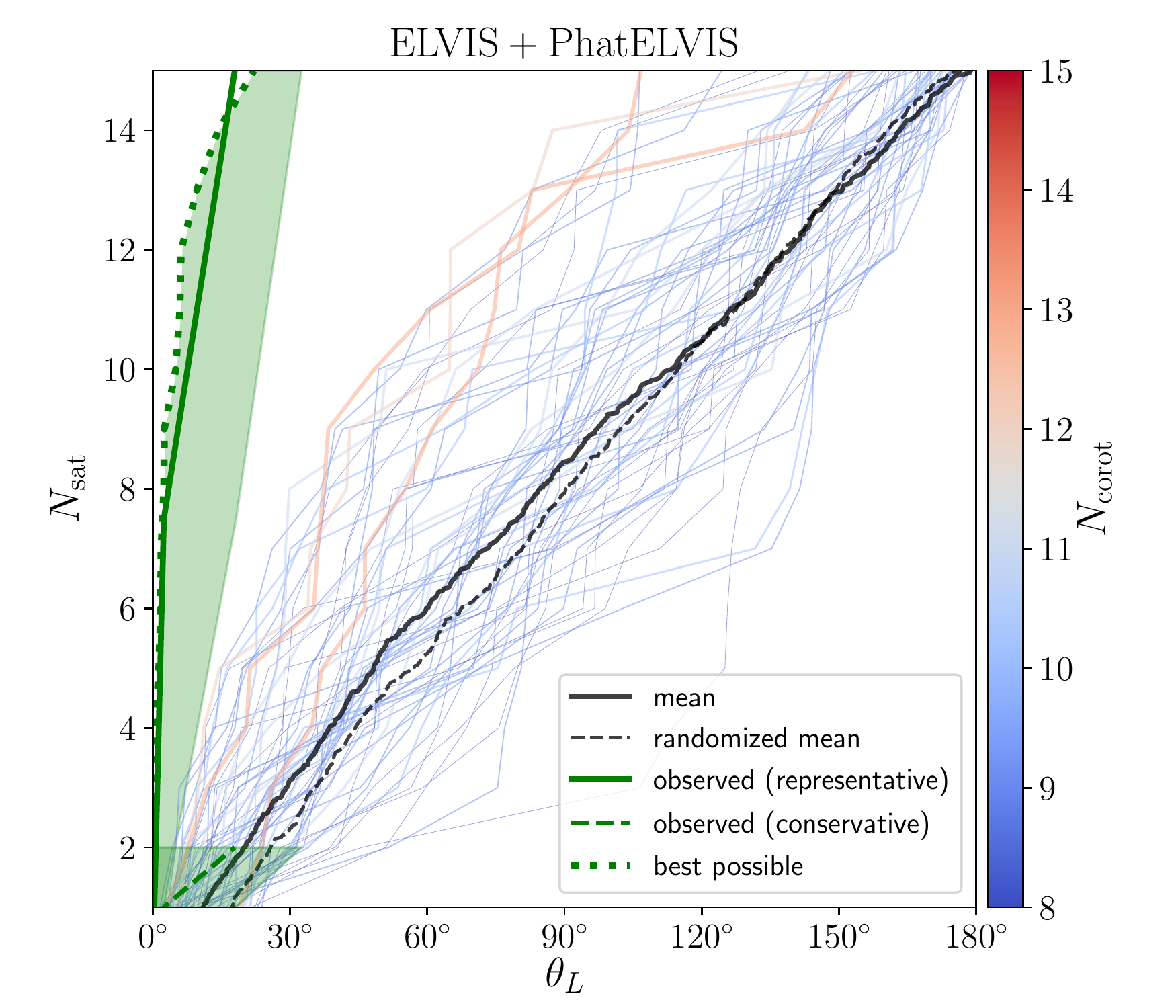}{0.33\textwidth}{(c)}
          }
\gridline{\fig{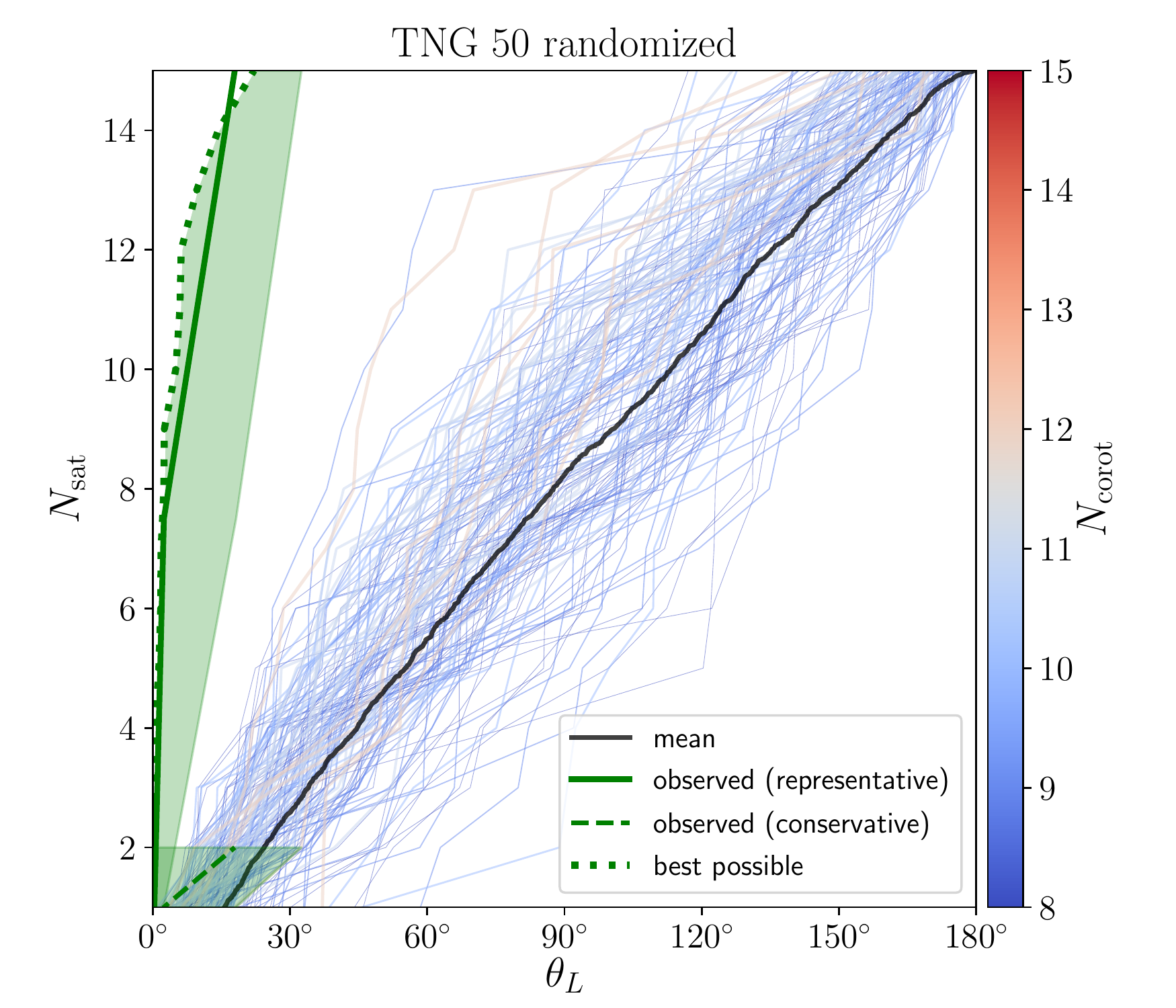}{0.33\textwidth}{(d)}
          \fig{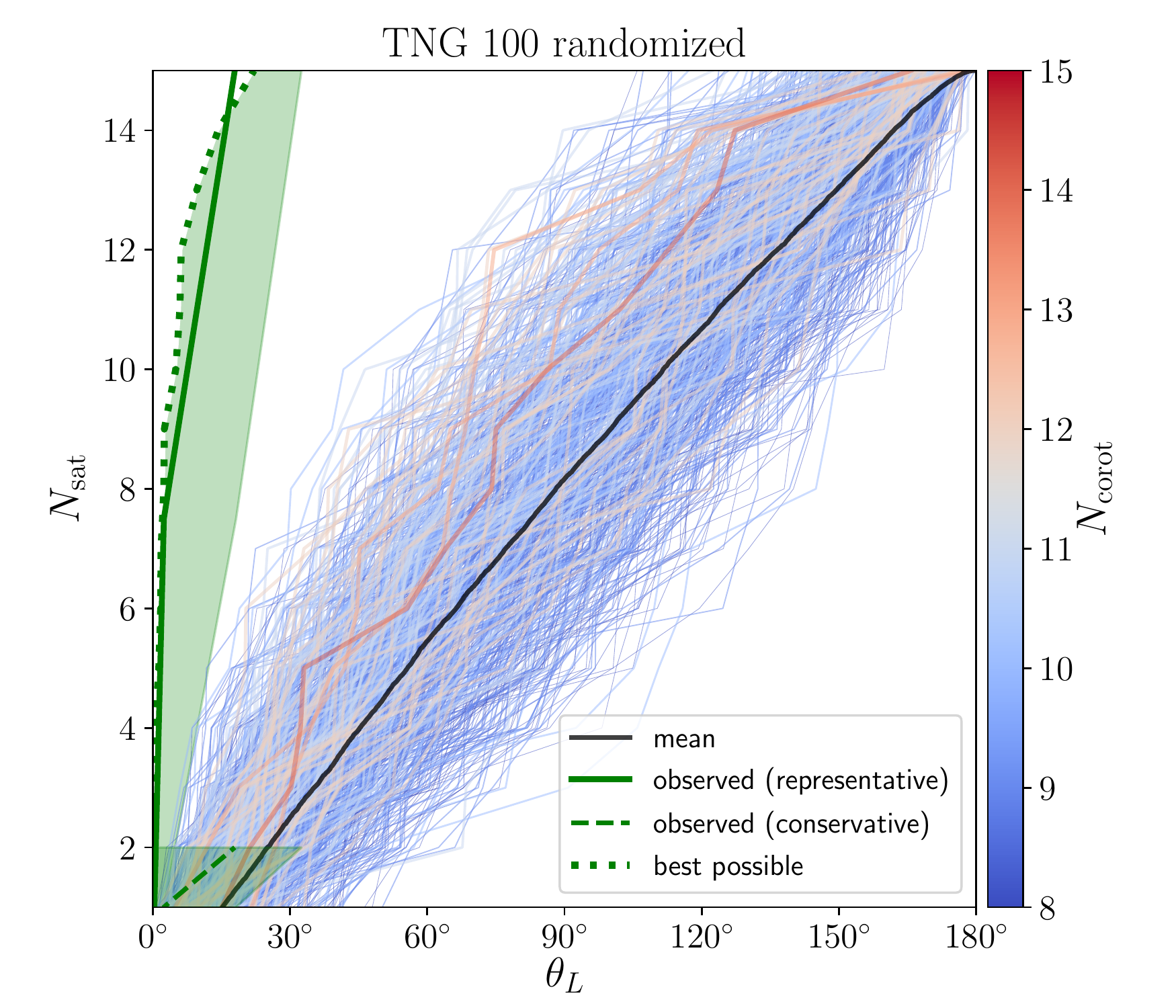}{0.33\textwidth}{(e)}
          \fig{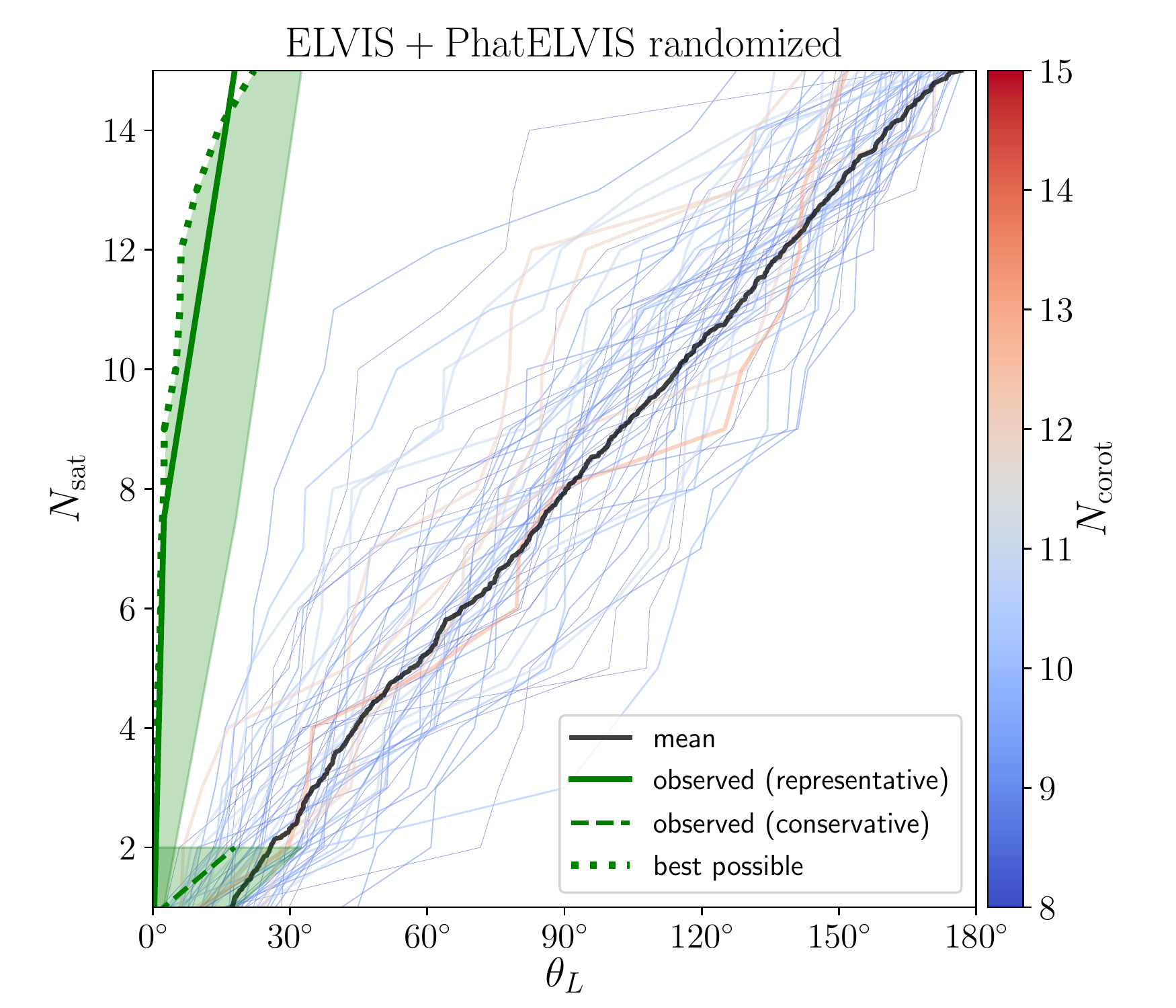}{0.33\textwidth}{(f)}
          }
\caption{Cumulative distributions of alignment angle $\theta_\mathrm{L}$\ between the orbital poles of the 15 on-plane satellites and the normal vector to their best-fit plane. The top panels (a to c) show satellite systems in simulations (from left to right: Illustris TNG 50, Illustris TNG 100, ELVIS and PhatELVIS), while the bottom panels (d and f) show the corresponding lines after randomizing the positions and velocity directions of the satellite galaxies. The black solid line shows the mean cumulative distribution for each panel, the randomized (bottom) panel's corresponding lines are repeated in the left panels as dashed lines to ease comparison. 
Each line corresponds to one simulated satellite galaxy system, and is color-coded by the kinematic correlation of its on-plane satellites as measured by $N_\mathrm{corot}$. Satellites in systems with stronger kinematic correlation show a preference for smaller $\theta_\mathrm{L}$\ (more steeply rising cumulative distributions).
The green  line (shaded region) indicates the measured most-likely ($1\sigma$) alignment angle for the observed M31 satellites from the GPoA normal, with a solid line assuming that these two satellites are representative for all 15, and the dashed line indicating the most conservative assumption that these two satellites are the best-aligned of all 15 observed on-plane satellites. The dotted lines limit the best-possible aligned orbital poles based on the spatial angular offset given the current positions of the GPoA members (see $\theta_\mathrm{GPoA}$\ in column 6 of Table \ref{tab:prediction}).
\label{fig:cumulativeangle}}
\end{figure*}

\begin{figure*}
\gridline{\fig{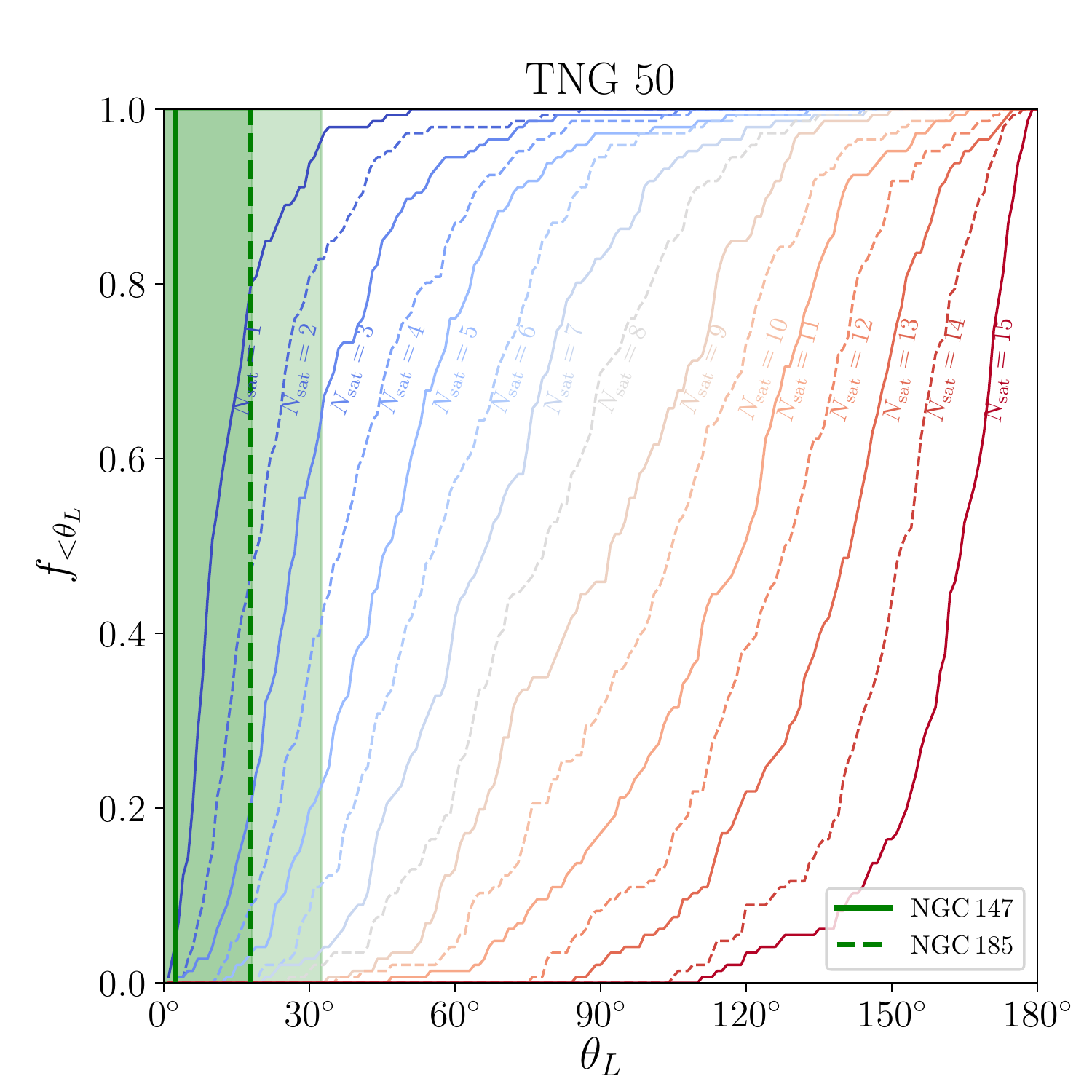}{0.33\textwidth}{(a)}
          \fig{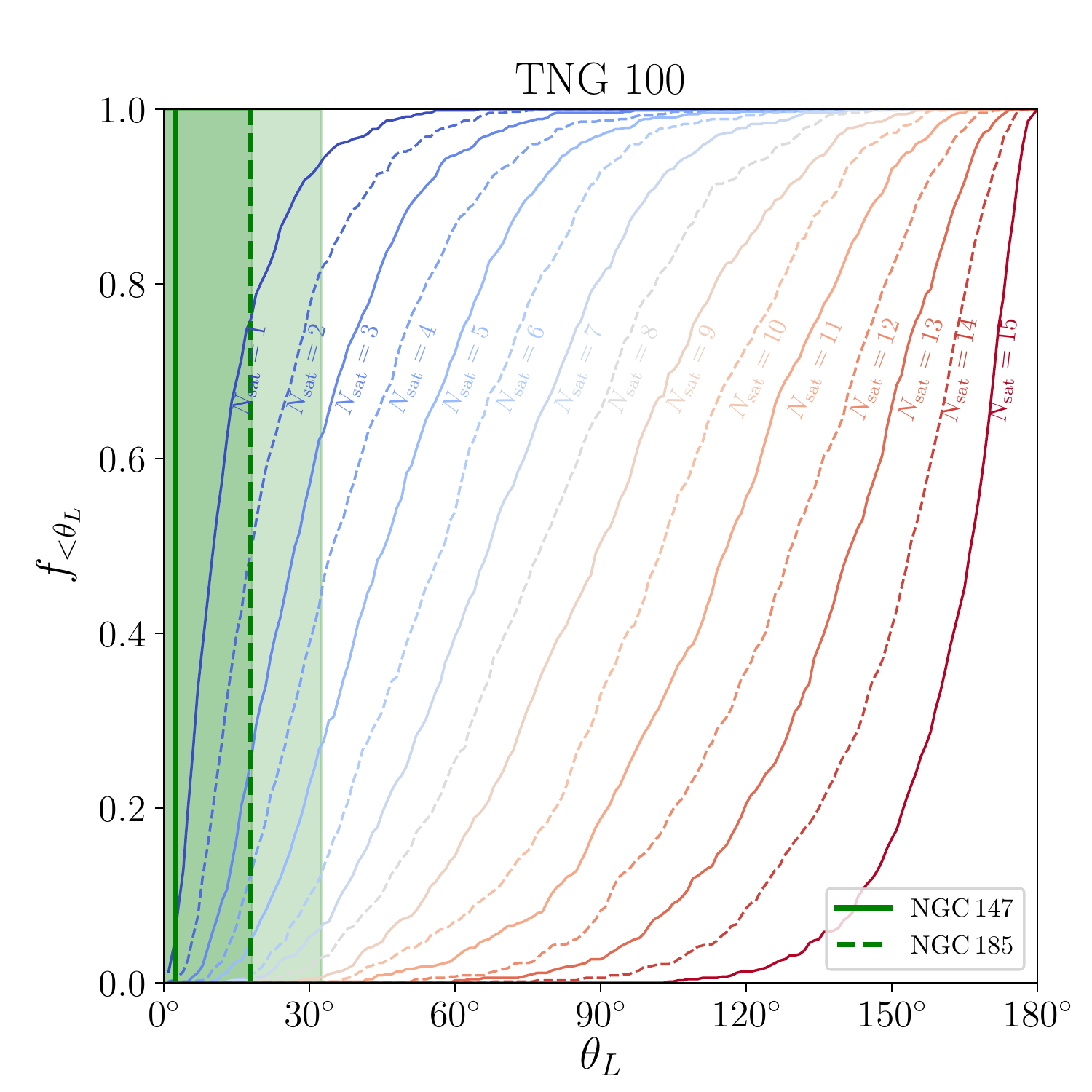}{0.33\textwidth}{(b)}
          \fig{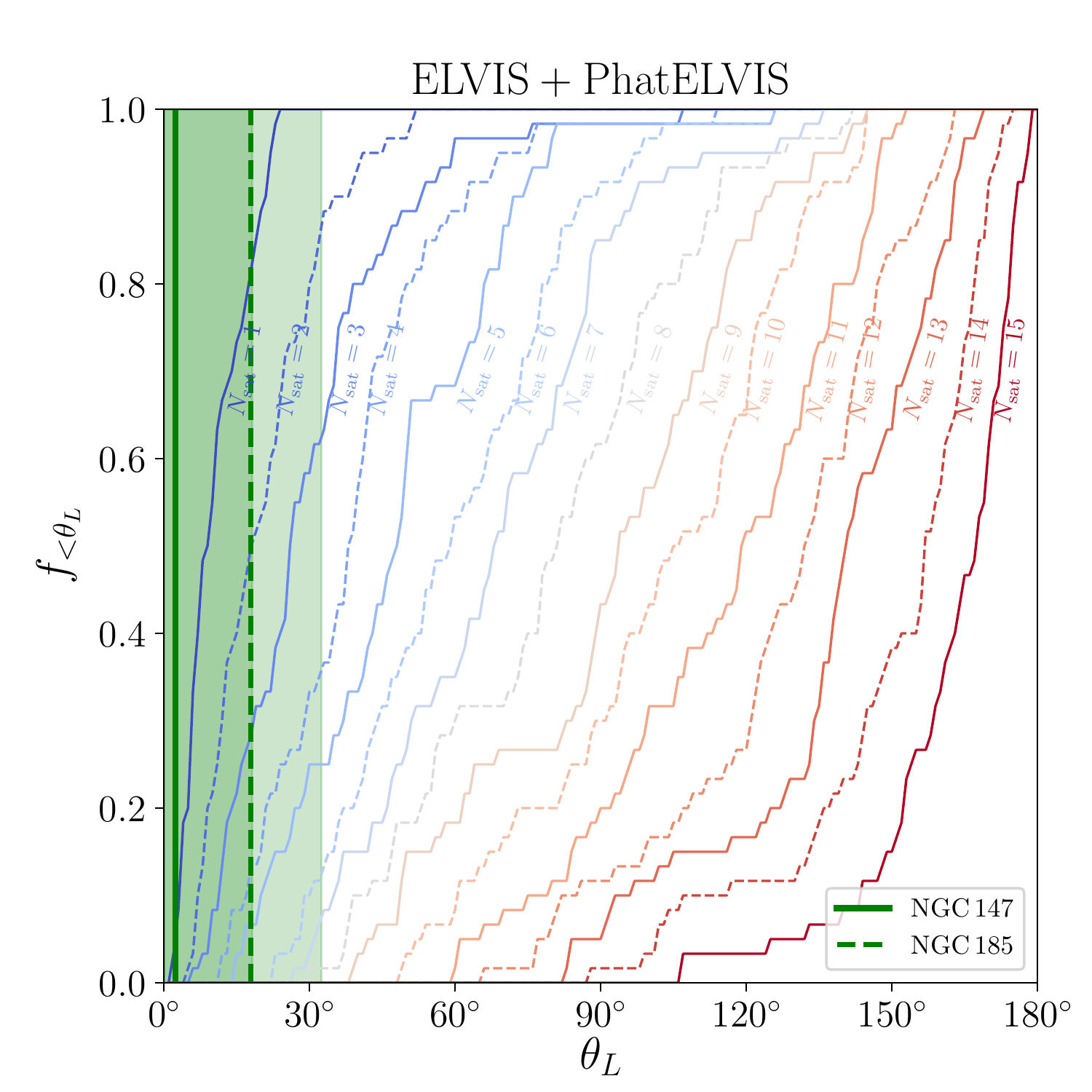}{0.33\textwidth}{(c)}
          }
\gridline{\fig{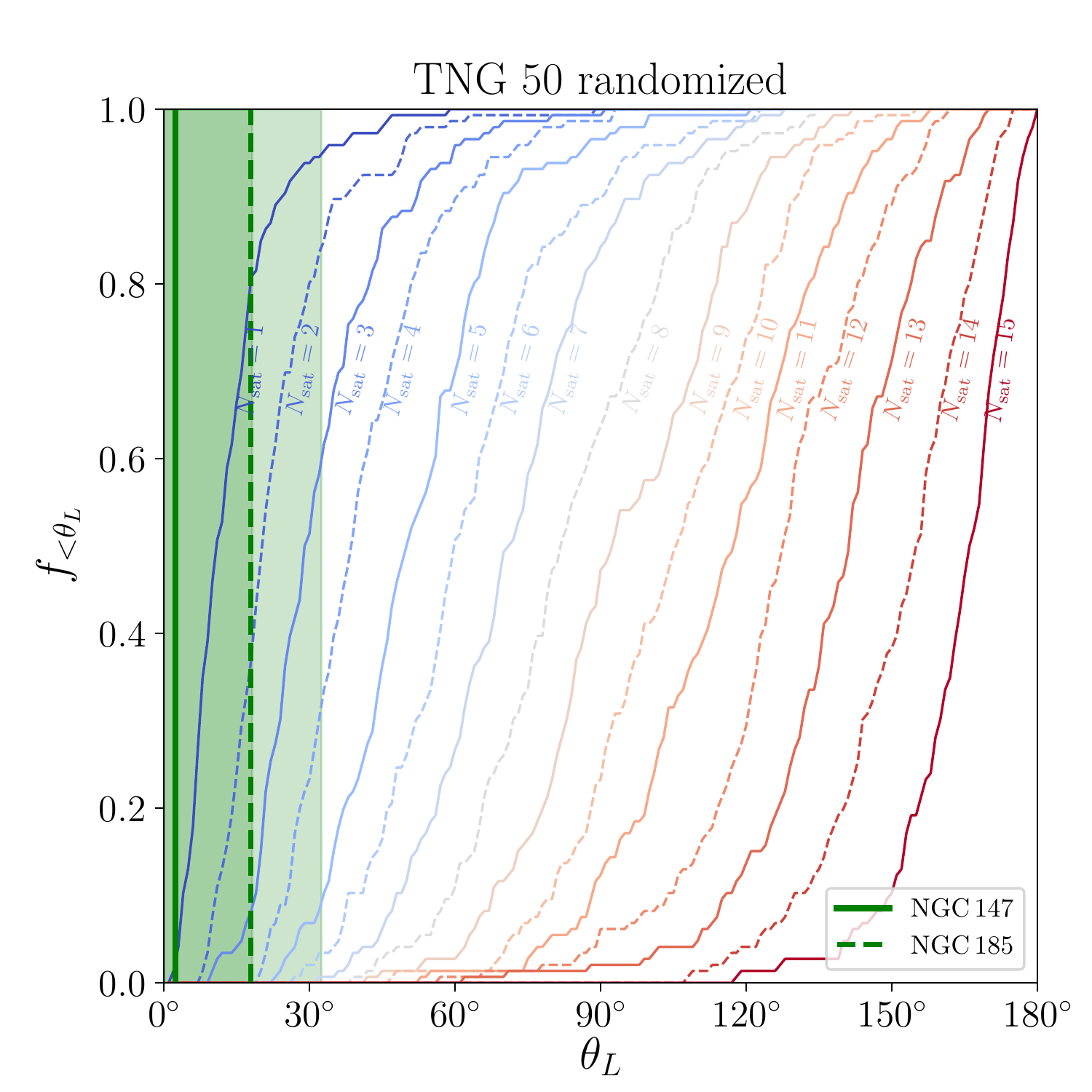}{0.33\textwidth}{(d)}
          \fig{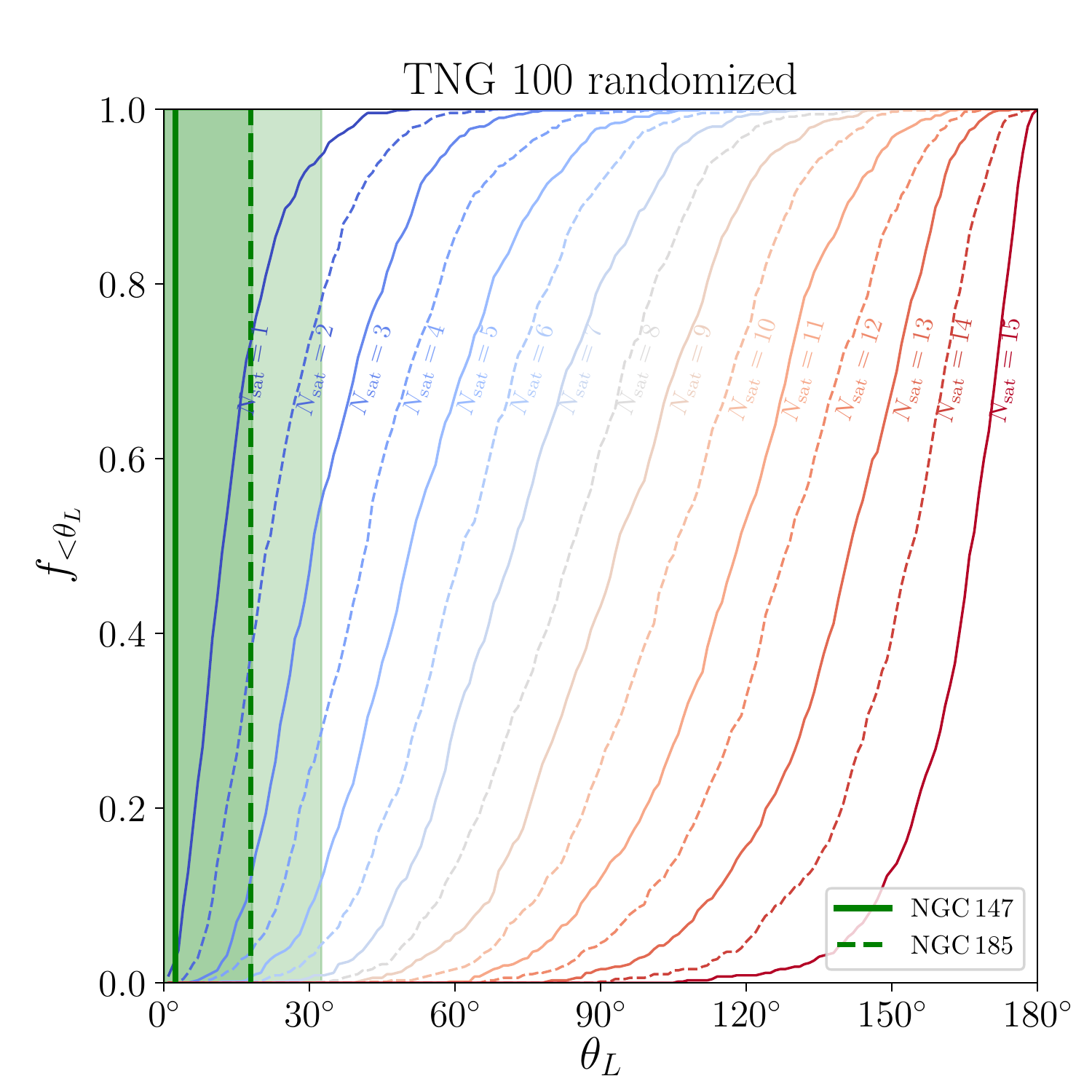}{0.33\textwidth}{(e)}
          \fig{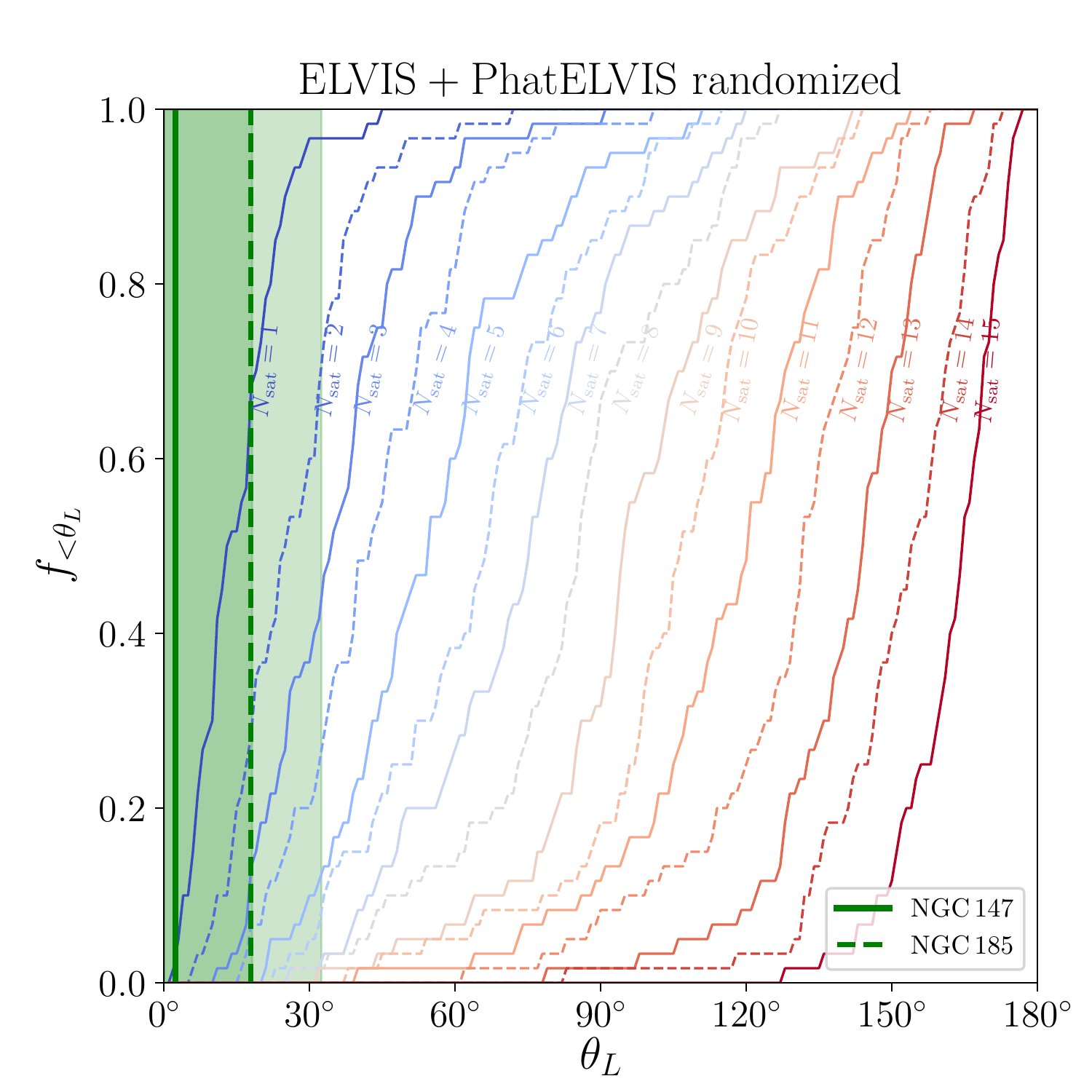}{0.33\textwidth}{(f)}
          }
\caption{Cumulative distributions in the alignment angels $\theta_\mathrm{L}$\ between the satellite orbital poles and the (co-orbiting) normal vector of their corresponding satellite plane, for the on-plane model satellites. The satellites are ranked by their alignment, ranging from the cumulative distribution for the best-aligned satellite in each system ($N_\mathrm{sat} = 1$)\ to that of the least-aligned (or most anti-aligned, $N_\mathrm{sat} = 15$). The vertical green lines indicate the most-likely alignment angles $\theta_\mathrm{L}$\ for the observed M31 satellites NGC\,147 (solid) and NGC\,185 (dashed), and the shaded areas indicate their corresponding $1\sigma$\ limits (the limit of NGC\,147 coincides with the most-likely value of NGC\,185).
\label{fig:individualangle}}
\end{figure*}

The analysis thus far has only considered the three-dimensional positions as well as the line-of-sight component of the mock-observed simulated satellite galaxy systems. This is analogous to the observational situation thus far. Now that first proper motion measurements for on-plane satellites exist, we can expand beyond those limitations and compare how likely it is to find simulated satellites whose orbits align as well with the satellite plane as NGC\,147 and NGC\,185 align with the GPoA, much like has become common practice for the Milky Way satellite galaxies \citep{2008ApJ...680..287M,2013MNRAS.435.2116P,2018A&A...619A.103F,2019MNRAS.488.1166S,2020MNRAS.491.3042P}.

To quantify the orbital alignment of satellite galaxies with their corresponding satellite galaxy plane, we have measured $\theta_\mathrm{L}$\ for each of the 15 on-plane satellites in each of the simulated satellite galaxy systems. The same was repeated in randomized satellite systems to generate a baseline-expectation for satellite systems that do not have any intrinsic correlations, while accounting for the PAndAS footprint constraints. The randomized satellite systems do have a spatial bias introduced by the selection volume that is based on the PAndAS footprint (to ensure that any survey-volume based biases are accounted for), but are otherwise random within it and have random velocity directions. Figure \ref{fig:cumulativeangle} plots the cumulative distributions of this alignment angle $\theta_\mathrm{L}$\ for all simulated systems, color-coded by their kinematic correlation as measured by $N_\mathrm{corot}$.

Satellites in the simulated systems tend to be on average slightly better aligned with their spatial planes than completely random systems, their mean line lies above that of randomized systems until $\theta_\mathrm{L} \approx 120^\circ$. This indicates that, as expected, there is some phase-space correlation present in the simulated satellite galaxy systems. However, this is only a very minor difference, with less than one additional satellite being aligned at any given angle.
This seems to be largely driven by a few simulated systems with steeply rising cumulative distributions, while the majority of simulated satellite systems cover a similar region as their randomized conterparts.

For systems with higher $N_\mathrm{corot}$\ (red in Fig. \ref{fig:cumulativeangle}), the cumulative curves in $\theta_\mathrm{L}$\ tend to rise more steeply. This is expected, because $N_\mathrm{corot}$, which measures the number of on-plane satellites that have coherent line-of-sight velocities, and $\theta_\mathrm{L}$, which measures the angle between the orbital angular momentum and the normal vector to the satellite plane, are related. For example, in the idealized case of circular orbits, $N_\mathrm{corot} = 15$\ would mean that all satellites have a common orbital sense around their host, such that all 15 $\theta_\mathrm{L}$\ angles for the satellites would be confined to $\theta_\mathrm{L} \leq 90^\circ$. Non-circular orbits and smaller $N_\mathrm{corot}$\ introduce scatter, but the overall trend remains and is also present in the randomized satellite systems (lower panels of Fig. \ref{fig:cumulativeangle}). Finding a high degree of correlation among the line-of-sight velocities thus indicates that some orbital alignment is more likely, but a majority of satellites could still have orbits almost perpendicular to the respective satellite plane (i.e. $\theta_\mathrm{L} \approx 90^\circ$). Consequently, a high $N_\mathrm{corot}$\ does not ensure mostly small $\theta_\mathrm{L}$, and direct measurements of proper motions to obtain 3D velocities and thereby measure the orbital alignments with the identified satellite plane are necessary to learn more about which satellites are associated, and how dynamically stable the overall structure might be.

These results hold true for the Illustris TNG50 and TNG100, as well as the ELVIS+PhatELVIS simulations, even though the latter are somewhat more noisy due to their smaller sample size. The mean distributions, as well as the overal scatter in the cumulative distributions, agree very well between the three sets of simulated satellite systems. This is despite their substantial differences in resolution, implemented physics, and radial distribution of the considered satellite galaxy systems (see Sect. \ref{sec:simresults} ).

If we assume that the close orbital alignment of the two observed satellite galaxies is representative (solid green line in Fig. \ref{fig:cumulativeangle}, with shaded region indicating the $1\sigma$ limits) for the 15 observed on-plane satellites of M31, then none of the simulated systems show a similar degree of orbital alignment. At most about half of the simulated satellites are similarly closely aligned. Thus, if half of the remaining on-plane satellites of the GPoA are found with future proper motion measurements to have similarly closely aligned orbital poles as NGC\,147 and NGC\,185, then this alone would pose an extreme problem for $\Lambda$CDM, irrespective of the rarity of the spatial and line-of-sight correlated nature of the GPoA.

However, we can also adopt the {\it most conservative} assumption that these first two satellites with measured proper motions turn out to be the two with the best-aligned orbital poles out of all 15 on-plane satellites of M31. In that case (dashed green line in Fig. \ref{fig:cumulativeangle}), no clear tension with $\Lambda$CDM expectations is apparent. Many simulated satellite galaxy systems have one or two satellites which are similarly closely aligned as the two observed objects, and the observed satellites are consistent within their $1\sigma$\ limits with the mean distribution of $\theta_\mathrm{L}$\ as deduced for the simulated systems.

Fig. \ref{fig:individualangle} plots the cumulative distribution of $\theta_\mathrm{L}$\ for the 15 on-plane satellites ranked by their alignment with their corresponding satellite plane. In other words, the line for $N_\mathrm{sat} = 1$\ indicates the distribution of alignment angles for the best-aligned satellite in any simulated system, that for $N_\mathrm{sat} = 2$\ for the second-best aligned satellite in any system, and so on. Again, the distributions are consistent between the three sets of simulations, and the randomized systems qualitatively follow a similar distribution, too.

To quantify the preceding discussion, we can ask: how many of the satellite systems can be excluded based merely on orbital pole alignment of any two satellites? 
For this, we count for how many simulated systems we find the best-aligned satellite's orbital pole to be as close as that of NGC\,147 is to its satellite plane, and the second-best aligned as close as that of NGC\,185. Specifically, we check whether the smallest $\theta_\mathrm{L}$\ of the set of on-plane satellites in a simulated system is smaller than $2^\circ.4$\ (or $18^\circ.2$\ for the $1\sigma$\ limit of the Monte-Carlo samplings), and the second-smallest is below $17^\circ.9$ ($32^\circ.5$). 
Adopting the most-likely orbital pole of NGC\,147, only 6.2, 5.1, and 5.0 \% of simulated systems in Illustris TNG50, TNG100, and ELVIS+PhatELVIS, respectively, have their best-aligned satellite as close. Finding the second-best-aligned satellite as close as NGC\,185 is a lot more common (46.6, 49.2, and 50.0 \%, respectively). However, simulateneously finding one satellite as well aligned as NGC\,147 and one as well aligned as NGC\,185 is somewhat more rare than focussing only on NGC\,147. It occurs in 4.1, 4.1, or 3.3 \% of the simulated systems, respectively.
This shows that, if the most-likely current proper motions of the two M31 satellites as well as of M31 itself are confirmed, these two systems alone already put substantial difficulty on the $\Lambda$CDM model to explain the observed GPoA dynamics. In particular because this comparison assumes that the two observed satellites are the best-aligned ones out of all 15 observed on-plane satellite galaxies of M31, which a-priori is not necessarily correct and in fact rather unlikely (see Sect. \ref{subsec:chancetofindpair} ).

However, the current measurements come with substantial uncertainties. It thus might be more prudent to compare to the $1\sigma$\ limits on the alignment angle $\theta_\mathrm{L}$. In this case, it is likely that the two best-aligned satellites in a simulated system are as closely aligned as both NGC\,147 and NGC\,185: this happens for 75.3, 69.6, and 71.7 \% for the TNG50, TNG100, and ELVIS+PhatELVIS simulations, respectively. Conversely, however, this means that about 30 \% of all simulated systems can {\it already} be excluded as analogs of the M31 GPoA, based only on the orbital poles of two satellite galaxies.

The above numbers are somewhat smaller for the randomized systems, though not substantially. Overall, the fraction of randomized systems for which we find the two closest angles smaller than the $1\sigma$\ limit of observed angles for NGC\,147 and NGC\,185 is 76.0\% for TNG50, 66.1\% for TNG100, and 55.0\% for the ELVIS+PhatELVIS simulations, an identical fraction for the TNG50 systems, and a drop of merely 3 and 17 per cent points for the other two simulation sets.

The discussion thus far was based on the (highly conservative) assumption that the two observed GPoA satellites correspond to the two best-aligned satellites in each simulated satellite system. But what is the chance to observe the best-aligned two satellites first, and find them as well aligned as the two observed objects?

\subsection{Chance to find two satellite orbits as closely aligned}
\label{subsec:chancetofindpair}

\begin{deluxetable*}{lcccc}
\tablenum{4}
\tablecaption{Frequencies of simulated satellite galaxy pairs with orbital poles as closely aligned with their respective satellite plane normal as the observed M31 system. \label{tab:frequencies}}
\tablewidth{0pt}
\tablehead{
\colhead{} & \multicolumn2c{Simulation} & \multicolumn2c{Randomized} \\
\colhead{Satellite selection} & \colhead{$f_\mathrm{147}$} & \colhead{$f_\mathrm{147+185}$} & \colhead{$f_\mathrm{147}$} & \colhead{$f_\mathrm{147+185}$} \\
\colhead{} & \colhead{(\%)} & \colhead{(\%)} & \colhead{(\%)} & \colhead{(\%)} \\ 
}
\startdata
Illustris TNG50 (146 systems) & & & & \\
~~Any two on-plane satellites & 0.91 (20.45) & 0.14 (3.72) & 0.27 (16.63) & 0.01 (2.39)  \\
~~$v_\mathrm{los}$\ correlated  & 1.66 (27.70) & 0.34 (7.07) & 0.45 (23.00) & 0.02 (4.73)  \\
~~$v_\mathrm{los}$\ correlated + 5 most massive only & 1.03 (26.99) & 0.27 (7.12)  & 0.27 (23.36) & 0.0 (4.93)  \\
~~$v_\mathrm{los}$\ correlated + $d > 110\,\mathrm{kpc}$ & 2.31 (30.59) & 0.56 (8.18) & 0.72 (27.14) & 0.0 (5.22) \\
\multicolumn5l{Only systems with $N_\mathrm{coorb} \geq 13$\ (simulations: 2, randomized: none):} \\
~~$v_\mathrm{los}$\ correlated &  14.74 (58.97) & 4.49 (19.23) & --- & ---  \\
\hline
Illustris TNG100 (703 systems) & & & & \\
~~Any two on-plane satellites & 0.75 (20.90) & 0.11 (4.15) & 0.45 (16.22) & 0.03 (2.36) \\
~~$v_\mathrm{los}$\ correlated  & 1.02 (27.75) & 0.19 (7.44) & 0.58 (21.62) & 0.05 (4.42) \\
~~$v_\mathrm{los}$\ correlated + 5 most massive only & 1.01 (27.70) & 0.14 (7.80) & 0.63 (21.42) & 0.04 (4.85)  \\
~~$v_\mathrm{los}$\ correlated + $d > 110\,\mathrm{kpc}$ & 1.15 (29.69) & 0.21 (8.12)  & 0.72 (23.83) & 0.08 (5.02) \\
\multicolumn5l{Only systems with $N_\mathrm{coorb} \geq 13$\ (simulations: 9, randomized: 4):} \\
~~$v_\mathrm{los}$\ correlated & 0.0 (27.67) & 0.0 (6.48)  & 0.0 (14.15) & 0.0 (2.46)  \\
\hline
ELVIS + PhatELVIS (60 systems) & & & & \\
~~Any two on-plane satellites & 0.67 (22.13) & 0.08 (4.00)  & 0.44 (14.27) & 0.02 (2.30) \\
~~$v_\mathrm{los}$\ correlated  & 1.01 (29.22) & 0.09 (7.93)  & 0.70 (19.24) & 0.04 (4.27)  \\
~~$v_\mathrm{los}$\ correlated + 5 most massive only & 2.0 (30.17) & 0.0 (10.5) & 1.33 (21.00) & 0.0 (4.67) \\
~~$v_\mathrm{los}$\ correlated + $d > 110\,\mathrm{kpc}$ & 2.18 (32.73) & 0.26 (9.50)  & 1.58 (23.16) & 0.13 (5.92)  \\
\multicolumn5l{Only systems with $N_\mathrm{coorb} \geq 13$\ (simulations: 2, randomized: 1):} \\
~~$v_\mathrm{los}$\ correlated & 0.0 (21.15) & 0.0 (7.69) & 0.0 (15.38) & 0.0 (1.28)  \\
\enddata
\tablecomments{Frequencies are calculated considering all possible combinations of two satellites of the same host, from the sample fulfilling the respective selection criteria specified in the leftmost column. Values correspond to the alignment of the orbital pole based on the observed satellites' most-likely proper motion, while values in brackets correspond to the $1\sigma$\ limit on the orbital pole alignment, both assuming the HST+Sat. proper motion of M31.
}
\end{deluxetable*}

The previous investigation is limited by the fact that thus far only two proper motions of on-plane satellites have been measured. This makes it difficult to draw firm conclusions on the overall alignment of satellite poles with their satellite plane. A different approach is to ask how likely it is to find two out of 15 on-plane satellites to be as closely aligned as the two observed ones if satellite systems in cosmological simulations are representative of the M31 satellite system.

To do this, we randomly draw two satellites from the on-plane satellites in our mock-observed systems and count how often they are at least as well aligned as the observed NGC\,147 and NGC\,185. We again repeat the same analysis on the randomized satellite systems to determine the degree to which satellites in $\Lambda$CDM are correlated.

Specifically, we determine how often:
\begin{enumerate}
\item any one of the two is at least as well aligned as the better-aligned NGC\,147: $f_\mathrm{147}$,
\item Any one is at least as well aligned as NGC\,147, while the other is also at least as well aligned as NGC\,185: $f_\mathrm{147+185}$,
\end{enumerate}
The results are summarized in Table \ref{tab:frequencies}. In the table as well as in the text below, the frequencies are reported for the most-likely alignment angle (based on the most-likely positions, distances, velocities, and proper motions of the satellites and M31), and in brackets for the one-sigma upper limit on the alignment angle from the Monte-Carlo samples in Sect. \ref{sec:observed}.

If we select any combination of two on-plane satellites, we find $f_\mathrm{147} = 0.91 (20.45) \%$\ for the Illustris TNG50, $f_\mathrm{147} = 0.75 (20.90) \%$\ for TNG100 and $f_\mathrm{147} = 0.67 (22.13) \%$\ for the ELVIS simulations. It is thus very unlikely to find, among random pairs of satellites in the simulations, one which is as well aligned as the most-likely phase-space coordinates of NGC\,147 suggest. However, taking the substantial measurement uncertainties into account, finding one similarly closely aligned satellite is not particularly unlikely: if the simulated satellite systems were representative of the observed M31 satellite system, there is a one-in-five chance to find such a close alignment for one of two satellites.

In reality, both NGC\,147 {\it and} NGC\,185 have well aligned orbital poles. If we require that both of their alignments are reproduced by the simulated satellites, only $f_\mathrm{147+185} = 0.14 (3.72) \%$\ (TNG50), $f_\mathrm{147+185} = 0.11 (4.15) \%$\ (TNG100), and $f_\mathrm{147+185} = 0.08 (4.00) \%$\ (ELVIS) of randomly drawn on-plane satellites align at least as well.
Finding as close an alignment between the orbital poles of both satellites and the best-fit plane normal as for the most-likely measured values is thus extremely rare. Only about one out of 1000 randomly drawn satellite pairs has one satellite at least as closely aligned as NGC\,147 and the other at least as closely aligned as NGC\,185. Considering instead the one-sigma upper limit, the frequency increases to 4\,per\,cent. This is by no means impossible, but a 1-in-25 chance to find the observed alignment, on top of the overal rarity of GPoA analogs in simulations, can neither be seen as reassuring for the validity of the $\Lambda$CDM models.
It is important to point out that these results are consistent between the IllustrisTNG and the ELVIS simulations. Thus, despite their differences in the radial distribution of the considered satellites, the orbital alignments are comparable, and thus seem to be a universial feature of satellite systems in $\Lambda$CDM.

If the satellite systems are randomized, the chance to draw a pair of satellites that is as closely aligned to the best-fit satellite plane is even lower, though the frequencies are typically only approximately 50\,per\,cent lower than for the simulated systems. This is in agreement with the fact that the satellite systems in cosmological simulations show some underlying correlation in their phase-space distribution. However, as has been demonstrated in the context of planes of satellite galaxies before, this additional correlation compared to randomized systems is not very strong \citep{2019ApJ...875..105P,2020MNRAS.491.3042P}.

One can argue that the observed satellite galaxies NGC147 and NGC185 are somewhat special among the on-plane GPoA satellites. They both follow the line-of-sight velocity trend of the overall GPoA, are both dE galaxies and among the most-massive satellites of M31, and they both are at distances beyond 110\,kpc from their host. It is thus conceivable that any of these properties could have an effect on the chance to have orbits that are more closely aligned with the satellite plane. To test this possibility, we also determine the frequency of similarly close orbital alignment for appropriately selected subsamples of the on-plane satellites in the mock satellite systems.

If we only draw from those satellites which are part of the dominant line-of-sight velocity trend (i.e. contribute to $N_\mathrm{corr}$), the frequencies indeed increase. They are $f_\mathrm{147+185} = 0.34 (7.07)$\,\% for the TNG50, $f_\mathrm{147+185} = 0.19 (7.44)$\,\% for the TNG100, and  $f_\mathrm{147+185} = 0.09 (7.93)$\,\% for the ELVIS simulations. Such an increase is to be expected, as now all satellites whose line-of-sight velocity makes it unlikely that they co-orbit along the satellite plane are excluded from the considered sample, increasing the chance to draw those satellites which show a better orbital alignment with the satellite plane.

If we additionally require that the satellites we draw from are among the five most-massive of the on-plane satellites for each given system, the determined frequencies do not change substantially. The mass of the satellites thus does not seem to have an noticable effect on their orbital alignment with the satellite plane in simulations. 
 Similarly, requiring the satellites to be beyond 110\,kpc, and thus ensuring that the considered simulated satellites are among the more distant part of the on-plane satellites, does not substantially change the frequencies. In this context it is worth noting that any small increase in the frequencies  when selecting only the more massive or more distant satellites is also reflected by a similar increase in the frequencies for randomized satellite systems. This indicates that any change is within the expected numerical scatter for completely uncorrelated systems, and thus not significant.

Finally, we can also restrict our comparison to those satellite systems which contain a satellite plane with a high $N_\mathrm{corr} \geq 13$. This is maybe the most fair comparison to the observed GPoA, which after all is characterized by a high degree of line-of-sight velocity coherence. Note, however, that this selection suffers from low number statistics due to the limited number of simulated satellite systems which reproduce the observed value of $N_\mathrm{corr}$.
For the TNG100 and ELVIS+PhatELVIS simulations, we find no evidence that satellite planes with a high degree of kinematic correlation show a tighter alignment between the individual satellite orbits and the orientation of their best-fit satellite planes.
For the TNG50, we indeed find a substantially enhanced chance to draw two satellite galaxies that are similarly closely co-orbiting with their respective satellite plane as found for the observed M31 system. In 4.5\% of the cases two satellites are drawn which have orbital poles aligned as well as the most-likely orbital poles of NGC\,147 and NGC\,185, while if one only requires them to match within the $1\sigma$\ limits this fraction rises to 19.2\%. This appears to be mainly driven by one of the two simulated satellite systems, which has a set of satellite plane properties that all come intriguingly close to the observed GPoA. This warrants some closer attention in the following subsection.

\subsection{Some Specific Simulated Systems}

\begin{figure*}
\gridline{\fig{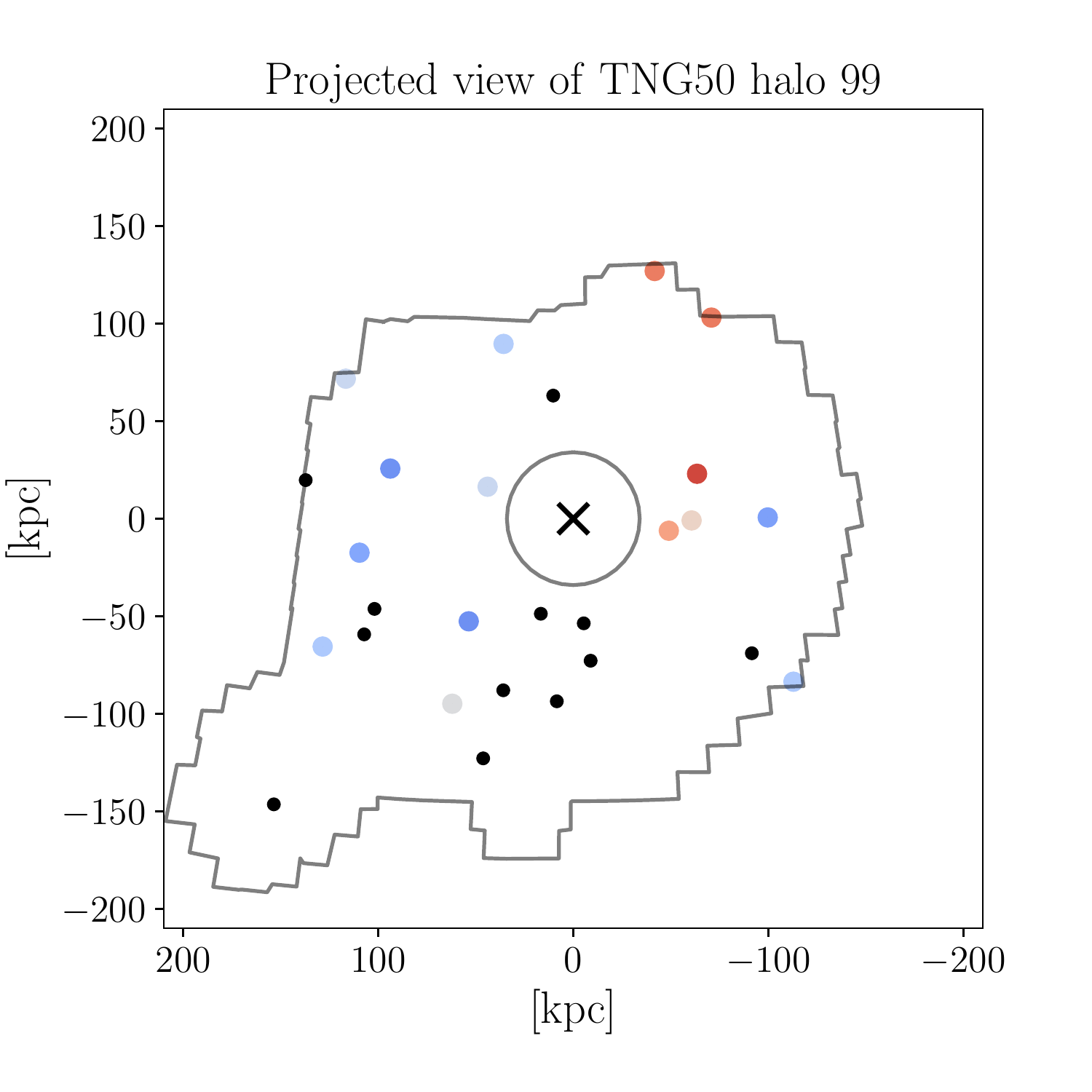}{0.33\textwidth}{(a)}
          \fig{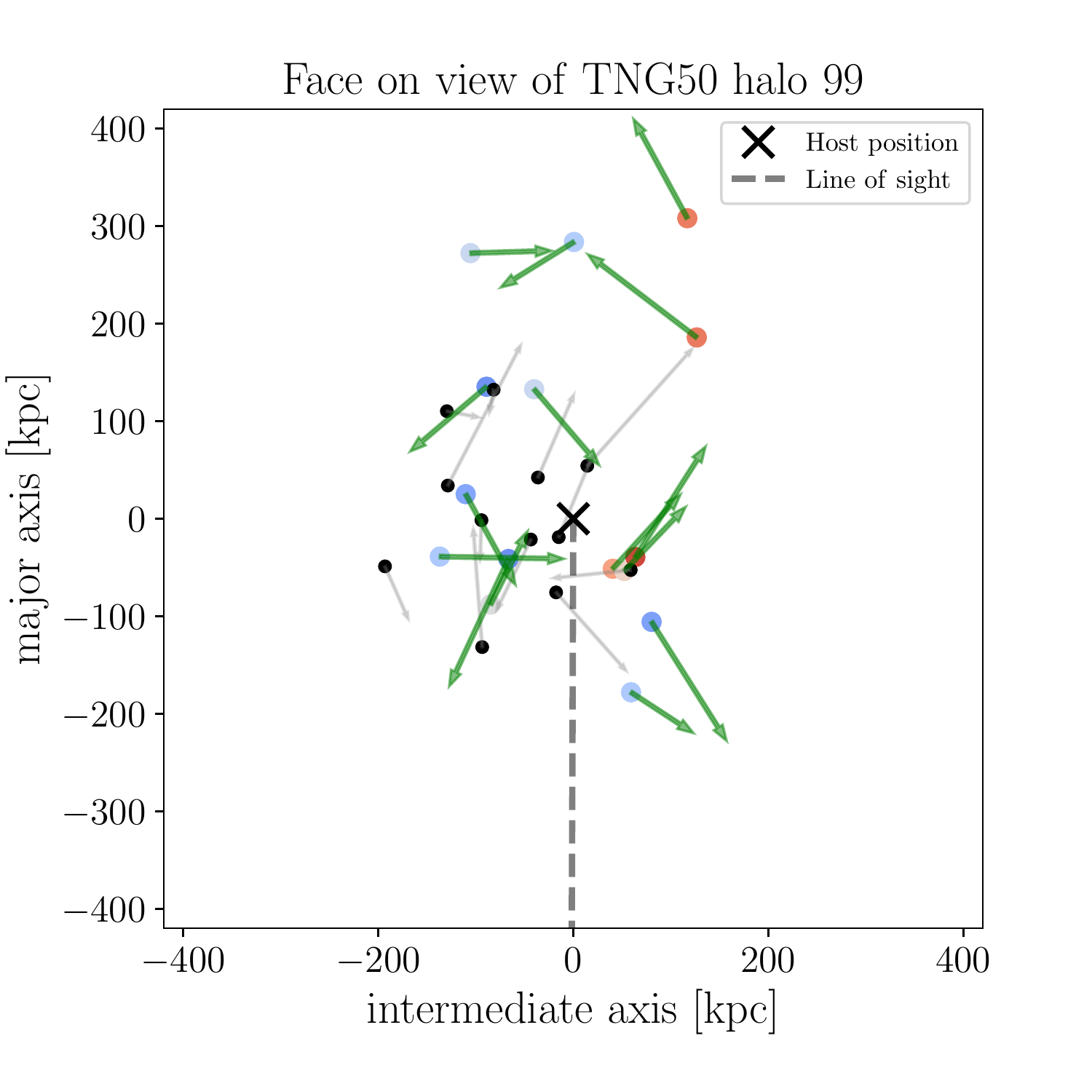}{0.33\textwidth}{(b)}
          \fig{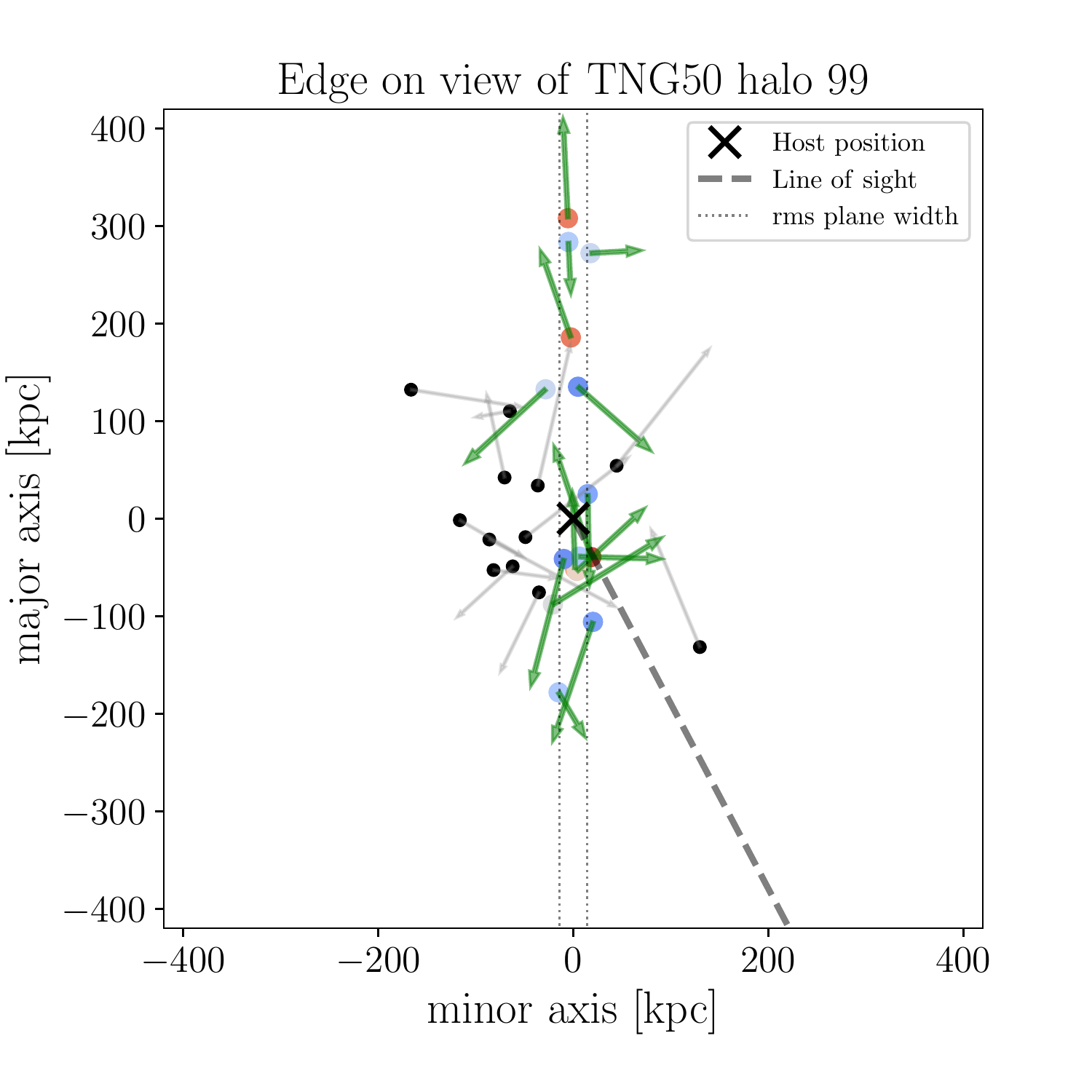}{0.33\textwidth}{(c)}
          }
\gridline{\fig{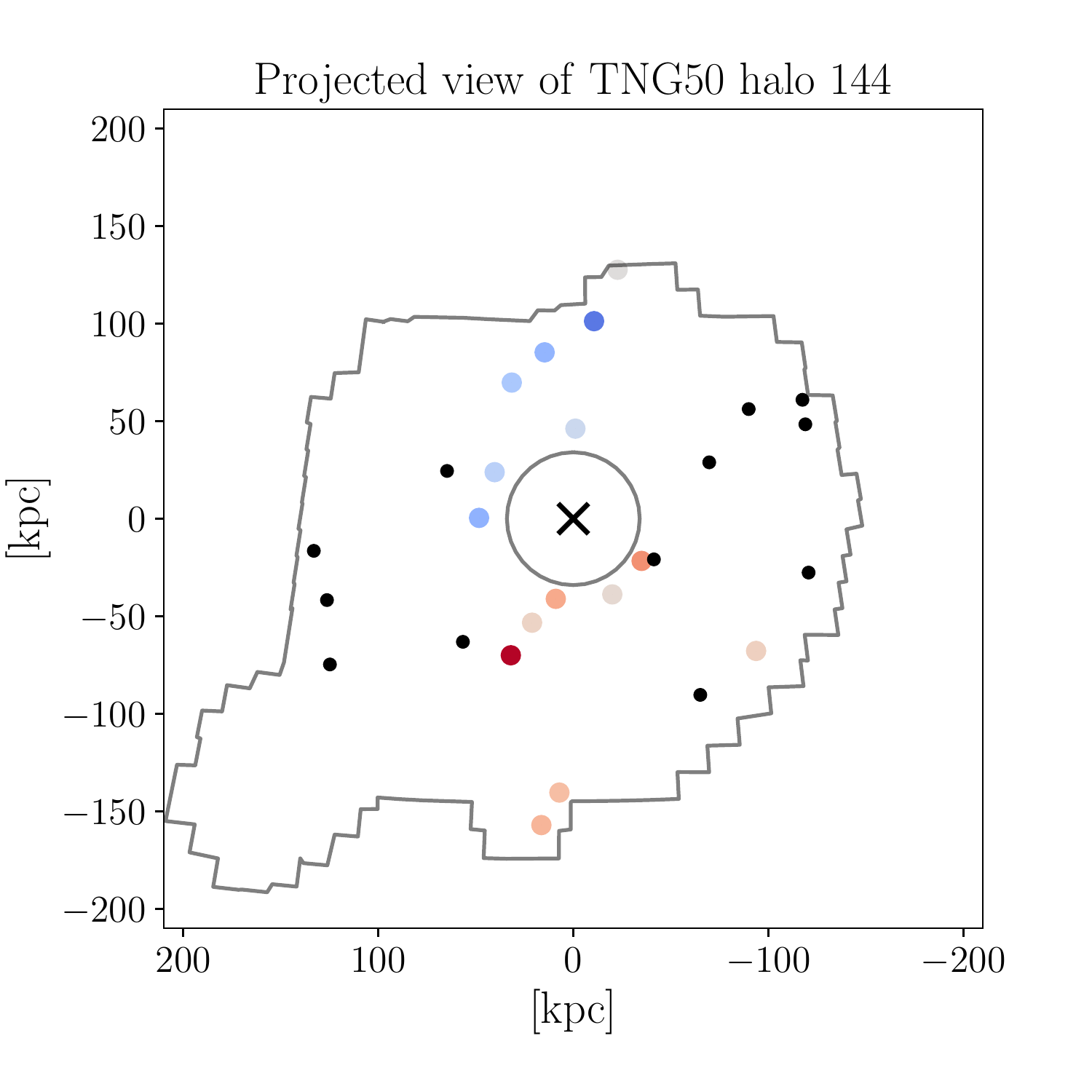}{0.33\textwidth}{(d)}
          \fig{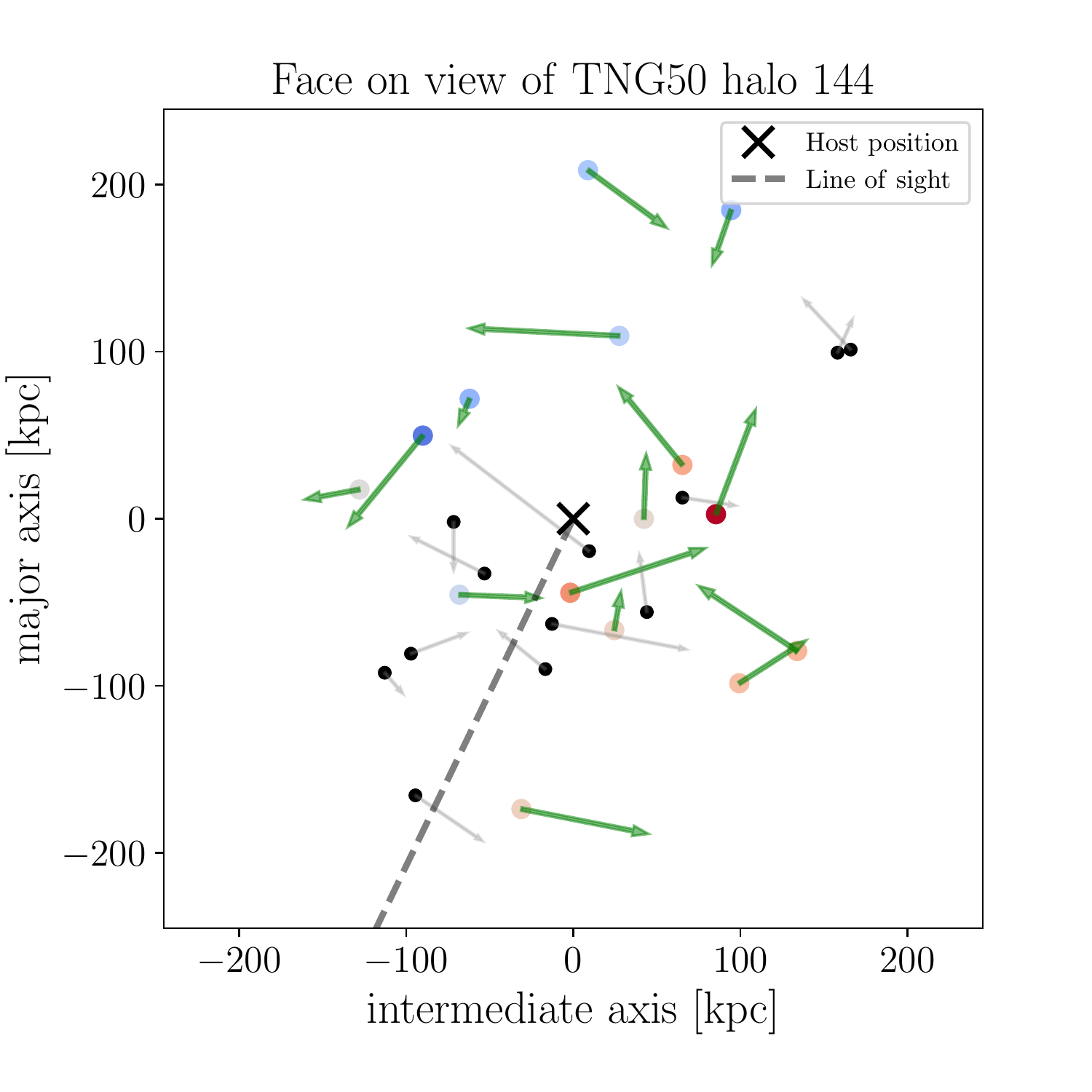}{0.33\textwidth}{(e)}
          \fig{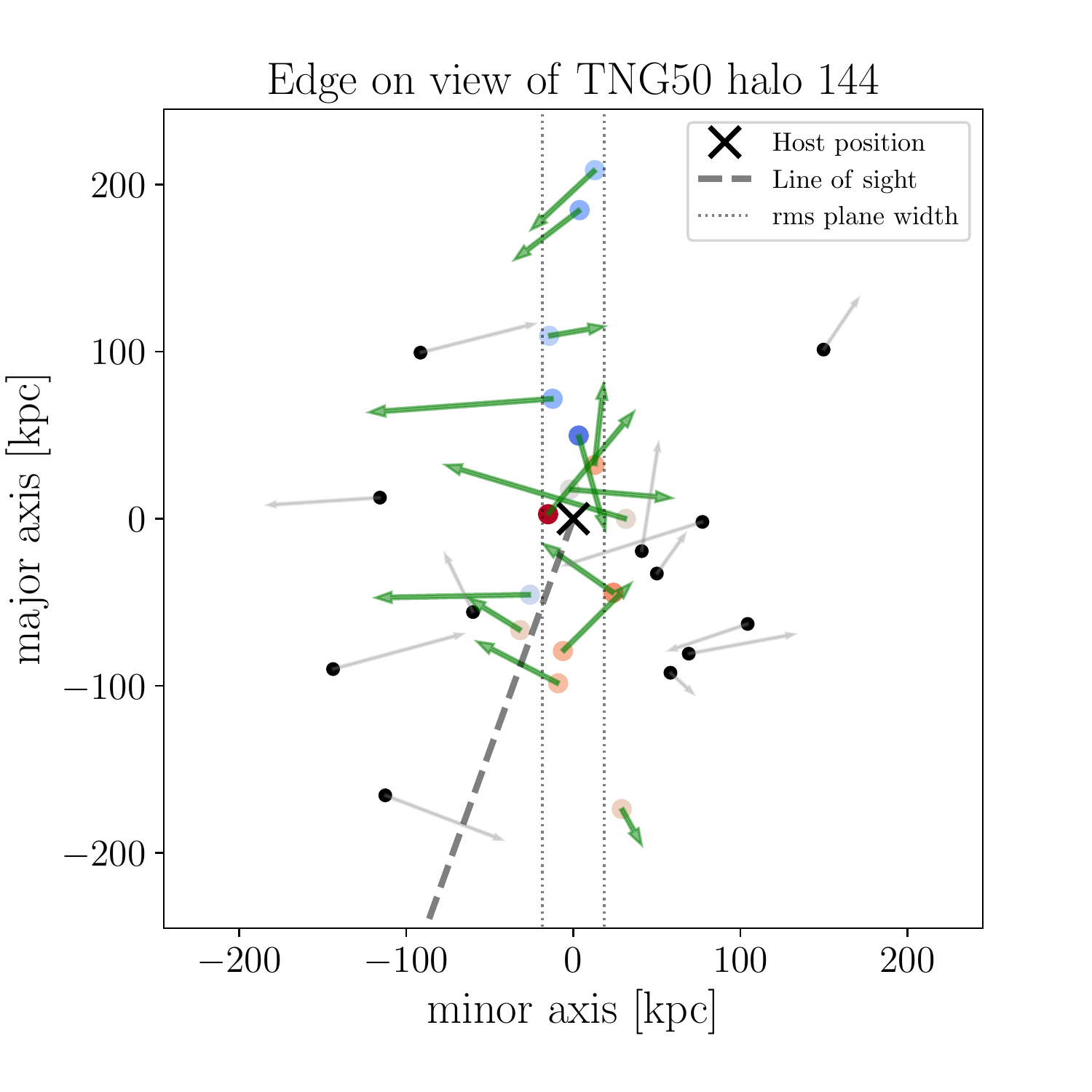}{0.33\textwidth}{(f)}
          }
\caption{Three projections of the spatial distribution of simulated satellite galaxy systems around two host halos in the TNG50 simulation, selected for having the strongest line-of-sight velocity correlation ($N_\mathrm{corr} = 13$). The left panels show the mock-observed systems projected along the line-of sight, with the PAndAS selection volume indicated by the poligon and the inner excludion region around the host (black cross) indicated by a circle. On-plane satellites are color-coded by their line-of-sight velocity (blue approching, red receding), non-members are plotted in smaller black symbols. The planes are significantly inclined in both cases, and thus not as pronounced as the observed GPoA is in this projection.
The middle panels plot face-on view of the satellite planes, rotated such that their major axis aligns is vertical and the intermediate axis is horizontal. The line-of-sight from the host to the mock-observer is indicated with the grey dashed line, and the satellite velocities projected into this view are indicated with arrows (green for plane members). The symbols are again color-coded according to the line-of-sight velocity relative to the mock-observer, as in the left panels.
The right panels are similar, but display the minor axis on the horizontal, and also indicate the width of the satellite planes with dotted lines. Especially halo\,144 (lower right panel) displays strong tangential motions outside of the identified satellite plane.
\label{fig:examplesystems}}
\end{figure*}

Two out of the 146 simulated satellite galaxy systems from the TNG50 simulation have satellite planes that display a high degree of line-of-sight velocity correlation with $N_\mathrm{corr} = 13$. We identify them by their TNG50 FoF halo IDs as halo\,99 and halo\,144. Figure \ref{fig:examplesystems} presents the spatial distribution and velocity directions of these two satellite galaxy systems, emphasizing the satellites identified as plane members.

For halo\,99 (top panels in Fig. \ref{fig:examplesystems}), the identified satellite galaxy plane has a root-mean-square height of $r_\perp(\mathrm{halo\,99}) = 14.1$\,kpc, which is close to the observed value of $r_\perp(\mathrm{M31}) = 12.6 \pm 0.6$. The radial extent of the satellite distribution of $r_\parallel(\mathrm{halo\,99}) = 181$\,kpc is also well comparable with that of the \citep{2013Natur.493...62I} satellites of $r_\parallel(\mathrm{M31}) = 191$\,kpc (assuming their adopted, far distance for Andromeda XXVII; for the closer alternative distance $r_\parallel(\mathrm{M31}) = 147$\,kpc). 
Of the 15 on-plane satellites, seven have orbital poles that align closely ($\theta_L < 25^\circ$) with the plane normal and can thus indeed be considered as co-orbiting along the satellite plane. In contrast, close-to-perpendicular motions ($60^\circ < \theta_L < 120^\circ$) relative to the satellite plane normal are found for only four of the 15 on-plane satellites. Yet, $N_\mathrm{corr} = 13$\ gives a misleading impression of the true number of satellites that have prograde orbits with respect to the satellite plane, as only 11 have $\theta_L < 90^\circ$. Intriguingly, of the 13 satellites that follow the common line-of-sight velocity trend, three have $\theta_L > 90^\circ$\ and are thus in fact counter-orbiting in 3D, with one having an orbital plane aligned to within $\approx 30^\circ$\ with the satellite plane. Meanwhile, one of the two satellites that do \textit{not} follow the line-of-sight velocity trend does in fact have a reasonably well aligned orbital pole ($\theta_L = 30^\circ$) in the co-orbiting direction. This means that there is a non-negligible chance that a satellite galaxy which is thought to co-orbit based on its line-of-sight velocity is instead found to counter-orbit once the full 3D velocity is measured, and based on these $\Lambda$CDM simulations one might in fact expect to find a few such cases among the M31 satellites.

The satellite plane identified around halo\,144 (bottom panels in Fig. \ref{fig:examplesystems}) is in less good agreement with the observed M31 satellite plane parameters. The plane is considerably wider ($r_\perp(\mathrm{halo\,144}) = 18.5$\,kpc), but also a lot more compact in its radial extent ($r_\parallel(\mathrm{halo\,144}) = 126$\,kpc). In line with the less pronounced spatial flattening, the satellites also display considerably higher velocity components outside of the plane (see the edge-on view in Fig. \ref{fig:examplesystems}). Only four out of 15 on-plane satellites have orbital poles aligned to with $\theta_L < 25^\circ$\ of the plane normal vector, while seven have highly perpendicular motions with orbital poles inclined by $60^\circ < \theta_L < 120^\circ$\ relative to the satellite plane normal. However, in contrast to halo\,99, $N_\mathrm{corr} = 13$\ here gives a realistic impression of the overall orbital sense, as 13 out of the 15 on-plane satellites are on prograde orbits ($\theta_L < 90^\circ$) relative to the satellite plane normal. Nevertheless, one of the 13 satellites following the line-of-sight velocity trend is in fact counter-orbiting (and has an orbital plane aligned with the spatial satellite plane to within $35^\circ$).

Neither of the two systems is oriented as close to edge-on as the observed GPoA. The simulated satellite planes are inclined by $\phi_\mathrm{los}(\mathrm{halo\,99}) = 27.8^\circ$\ and $\phi_\mathrm{los}(\mathrm{halo\,144}) = 18.0^\circ$\ with respect to the mock-observed line-of-sights. Since we do not model uncertainties, this affects the comparability of the identified simulated planes with the real structure around M31. Adpting a typical TRGB distance error of 5\% at the distance of M31 of $\approx 800$\,kpc corresponds to positional uncertainties of $\pm 40$\,kpc. For an inclined satellite plane, a fraction $\sin(\phi_\mathrm{los})$\ of this error is projected into the direction perpendicular to the plane. For $\phi_\mathrm{los} = 30^\circ$, we can therefore expect an additional contribution of $\pm 20$\,kpc to the inferred plane height, which is of order of the intrinsic plane heights and adds to them in quadrature. Thus, the simulated satellite plane heights are underestimating the expected measured plane heights by about a factor of $\sqrt{2}$. One can thus not unreservedly consider these systems as analogs of the GPoA. Nevertheless, and while beyond the scope of the present study, it will be of interest in the future to study the evolution of these and other satellite systems resembling the observed M31 satellites and their planar structure, to learn more about the influence of group accretion events, past minor or major mergers, and the host's environment and surrounding cosmic web.

\section{Conclusion} \label{sec:conclusion}

We have compared the observed Great Plane of Andromeda (GPoA) with satellite galaxy systems from a range of $\Lambda$CDM cosmological simulations: the hydrodynamical Illustris TNG50 and TNG100 simulations, as well as the dark-matter-only ELVIS and PhatELVIS simulations. There is good agreement of results between these sets of cosmological simulations regarding the flattening and orbital correlation of on-plane satellite galaxies, despite their substantial differences in implemented physics, resolution, and resulting radial distribution of the selected satellite galaxies among the simulations. We therefore consider our results robust expectations of the underlying hierachical galaxy formation paradigm within the $\Lambda$CDM context, as they are not merely a result of one specific simulation.

As found in previous studies, the simulated satellite systems preferentially populate regions in spatial flattening and kinematic correlation that are well removed from the corresponding properties of the observed GPoA. Satellite planes as extreme as the observed GPoA are rare\footnote{Note, that this does not account for the look-elsewhere effect, and an alternative approach based on identifying the most-significant sub-sample satellite planes has been proposed by \citet{2015MNRAS.452.3838C}.}, with none of our 909 simulated satellite systems simultaneously reproducing the flattening and the kinematic correlation of the observed GPoA. This indicates that the frequency of similarly extreme systems is of order 1-in-1000 or lower, in line with previous results based on a different cosmological simulation \citep{2014ApJ...784L...6I, 2014MNRAS.442.2362P}.

The measurement of proper motions for the first two M31 satellite galaxies, NGC\,147 and NGC\,185 by \citet{2020ApJ...901...43S}, makes it possible to for the first time investigate the orbital alignment of M31 satellites with the GPoA, and compare to analogous measurements in the cosmological simulations. Specifically, we have evaluated the angular offset between the normal vector of the best-fit satellite plane, and the orbital pole (direction of angular momentum) of the satellite galaxies.  We find that within current measurement uncertainties, the orbital poles of NGC\,147 and NGC\,185 align to better than $\theta_{1\sigma} = 20^\circ$\ and to about $\theta_{1\sigma} = 30^\circ$, respectively, and the most-likely proper motions result in considerably tighter alignments.

In the simulated systems, the orbital poles of satellites are on average slightly better aligned with their spatial planes than expected from completely random systems, in line with the expected presence of a mild degree of phase-space correlation in cosmological simulations. Many simulated satellite galaxy systems ($\approx 70$\%) have their two {\it best-aligned} satellites similarly closely aligned as the $1\sigma$\ uncertainty range of the two observed satellite galaxies. The alignment of these two satellites alone thus does not preclude a consistency with $\Lambda$CDM. 

Yet, it is by no means guaranteed, and would in fact constitute a puzzling coincidence, if the first two on-plane satellites of M31 for which proper motions were measured happen to be the two best-aligned ones. The chance to find an orbital pole alignment within the $1\sigma$ limits of NGC\,147 and NGC\,185 when drawing any two on-plane satellites from the simulated systems is only 4 to 8\%, and this number drops to the sub-percent level if compared to the tigher most-likely orbital pole alignments. Neither the mass nor the distance of a simulated satellite from their host appears to have an effect on their orbital alignment with the satellite plane, such that one can not claim such exceptional properties of NGC\,147 and NGC\,185 to explain their well aligned orbital poles.

If we assume that the close orbital alignment of NGC\,147 and NGC\,185 is representative of the typical on-plane satellite of M31, then none of the simulated systems show a similar degree of orbital alignment for a majority of its on-plane satellites. Thus, if half of the remaining on-plane satellites of the GPoA are found with future proper motion measurements to have similarly closely aligned orbital poles, then this alone would pose an extreme problem for $\Lambda$CDM, irrespective -- and in addition to -- of the rarity of the spatial and line-of-sight correlated nature of the GPoA.

In the simulated satellite systems, a high degree of correlation among the line-of-sight velocities of mock-observed satellites indicates that some orbital alignment is more likely, but a majority of satellites can still have orbits almost perpendicular to the respective satellite plane. Therefore, a strongly line-of-sight velocity correlated satellite plane does not guarantee tight orbital alignment. We also studied two specific simulated satellite galaxy systems in TNG50, selected as the closest to the observed GPoA in flattening and velocity correlation of their respective satellite planes. Both systems contain several satellite galaxies that are identified as satellite plane members and follow a common line-of-sight velocity trend, but that in fact have motions that make them move out of the plane. We also find that, within $\Lambda$CDM, one can expect that some of the on-plane satellites whose line-of-sight velocity is indicative of co-rotation, have in fact 3D velocities that result in their counter-orbit. Obtaining 3D velocities via proper motion measurements is thus necessary to determine which satellites orbit along the structure, and to thereby better understand the observed GPoA and its dynamical status as either a transient alignment, a kinematic structure, or a mixture of both.

Luckily, on the observational front for constraining 3D motions of M31 satellite galaxies, the future is bright. In the next few years, proper motion measurements for 7 additional satellite galaxies (Andromeda\,I, II, III, VII, XV, XVI, and XXVIII), of which 3 belong to the GPoA, will be available through the ongoing HST program GO-16273 (PI: S. T. Sohn). Meanwhile, the uncertainty of the M31 proper motion will be improved by a factor of $\sim 2.5$ over previous HST measurements (Sohn et al., in prep.) based on the data obtained through the HST program GO-15658 (PI: S. T. Sohn) allowing significantly improved velocity zero points for the M31 satellites. Finally, future additional-epoch observations with HST, JWST, and Roman Space Telescope will utilize the first-epoch deep imaging data obtained through GO-15902 (PI: D. Weisz) for 23 M31 satellites to determine their proper motions, ultimately increasing the total sample of M31 satellites with proper motion measurements to nearly 30.

\acknowledgments

We thank the anonymous referee for their helpful comments and questions.
M.S.P. acknowledges funding of a Leibniz-Junior Research Group via the Leibniz Competition, and thanks the Klaus Tschira Stiftung and German Scholars Organization for support via a KT Boost Fund. This research has made use of NASA's Astrophysics Data System.\\

%



\software{astropy \citep{2018AJ....156..123A, 2013A&A...558A..33A},  
            IPython \citep{PER-GRA:2007},
            matplotlib \citep{Hunter:2007}, 
            NumPy \citep{harris2020array},
            SciPy \citep{Virtanen_2020}.
          }



\appendix

\section{Predicted proper motions for other M31 satellite galaxies}
\label{sect:appendix}

\begin{figure*}
\gridline{\fig{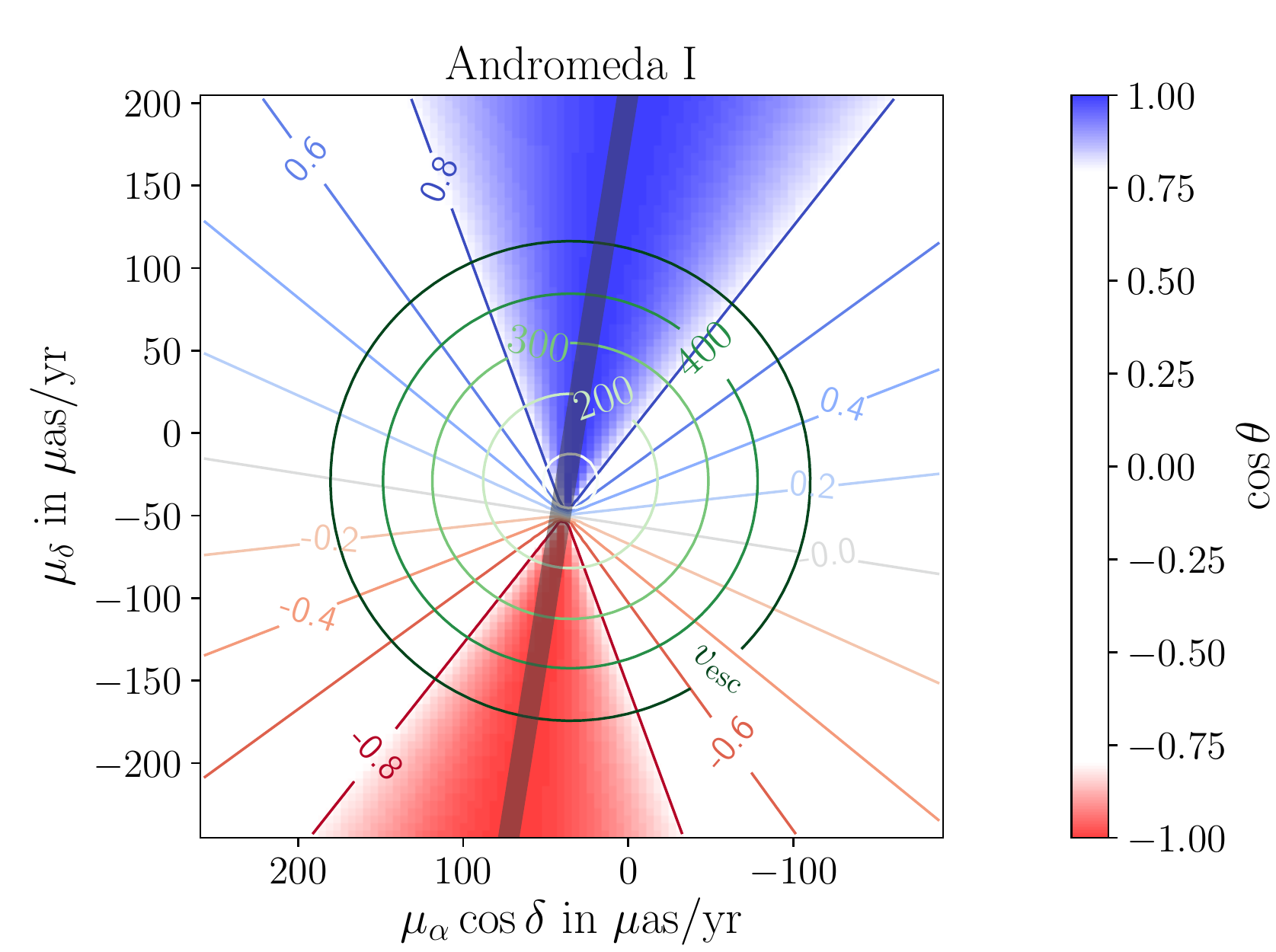}{0.33\textwidth}{(a)}
              \fig{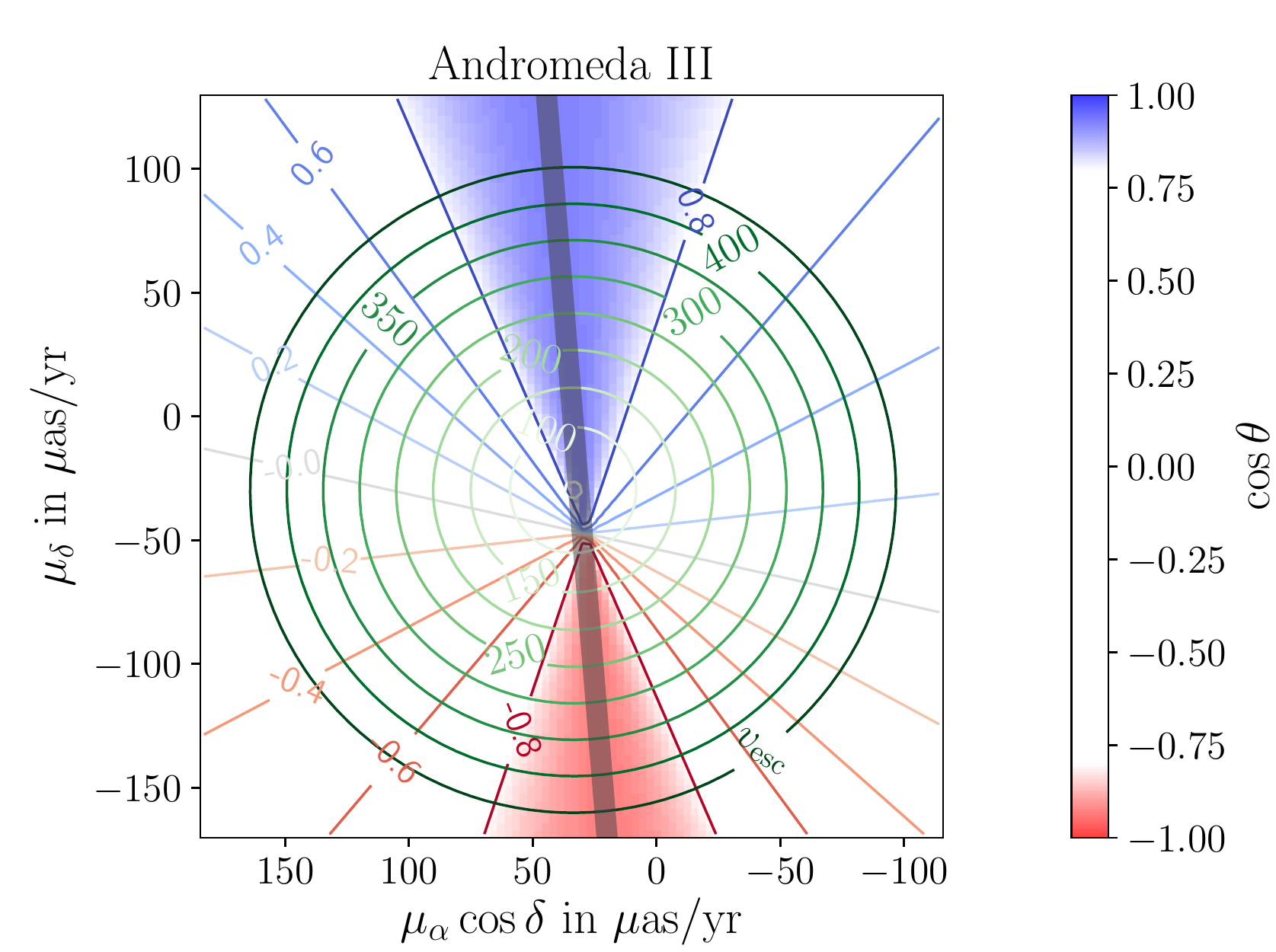}{0.33\textwidth}{(b)}
              \fig{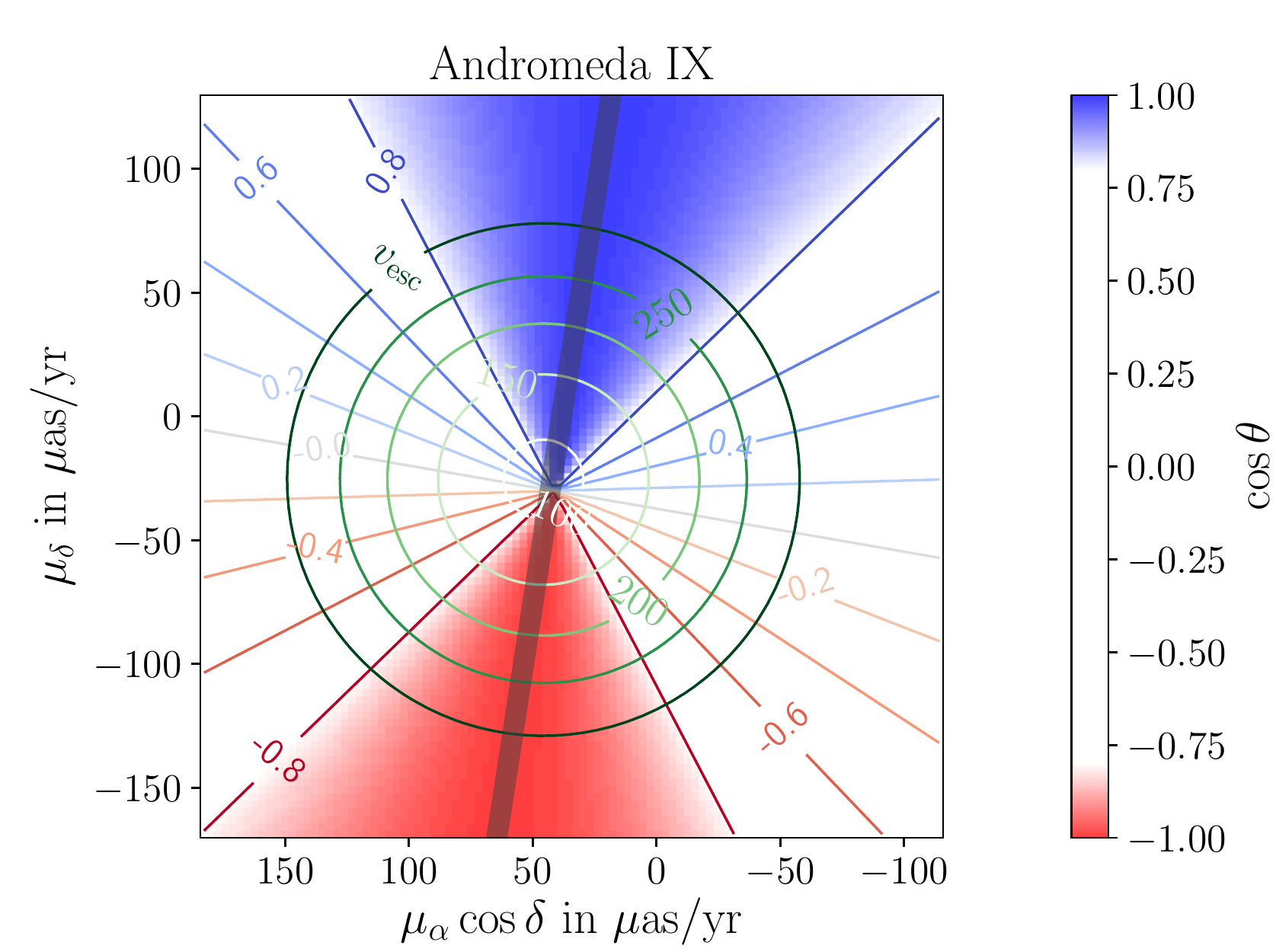}{0.33\textwidth}{(c)}
          }
\gridline{\fig{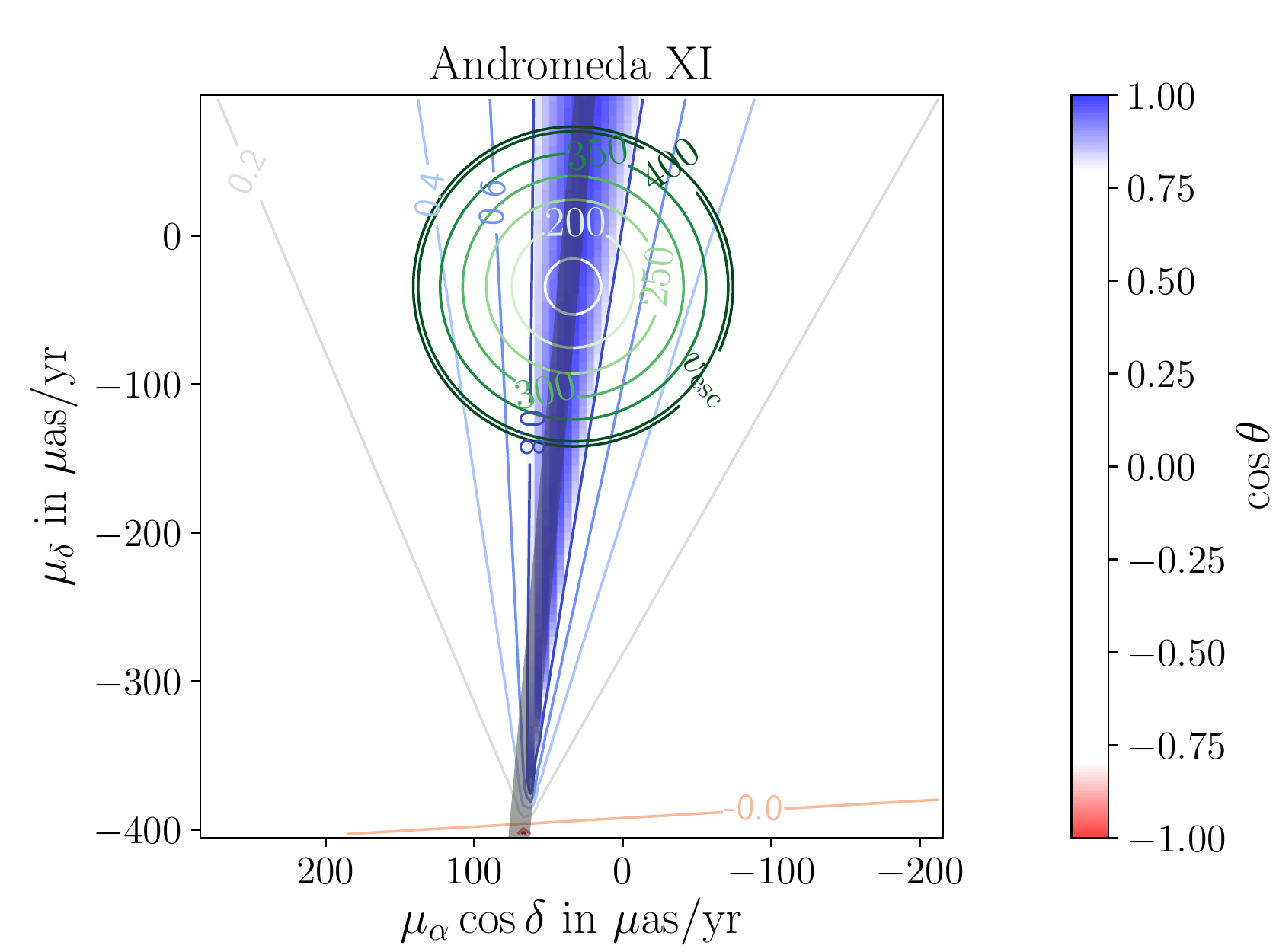}{0.33\textwidth}{(d)}
              \fig{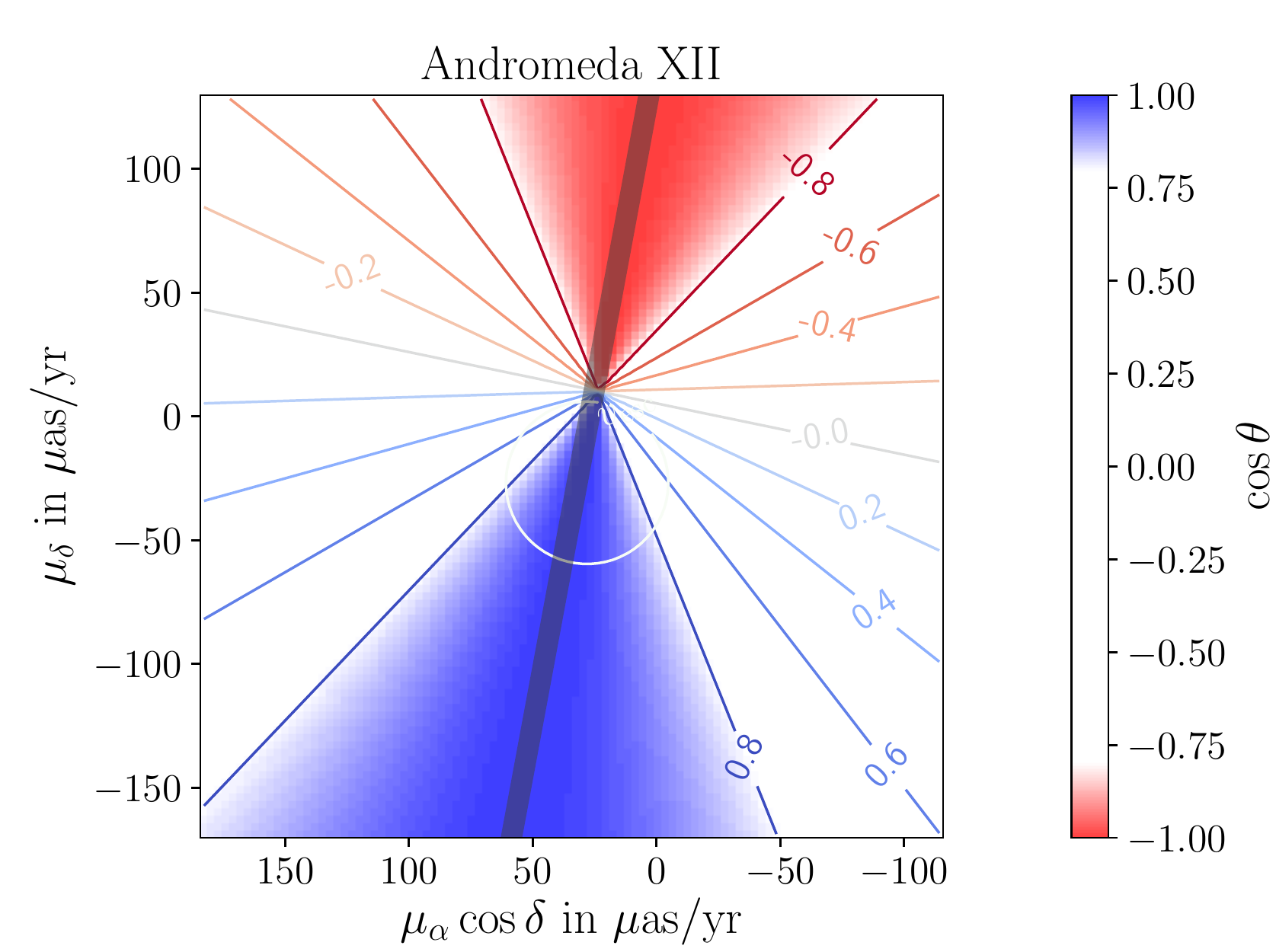}{0.33\textwidth}{(e)}
              \fig{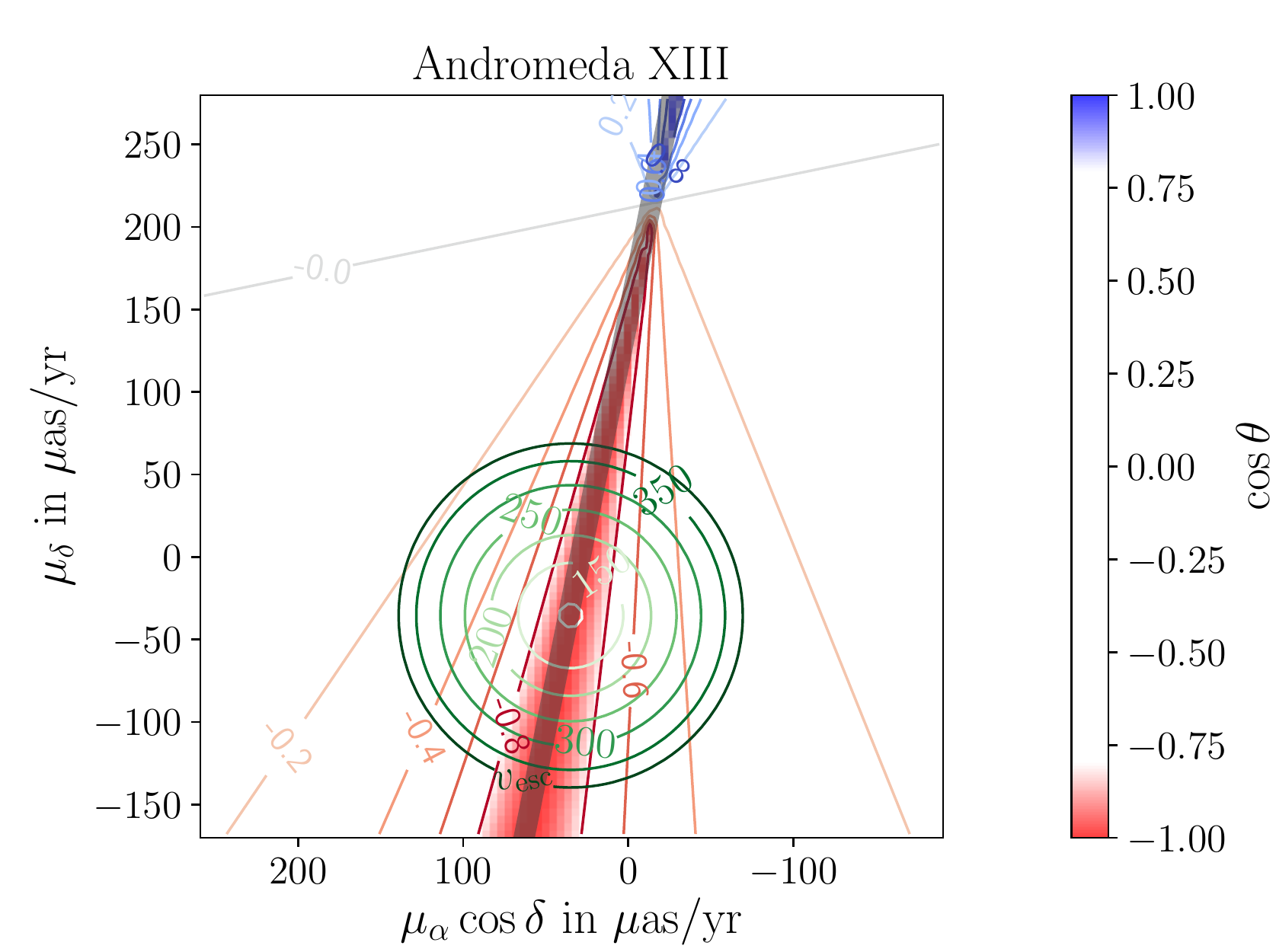}{0.33\textwidth}{(f)}
          }
\gridline{\fig{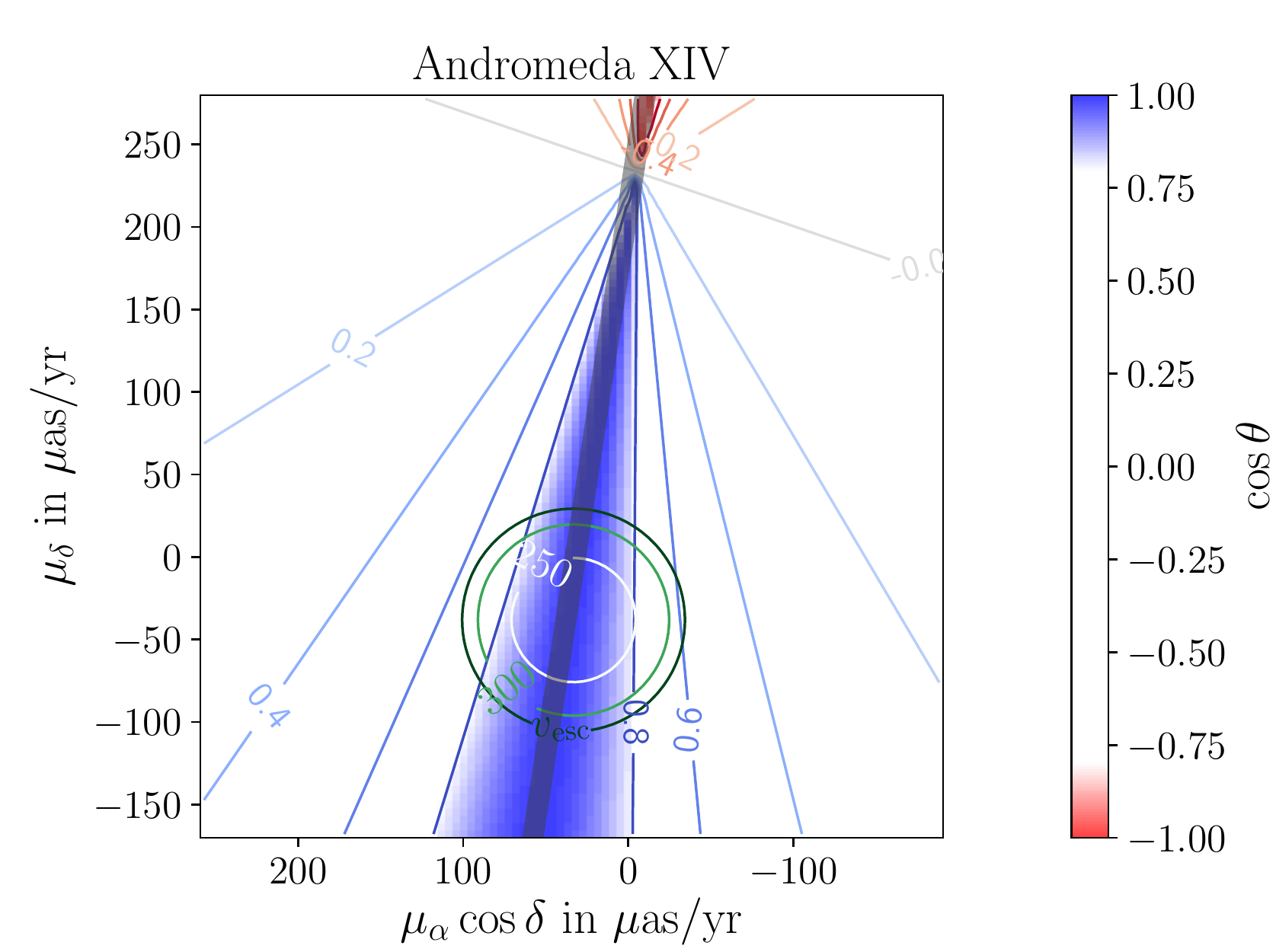}{0.33\textwidth}{(g)}
              \fig{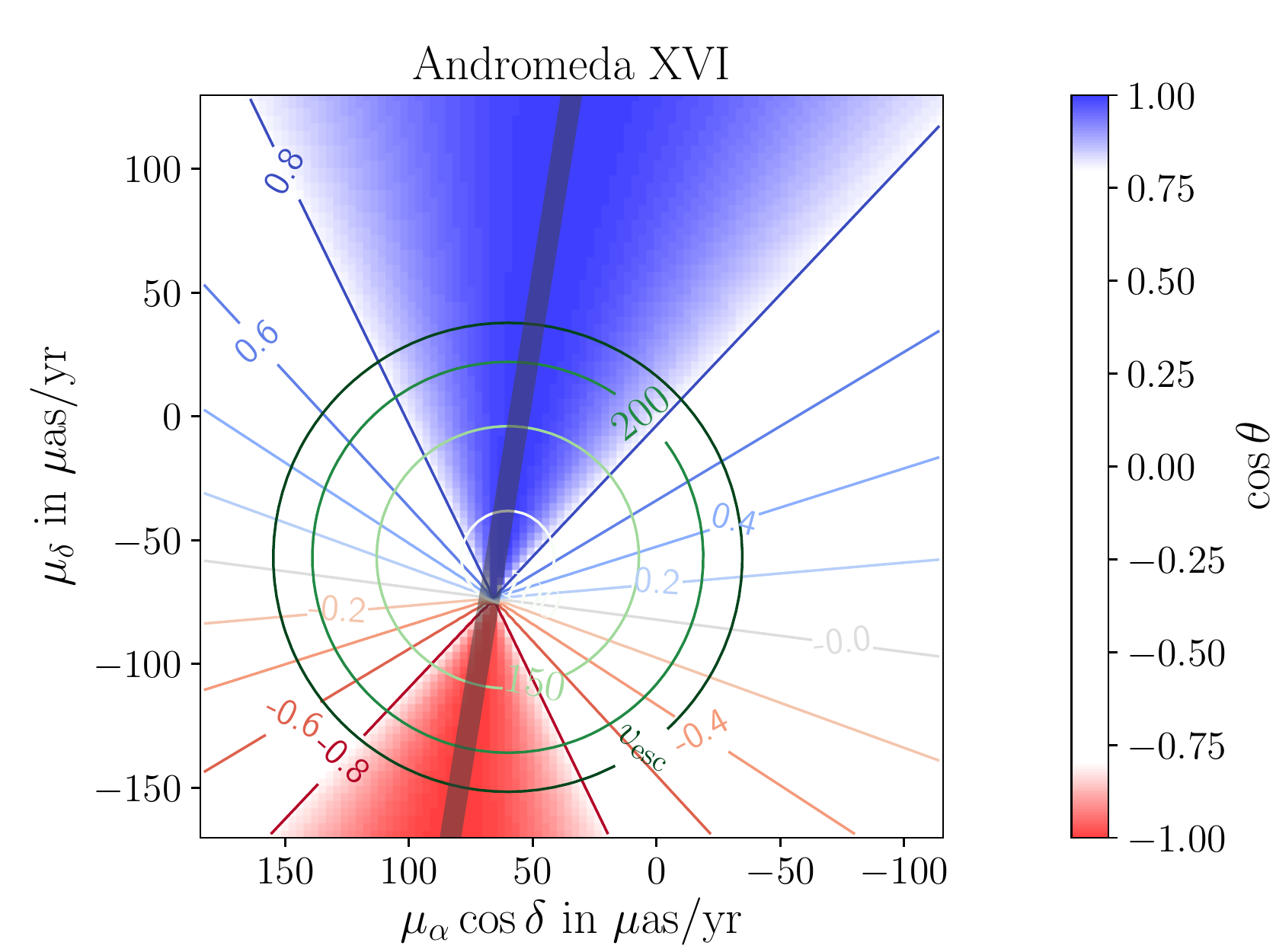}{0.33\textwidth}{(h)}
              \fig{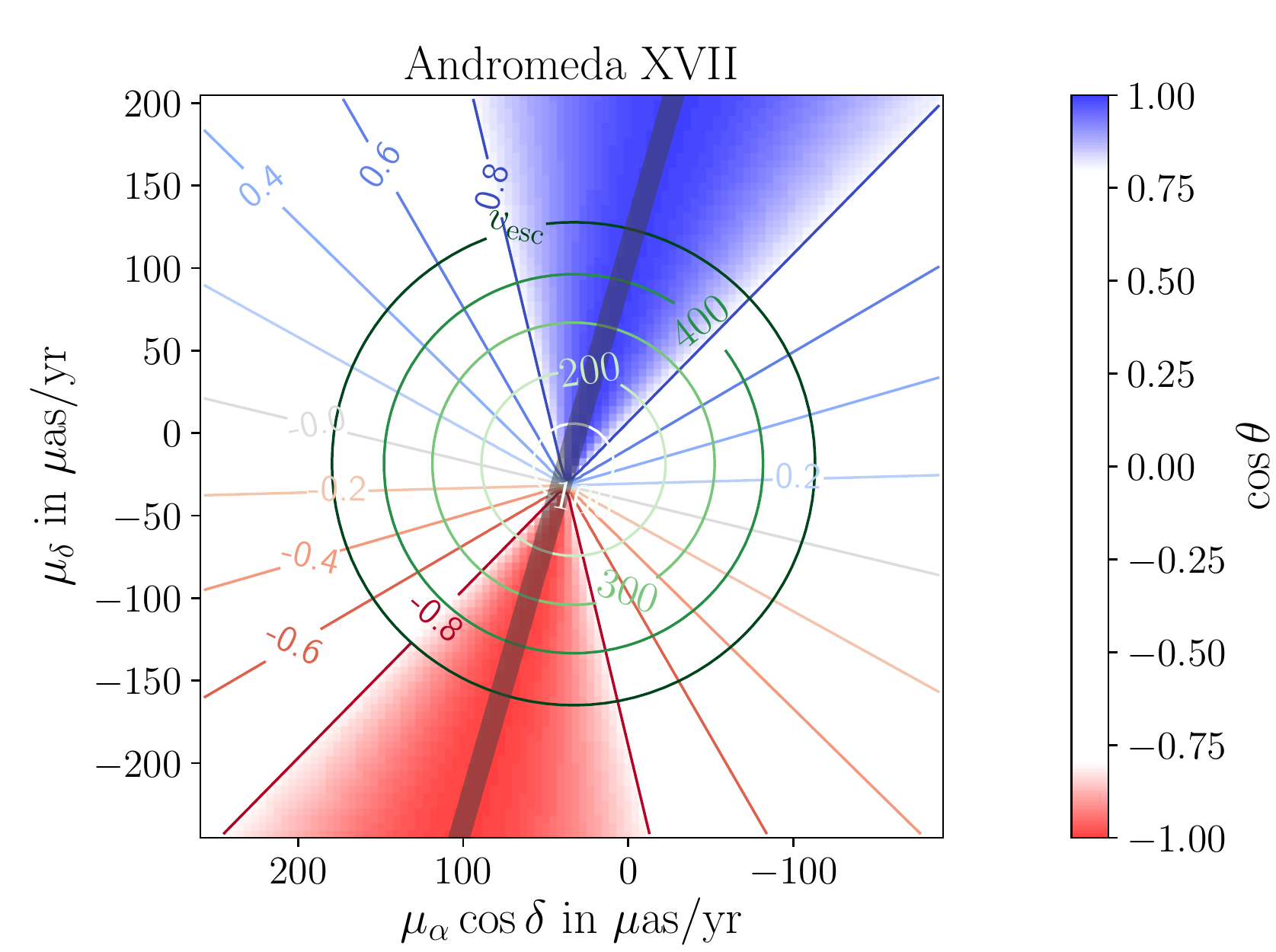}{0.33\textwidth}{(i)}
          }
\gridline{\fig{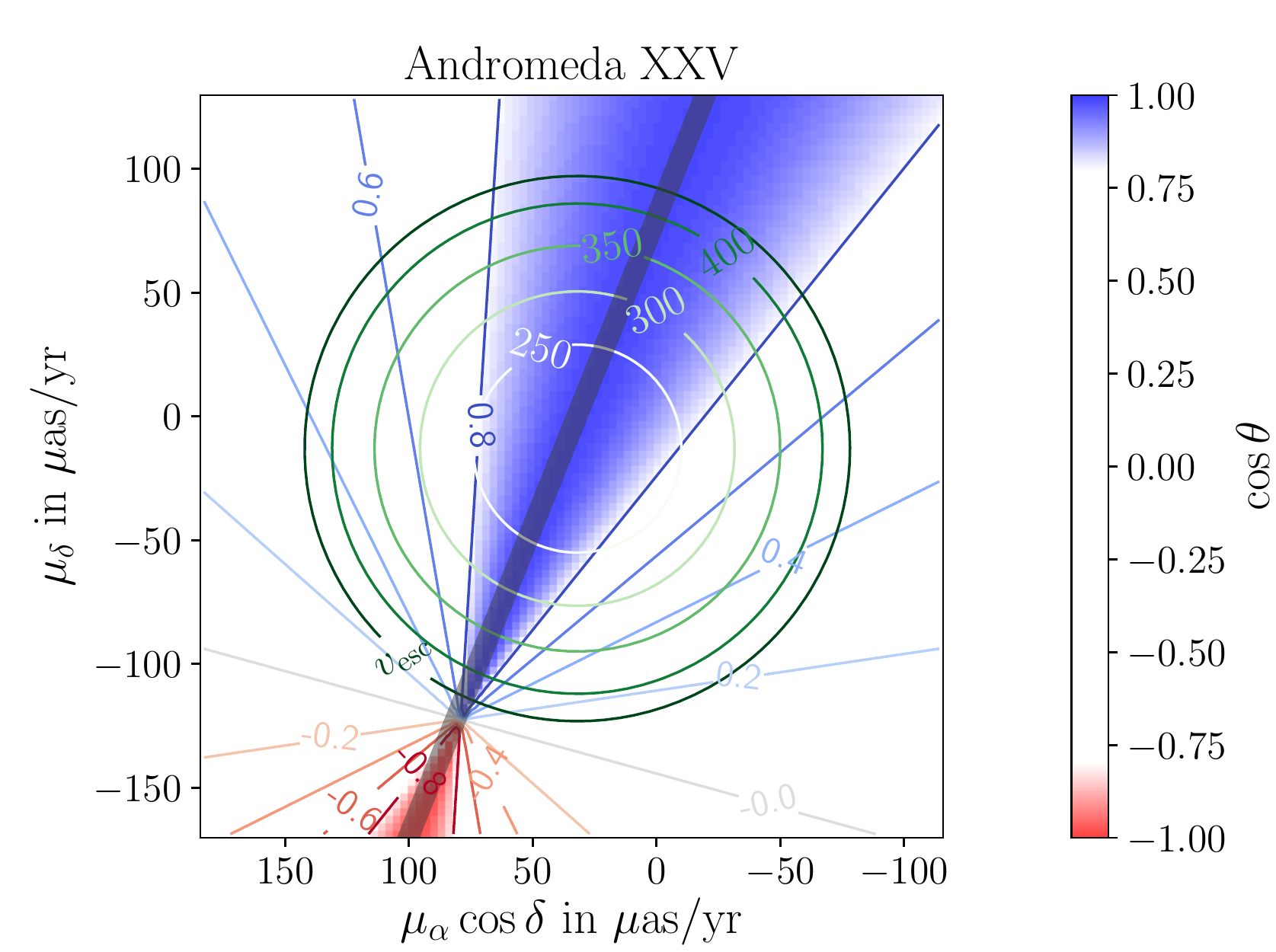}{0.33\textwidth}{(j)}
              \fig{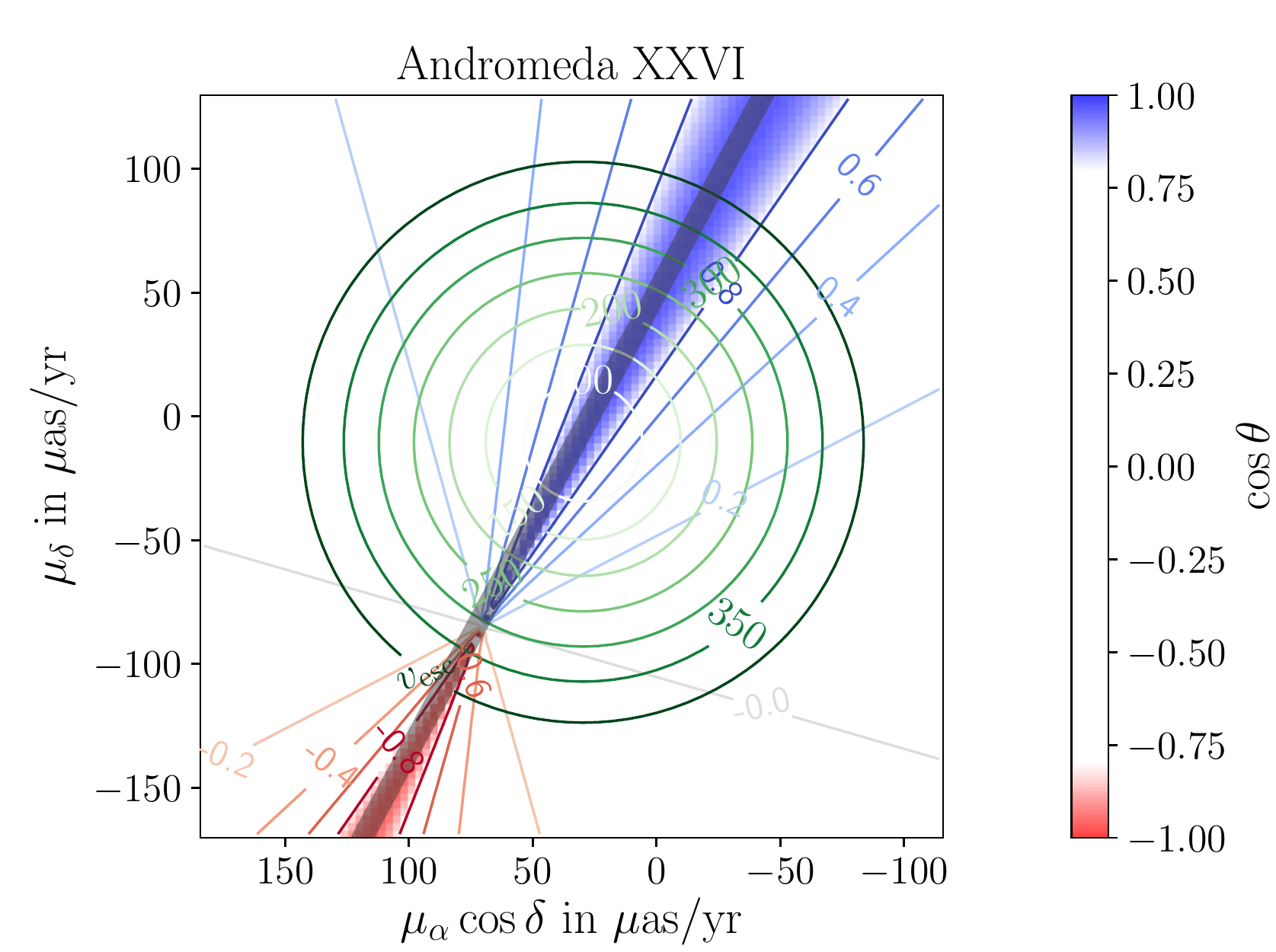}{0.33\textwidth}{(k)}
              \fig{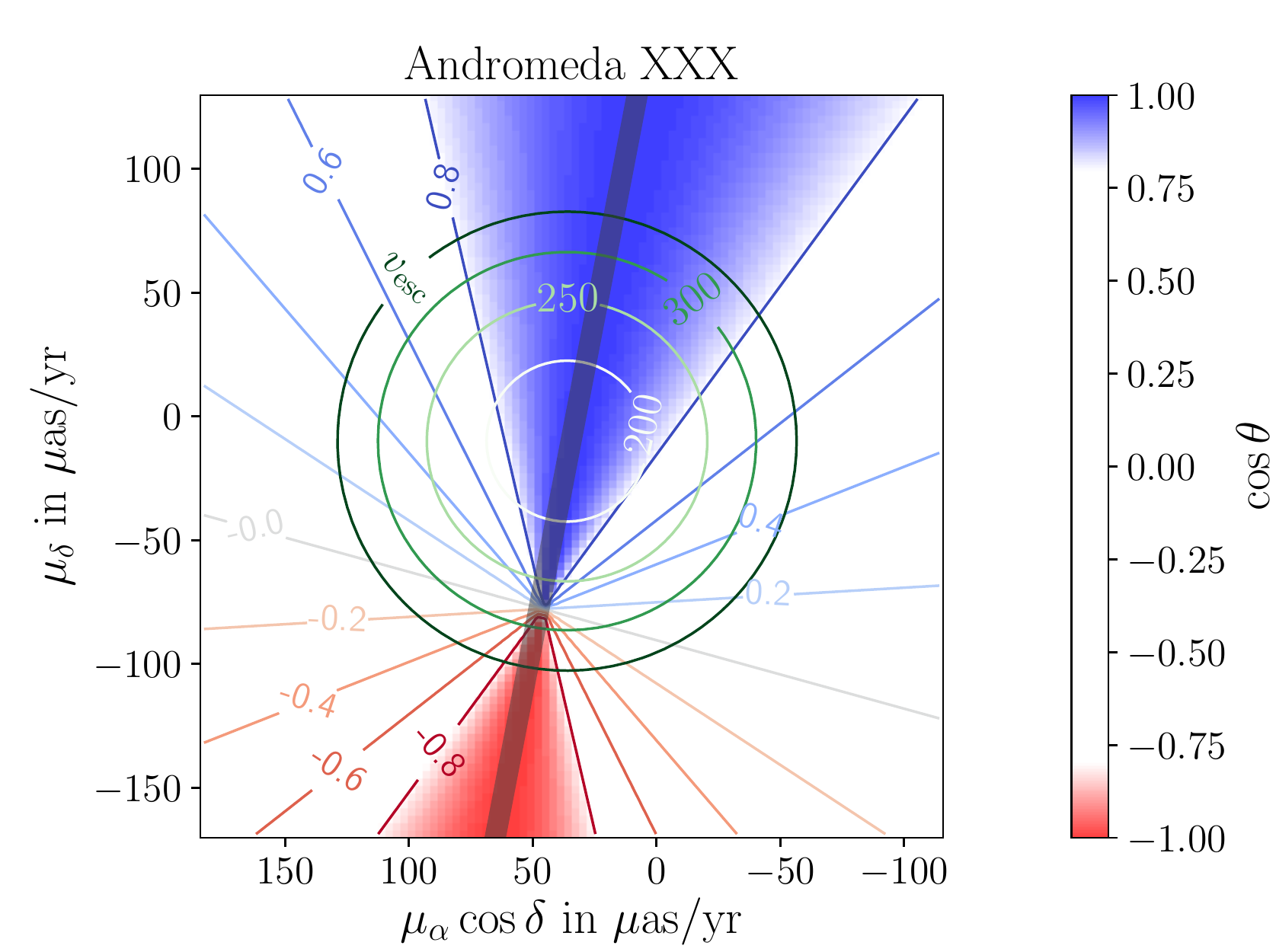}{0.33\textwidth}{(l)}
          }
\caption{Predicted proper motions for the other on-plane M31 satellites in the GPoA, same style as Fig. \ref{fig:PMprediction}.
\label{fig:PMprediction_others}}
\end{figure*}

\begin{figure*}
\gridline{\fig{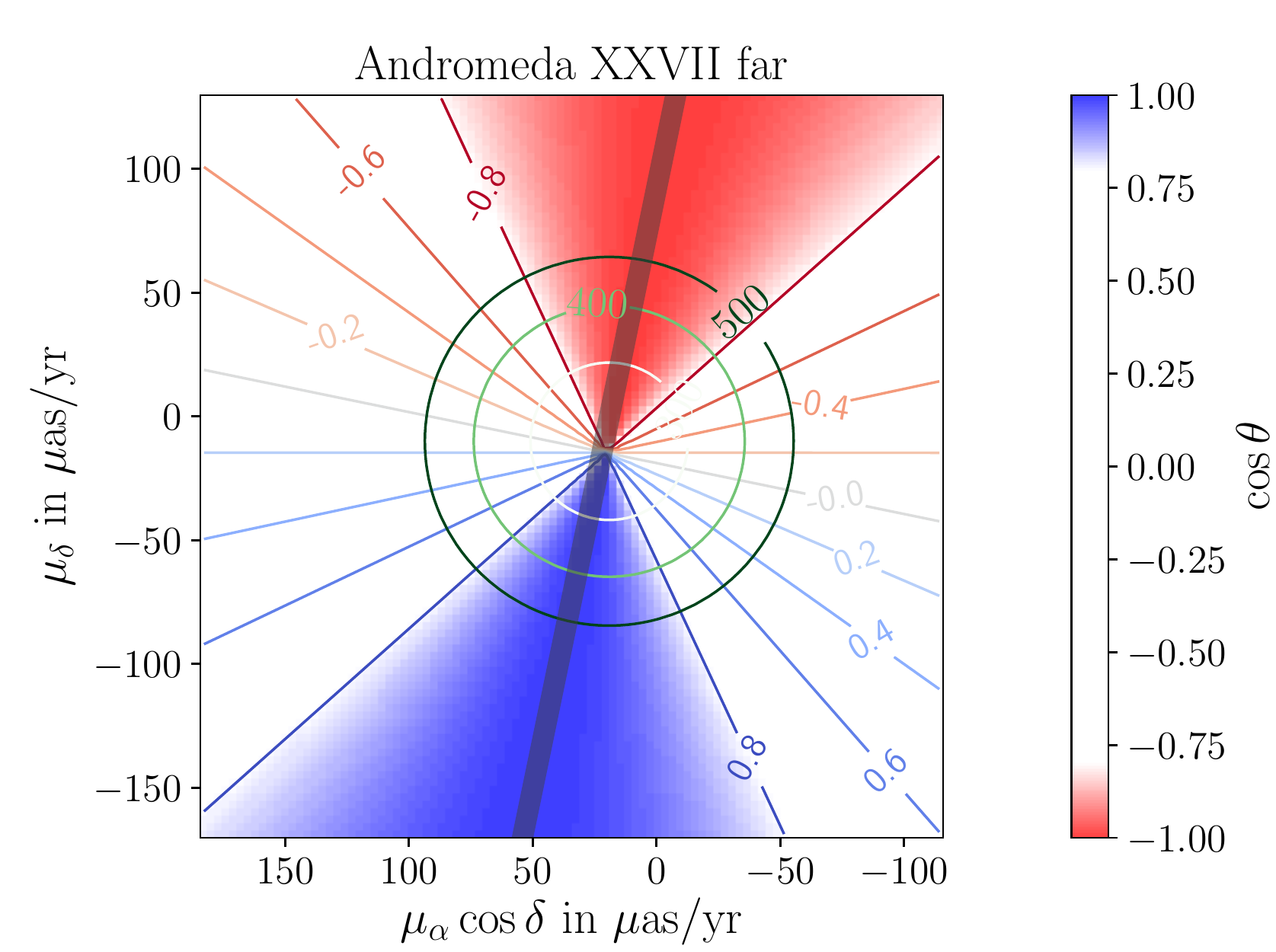}{0.4\textwidth}{(a)}
              \fig{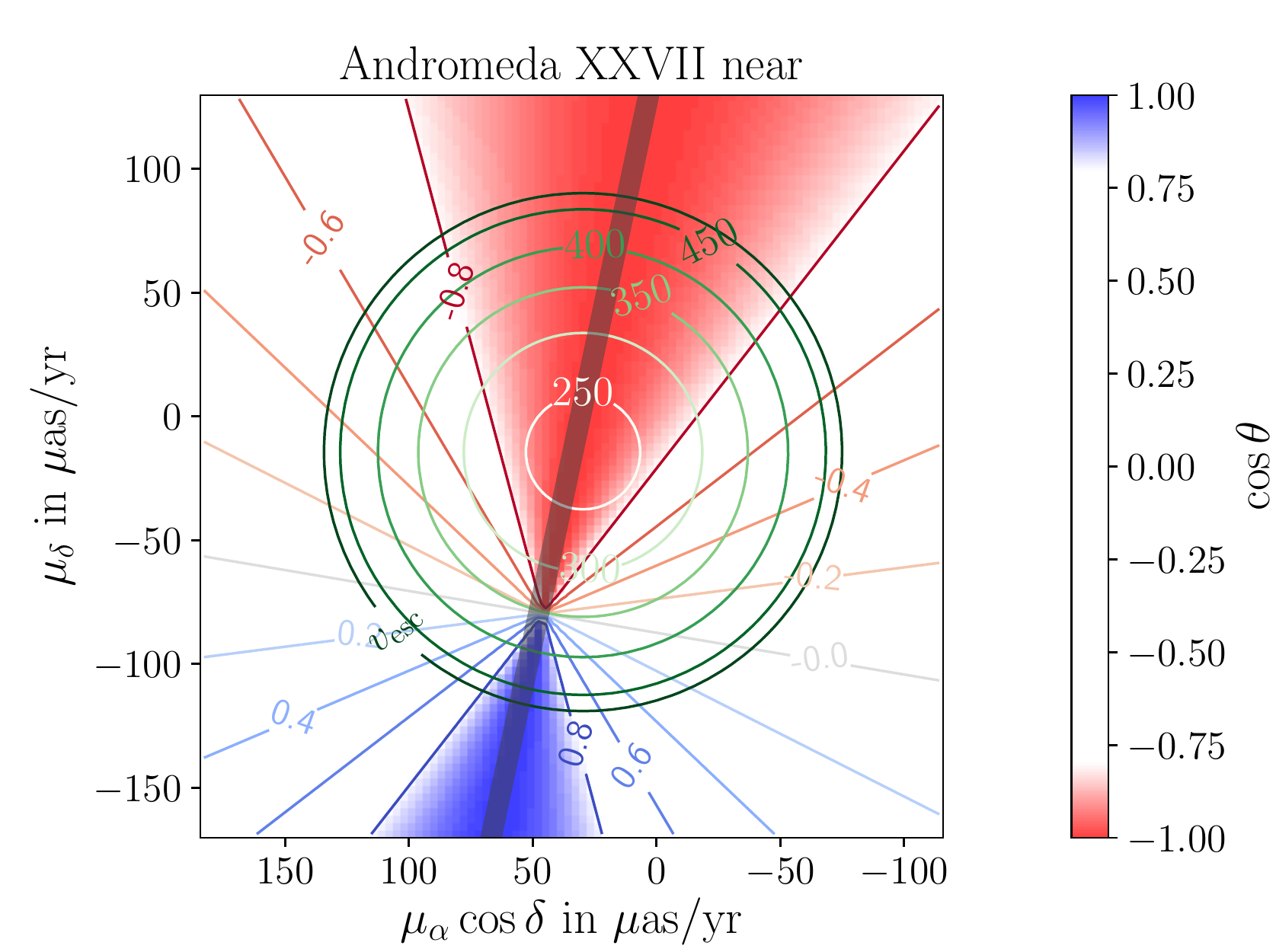}{0.4\textwidth}{(b)}
          }
\caption{Predicted proper motions for Andromeda\,XXVI, for two different assumed distances (see Table \ref{tab:prediction}). Assuming a far distance from M31 implies that the dwarf galaxy is unbound for any proper motion, but could co- or counter-orbit along the GPoA. For a smaller assumed distance (placing the dwarf closer to M31), its line-of-sight velocity makes it more likely to be counter-orbiting. Same style as Fig. \ref{fig:PMprediction}.
\label{fig:PMprediction_AndXXVII}}
\end{figure*}

Following the method described in Section \ref{sec:predicted}, we present predicted proper motions for all 15 on-plane satellites of M31 in Table \ref{tab:prediction}, as well as in Figures \ref{fig:PMprediction_others}. Figure \ref{fig:PMprediction_AndXXVII} compares the predicted proper motions for the two adopted distances of Andromeda\,XXVII. These all assume the HST+Sat. proper motion for M31.

\vspace{8mm}

\begin{deluxetable*}{lccccccc}
\tablenum{5}
\tablecaption{Predicted PMs of M31's on-plane satellite galaxies, assuming them to orbit along the GPoA \label{tab:prediction}}
\tablewidth{0pt}
\tablehead{
\colhead{Name} & \colhead{$l$} & \colhead{$b$} & \colhead{$r_{\sun}$} & \colhead{$v_{\mathrm{los}}$} & \colhead{$\theta_{\mathrm{GPoA}}$} & 
\colhead{$[\mu_{\alpha} \cos \delta, \mu_{\delta}]_{\mathrm{co}}$} & \colhead{$[\mu_{\alpha} \cos \delta, \mu_{\delta}]_{\mathrm{counter}}$} \\
\colhead{} & \colhead{($^\circ$)} & \colhead{($^\circ$)} & \colhead{(kpc)} & \colhead{($\mathrm{km\,s}^{-1}$)} & \colhead{($^\circ$)} & \colhead{($\mathrm{{\mu}as\,yr}^{-1}$)} & \colhead{($\mathrm{{\mu}as\,yr}^{-1}$)}
}
\startdata
Andromeda I &  121.7 & -24.8 & 728 & $ -376 $ & $ 2.5 $ & $[+41,~-47]\ \mathrm{to}\ [+14,~+117]$ & $[+41,~-47]\ \mathrm{to}\ [+60,~-170]$\\
Andromeda III &  119.4 & -26.3 & 724 & $ -344 $ & $ 22.1 $ & $[+30,~-46]\ \mathrm{to}\ [+42,~+102]$ & $[+30,~-46]\ \mathrm{to}\ [+21,~-158]$\\
Andromeda IX &  123.2 & -19.7 & 600 & $ -209 $ & $ 5.0 $ & $[+43,~-29]\ \mathrm{to}\ [+26,~+78]$ & $[+43,~-28]\ \mathrm{to}\ [+58,~-127]$\\
Andromeda XI &  121.7 & -29.1 & 762 & $ -420 $ & $ 5.9 $ & $[+46,~-139]\ \mathrm{to}\ [+27,~+76]$ & not possible (unbound) \\
Andromeda XII &  122.0 & -28.5 & 929 & $ -558 $ & $ 0.3 $ & $[+25,~+12]\ \mathrm{to}\ [+38,~-57]$ & $[+25,~+12]\ \mathrm{to}\ [+26,~+7]$\\
Andromeda XIII &  123.0 & -29.9 & 759 & $ -185^\ast $ & $ 1.8 $ & not possible (unbound) & $[+15,~+69]\ \mathrm{to}\ [+56,~-136]$\\
Andromeda XIV &  123.0 & -33.2 & 794 & $ -481 $ & $ 1.7 $ & $[+27,~+31]\ \mathrm{to}\ [+47,~-102]$ & not possible (unbound) \\
Andromeda XVI &  124.9 & -30.5 & 476 & $ -367 $ & $ 0.5 $ & $[+67,~-72]\ \mathrm{to}\ [+49,~+39]$ & $[+67,~-72]\ \mathrm{to}\ [+80,~-148]$\\
Andromeda XVII &  120.2 & -18.5 & 728 & $ -252 $ & $ 6.2 $ & $[+40,~-29]\ \mathrm{to}\ [-4,~+125]$ & $[+40,~-29]\ \mathrm{to}\ [+77,~-157]$\\
Andromeda XXV &  119.2 & -15.9 & 735 & $ -108 $ & $ 9.6 $ & $[+81,~-121]\ \mathrm{to}\ [-5,~+92]$ & not possible (unbound) \\
Andromeda XXVI &  118.1 & -14.7 & 755 & $ -262 $ & $ 14.3 $ & $[+72,~-83]\ \mathrm{to}\ [-22,~+91]$ & $[+72,~-83]\ \mathrm{to}\ [+85,~-108]$\\
Andromeda XXVII far &  120.4 & -17.4 & 1253 & $ -540^\ast $ & $ 1.8 $ & not possible (unbound) & not possible (unbound) \\
Andromeda XXVII near &  120.4 & -17.4 & 828 & $ -540^\ast $ & $ 0.1 $ & $[+47,~-78]\ \mathrm{to}\ [+55,~-115]$ & $[+47,~-78]\ \mathrm{to}\ [+12,~+90]$\\
Andromeda XXX &  120.5 & -13.2 & 682 & $ -140 $ & $ 1.3 $ & $[+47,~-76]\ \mathrm{to}\ [+17,~+82]$ & $[+47,~-76]\ \mathrm{to}\ [+52,~-100]$\\
NGC 147 &  119.8 & -14.3 & 711 & $ -193 $ & $ 2.4 $ & $[+49,~-65]\ \mathrm{to}\ [+8,~+95]$ & $[+49,~-65]\ \mathrm{to}\ [+62,~-114]$\\
NGC 185 &  120.8 & -14.5 & 619 & $ -204 $ & $ 0.5 $ & $[+44,~-34]\ \mathrm{to}\ [+21,~+84]$ & $[+44,~-34]\ \mathrm{to}\ [+59,~-109]$\\
\enddata
\tablecomments{
$l$, $b$: Heliocentric position of the M31 satellite galaxies in Galactic longitude and latitude.
$r_{\sun}$: heliocentric distance, adopted from the most-likely values reported in \citet{2012ApJ...758...11C}, except for Andromeda XXVII for which the additional {\it near} entry also includes a closer distance as reported by \citet{ 2011ApJ...732...76R}.
$v_{\mathrm{los}}$: heliocentric line-of-sight velocity. Satellites which do not follow the common kinematic trend (i.e. suspected to be counter-orbiting) are marked with a $^\ast$.
$\theta_{\mathrm{GPoA}}$: The angle between the satellite galaxy position as seen from the M31 centre and the GPoA, which is the smallest-possible inclination between the orbital plane of the satellite galaxy and the GPoA.
$[\mu_{\alpha} \cos \delta, \mu_{\delta}]_{\mathrm{co}}$: predicted range in PMs assuming the satellite galaxy to co-orbit with respect to the rotational sense of the GPoA inferred from the dominant line-of-sight velocity trend of the on-plane satellites.
$[\mu_{\alpha} \cos \delta, \mu_{\delta}]_{\mathrm{counter}}$: predicted range in PMs assuming the satellite galaxy to be counter-orbiting.
}
\end{deluxetable*}

\section{Effect of Inclination on Line-of-Sight Velocity Coherence}
\label{sect:appendix2}

\begin{figure*}
\gridline{\fig{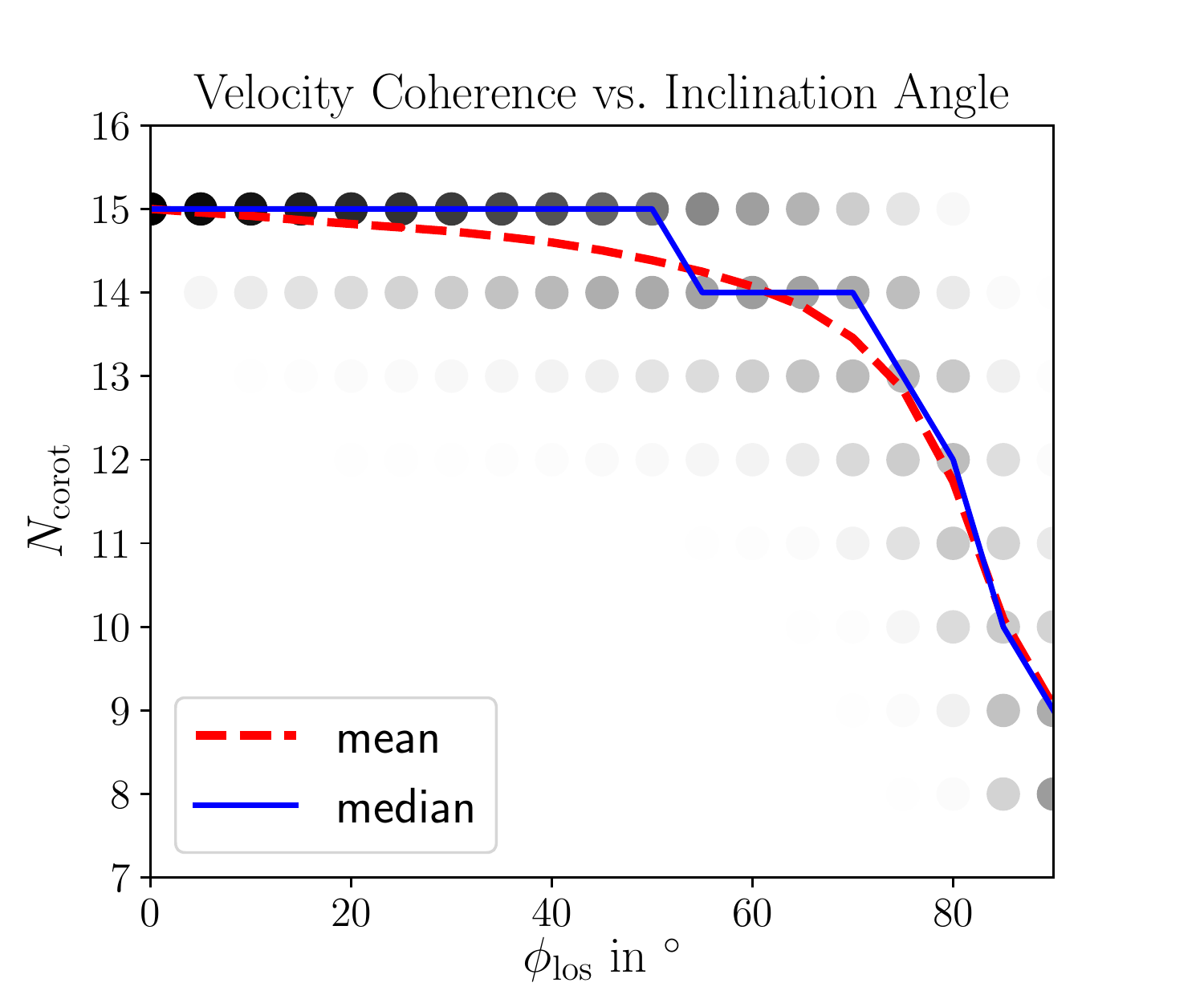}{0.45\textwidth}{(a)}
\fig{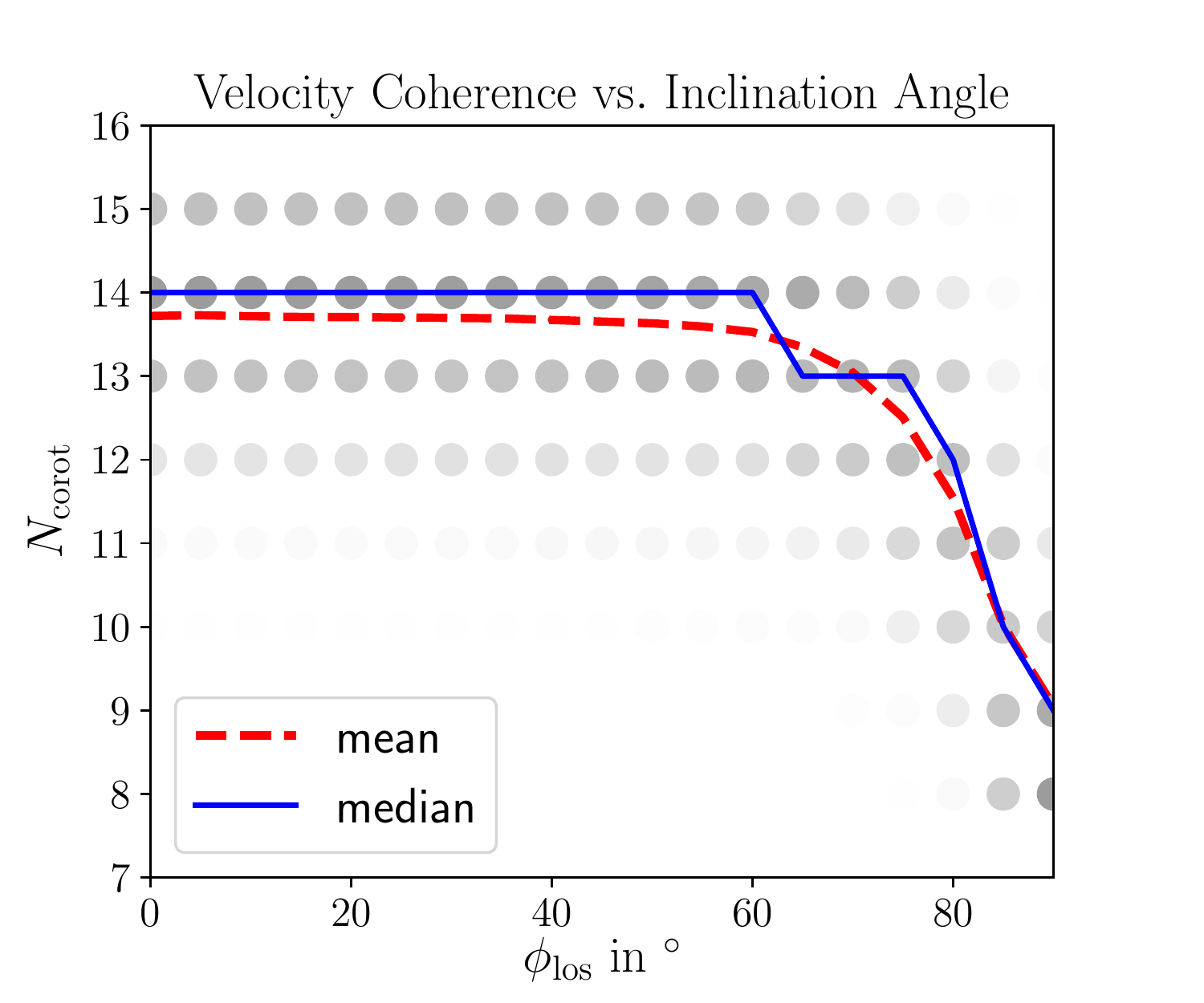}{0.45\textwidth}{(b)}
          }
\caption{
Strengths of line-of-sight velocity coherence $N_\mathrm{corot}$\ for 1000 artificial satellite galaxy planes consisting of 15 co-orbiting satellites each, as mock-observed from different angles $\phi_\mathrm{los}$. In the left panel (a), the mock satellites are given a purely tangential in-plane velocity, in the right panel (b) their velocity direction are drawn from a flat distribution allowing up to 50\% of their kinetic energy to be within the radial direction towards or away from the host. The shade of the grey dots represents the fraction of mock systems at a given view direction which results in the corresponding $N_\mathrm{corot}$, where black is 100\% and white is 0\%.
\label{fig:incl_vs_los}}
\end{figure*}
 
If a satellite galaxy plane is seen perfectly edge-on, it is intuitive to infer that a strong line-of-sight velocity trend could be due to in-plane co-orbitation of the satellites. However, if satellite planes are observed with increasing degrees of inclination (i.e. larger $\phi_\mathrm{los}$), an increasing amount of the velocity component perpendicular to the plane contributes to the measured $v_\mathrm{los}$. At $\phi_\mathrm{los} = 90^\circ$, the observable velocities are entirely measured perpendicular to the plane, and thus no rotational signature along plane could be inferred. This raises the question of up to which inclination (or $\phi_\mathrm{los}$) we can expect to reliably interpret a line-of-sight velocity coherence as a satellite plane's rotation signature, assuming that we are indeed dealing with stable, rotating plane of satellite galaxies.

To test this, we set up 1000 realizations of artificial satellite planes consisting of 15 co-orbiting satellites each. We then mock-observed each of them under different inclinations ranging from $\phi_\mathrm{los} = 0^\circ$\ to $90^\circ$\ in steps of $5^\circ$. For each inclination, $N_\mathrm{corot}$\ is determined solely based on the $v_\mathrm{los}$\ accessible to this view direction. The parameters of our artificial planes are motivated by \citet{2017MNRAS.465..641F}, who studied under which conditions a plane of satellites around M31 could be dynamically stable. We follow their general plane setup parameters. Specifically, our satellites are drawn from a flat distribution in radius between 50 and 250\,kpc, with a perpendicular scatter of $\pm 5\,\mathrm{kpc}$, and have in-plane velocities between 100 and 200\,$\mathrm{km\,s}^{-1}$. For the velocity dispersion perpendicular to the plane we adopt a value of $20\,\mathrm{km\,s}^{-1}$, which \citet{2017MNRAS.465..641F} find to be the largest to still result in about half of the original on-plane satellites remaining on the original satellite plane over several Gyr.

In Fig. \ref{fig:incl_vs_los} we plot the measure of the line-of-sight velocity coherence $N_\mathrm{corot}$\ for these 1000 artificial satellite galaxy planes, as mock-observed from different angles $\phi_\mathrm{los}$. In the left panel (a), the mock satellites are given a purely tangential in-plane velocity. However, for more eccentric orbits the velocity is not necessarily purely tangential, so we also show in the right panel (b) a case where the in-plane velocity direction is drawn from a flat distribution that allows up to 50\% of their kinetic energy to be within the radial direction towards or away from the host. In the latter case, even for the edge-on view direction in some cases $v_\mathrm{los}$\ does not correspond to a satellite's orbital direction, such that $N_\mathrm{corot} < 15$. However, in both cases $v_\mathrm{los}$\ is indicative of the strong underlying correlation in orbital directions irrespective of the exact angle under which the artificial satellite planes are observed until very high inclinations ($\phi_\mathrm{los} > 60^\circ$). Thus, only for large angles between the line-of-sight and the plane of the satellites does the random velocity perpendicular to the satellite plane of $20\,\mathrm{km\,s}^{-1}$\ become dominant.  As shown in \ref{fig:cumulativeplaneproperties}, the planes identified in the simulations analysed in this work predominantly have small $\phi_\mathrm{los} < 30^\circ$, and only extend up to $\phi_\mathrm{los} \approx 60^\circ$. We are thus confident that if a stable, rotating satellite plane (following the definition and findings of \citealt{2017MNRAS.465..641F}) is identified among the simulations, its line-of-sight velocity components alone allows to infer the corresponding rotational signature.


\bibliography{M31_PMs}{}
\bibliographystyle{aasjournal}


\end{document}